\documentclass[leqno, a4paper, 11pt]{article}
\usepackage{geometry}[margin=1in]
\usepackage{graphicx,subcaption}
\usepackage{amsmath}
\usepackage{amsfonts}
\usepackage{algorithm}
\usepackage{algpseudocode}
\usepackage{comment}
\usepackage{nicefrac}
\DeclareMathOperator*{\argmin}{arg\,min}
\usepackage{amsthm}
\usepackage{bbm}

\usepackage{setspace}
\usepackage{natbib}
\usepackage{xcolor}
\usepackage{amssymb}
\usepackage{booktabs}
\usepackage{tabularx}
\usepackage[shortlabels]{enumitem}
\usepackage[colorlinks = true,
            linkcolor = blue,
            urlcolor  = blue,
            citecolor = blue,
            anchorcolor = blue]{hyperref}

\newcommand\independent{\protect\mathpalette{\protect\independenT}{\perp}}
\def\independenT#1#2{\mathrel{\rlap{$#1#2$}\mkern2mu{#1#2}}}

\newtheorem{rem}{Remark}
\newtheorem{assumption}{Assumption}

\newtheorem{proposition}{Proposition}
\newtheorem{definition}{Definition}

\newtheorem{theorem}{Theorem}
\newtheorem{example}{Example}

\newtheorem{lemma}{Lemma}

\newcommand\citeapos[2]{\citeauthor{#1}'s (\citeyear{#1}) }

\def\independenT#1#2{\mathrel{\rlap{$#1#2$}\mkern2mu{#1#2}}}

\title{Linear estimation of global average treatment effects\footnote{We thank Eric Auerbach, Graham Elliott, Robert Gonzalez, Michael Leung, Richard Mansfield, Andres Santos, Davide Viviano, Vedant Vohra, Kaspar W\"{u}thrich, Karen Yan, and participants at the CEPR Development Symposium, the Microeconometrics Class of 2024 Conference, and the Economic geography Session of the Urban Economics Association Annual Meetings for very helpful comments on early drafts of this paper, and Aditi Acharya, Thomas
Brailey, Wilson King, and Erick Rosas Lopez for exceptional research assistance. We are also grateful to Karthik Muralidharan and Sandip Sukhtankar, whose collaboration on the design of several large-scale experiments helped motivate this question, and to the Bill and Melinda Gates Foundation for financial support for those projects.}}
\author{ Stefan Faridani\footnote{Georgia Institute of Technology. sfaridani6@gatech.edu.} \and Paul Niehaus\footnote{UC San Diego, JPAL, NBER, and BREAD. pniehaus@ucsd.edu.}}
\date{\today}

\begin{document}
\begin{titlepage}

\maketitle
\thispagestyle{empty}
\begin{abstract}

We study estimation of and inference for the average causal effect of treating every member of a population, as opposed to none, using an experiment that treats only some. Considering settings where spillovers can occur between any pair of units and decay slowly with distance, we derive the minimax rate over all linear estimators and experimental designs, which increases with the spatial rate of spillover decay. This rate of convergence can be achieved using an inverse probability weighting estimator when randomization clusters are large, but not otherwise. If the causal model is linear, however, an OLS-based estimator converges faster than IPW when clusters are small and is consistent even under unit-level randomization. We provide methods for radius selection and inference and apply these to the cash transfer experiment studied by \citet{EggeretalGE}, obtaining a 22\% larger estimated effect on consumption. 

\end{abstract}
\end{titlepage}

\clearpage
\pagenumbering{arabic}
\onehalfspacing
\section{Introduction}

Economists often study situations in which some units received a given treatment in order to estimate the consequences of treating more of them. For example, \citeapos{worms} eevaluation of deworming medication in a sample of 2,300 schoolchildren shaped subsequent debate over whether to administer similar medication to all schoolchildren in Kenya and beyond. A key quantity for policy-making in such settings is the global average treatment effect (GATE), or average causal effect of treating all (relevant) units.

Learning the GATE is challenging, even when treatment is randomly assigned, because it compares two counterfactual states which are never directly observed for \emph{any} unit: one in which it, and all other units, are treated, and another in which none are. The econometrician must therefore extrapolate from, for example, the observed outcome for a unit with many treated neighbors to its counterfactual outcome had everyone been treated. Such extrapolation requires an assumption controlling interference or ``spillovers" between units.

The predominant approach to dealing with spillovers has been to assume they are local via exposure mappings  \citep{Manski13} which imply that treating one unit affects only a small number of neighbors, such as those within the same classroom or village. (The Stable Unit Treatment Value Assumption is an extreme example, implying that \emph{no} other units are affected.) But such assumptions are sometimes in uncomfortable tension with physical or economic logic. Parasitic worms, as \citet{worms} emphasized, need not stay neatly confined within a classroom. Economic models often imply that everyone's behavior affects everyone else to some degree through prices, strategic interactions, and other mechanisms. Recent, large-scale field experiments have illustrated these kinds of far-reaching general-equilibrium effects \citep{MuralidharanNiehaus2017}. Reforming a public workfare program in India, for example, affected market wages and employment, land rents, and firm entry, and these effects did not stop at sub-district boundaries \citep{Muralidharanetal2021GE}. Causal channels like these are difficult to reconcile with strictly local spillovers. 

In this paper we study the estimation of the GATE via bounds on the magnitude of spillovers, rather than exposure mappings. Following \citet{Leung22}, the specific assumption we work with is that interference decays with at least a power $\gamma$ of Euclidean distance.\footnote{An earlier version of the paper \citep{arxivprevious} extends the results to ``nearly Euclidean'' spaces in which the relevant notion of distance can be asymmetric (e.g. if pollution travels from upwind to downwind sites) and need not exactly satisfy the triangle inequality (e.g. trade flows in a gravity model), or where the researcher is uncertain what notion of distance is most relevant and so uses the minimum of several proper metrics.} This yields a statistical framework better-suited to many economic settings: endogenous economic interactions decay with a power of distance in the gravity \citep{AllenArkolakisTakahashiUniversal} and market access \citep{DH16} traditions, for example. We allow for $d$-dimensional spaces, arbitrary experimental designs, and for any estimator that can be written as a linear function of observed outcomes---a class which includes those commonly used in both applied and theoretical work.

In this setting we show that no linear estimator, regardless of the experimental design, can be guaranteed to converge to the GATE faster than $n^{-\frac{1}{2 + d/ \gamma}}$. The fact that this is slower than the parametric rate $n^{-\frac{1}{2}}$ illustrates the intrinsic difficulty of estimating the GATE, while the fact that it is faster for large $\gamma$ illustrates the intuitive idea that the problem becomes less difficult when the population is economically less interconnected. The optimal rate can be achieved using inverse probability weighting (IPW) when the experimental design is cluster-randomized with clusters that grow with the population at the right rate. The key idea is to prevent the probability of any one unit having \emph{all} of its neighbors (within some growing radius) treated or untreated from decaying to zero.

These results build on and extend \citeauthor{Leung22}'s (\citeyear{Leung22}) important recent contribution, which characterizes the optimal rate when using a Horvitz-Thompson estimator in $\mathbb{R}^2$ and demonstrates that it is achievable using clusters that take the form of growing squares if $\gamma > 2$. We extend this result to $\mathbb{R}^d$, optimize over all linear estimators as well as all experimental designs, and allow $\gamma \in (0,\infty)$. The key to the last generalization in particular is showing that, while a phase change in the estimation error does occur at $\gamma = d$, the problematic terms this introduces shrink with the radius of the estimator and are never larger than the asymptotic bias. The phase change therefore does not affect the rate of convergence or disrupt inference.

Achieving the optimal rate does require assigning treatment in clusters that are ``large'' in the sense that they grow rapidly alongside the experiment. If instead they grow at a slower-than-optimal rate, then units with fully treated (``saturated'') or untreated (``dissaturated'') neighborhoods become too rare.\footnote{One can think of this as analogous to the ``limited overlap'' problem where individual propensity scores are close to zero or one \citep[e.g.,][]{imbens_limited_overlap}.} The IPW estimator can then capture spillovers only in small neighborhoods around each unit, so that it cannot achieve the optimal rate. Moreover, its bias will tend to dominate its variance, posing challenges for inference. These issues are exacerbated under ``random saturation'' designs, where spatially grouped units can have different treatment assignments; here the probability that even \emph{one} neighborhood is fully saturated or dissaturated vanishes, so that IPW cannot be guaranteed to consistently estimate the GATE at any polynomial rate. In short, when free to choose, researchers prioritizing the GATE should use large clusters. 

In practice, however, researchers have often used small-cluster or even unit-level randomization \citep{MuralidharanNiehaus2017}. There are several reasons for this. First, tradeoffs between estimands can arise; there may be scientific motives for separately identifying direct and indirect treatment effects, for example, as well as the policy-relevant GATE. Second, researchers may face practical constraints, such as a requirement to randomize treatment across pre-existing administrative units. For such scenarios we consider positing a linear causal model: potential outcomes are linear functions of treatment indicators. This assumption compensates for the scarcity of (dis)saturated neighborhoods in small-cluster designs by allowing learning about them from neighborhoods that are less than fully (dis)saturated. We view it as implicit in prominent applied papers which have reported results from linear regressions of outcomes on (among other things) the mean treatment status of nearby units \citep[e.g.,][]{worms, EggeretalGE, Muralidharanetal2021GE}.

The linear causal model does not change the optimal rate, which can still be achieved using IPW if the rate of cluster growth is unconstrained. If, on the other hand, clusters must be ``small'' in the sense above, then linearity lets the researcher improve on the IPW rate using an OLS estimator. This estimator is formed by regressing outcomes on the share of treated clusters intersecting a neighborhood around each unit---a regression related to those in the applied studies above, but distinct in that (among other things) they regress on the share of nearby treated \emph{units}. These regressions concentrate around a weighted average of the spillover effects, rather than the unweighted GATE.\footnote{The weights are non-negative, and in that sense less problematic than those which arise in two-way fixed effect designs \citep{TWFEChaise,GOODMANBACON2021254}. However, since the weights depend on both the treated unit and the affected unit, they have no welfarist interpretation.} The OLS estimator proposed here is consistent for the GATE, and remains so even if the true data-generating process is non-linear under weak conditions on the experimental design. Overall these results imply the following tradeoff for practitioners: if feasible, a large-cluster experimental design paired with the IPW estimator is recommended for the GATE. If large clusters are not available, and if it is plausible that the underlying DGP is not too non-linear, then an OLS approach may be attractive.\footnote{The OLS-based approach is also amenable to the introduction of prior information the researcher may have about the pattern of economic interactions, which can be incorporated in a TSLS setup. We omit this approach here for brevity but study it in an earlier working paper version \citep{arxivprevious}.} 

For small-cluster designs there remains the practical problem of choosing the ``radius" of an OLS estimator in a given, finite population. Should the researcher regress outcomes on the average treatment status of clusters within 1km, 10km, or some other distance? Choosing the radius that minimizes mean square error (MSE) is infeasible, since bias is always unknown. We propose a minimax procedure that is asymptotically guaranteed to select the radius that minimizes the worst-case MSE over a credible set of potential outcomes. This procedure does not require the use of outcome data, but only that the researcher specify a conservative lower bound on the rate of spatial decay of spillovers. This in turn facilitates inference. Inference faces two challenges: the bias is unknown, and the variance includes the usual unidentified component attributable to heterogeneous treatment effects in fixed populations. Undersmoothing is a natural consequence of this method because the researcher never knows the exact rate of spatial spillover decay and therefore will always use a conservative lower bound. This results in a wider radius than if they knew the exact rate for sure, addressing the first challenge. We provide a new asymptotically valid bound on the unidentified variance term to address the second. 

We apply these methods to data from the large-scale cash transfer experiment in rural Kenya studied by \cite{EggeretalGE} (henceforth, EHMNW). We first estimate the GATE on annualized household consumption. Using our OLS estimator with the minimax radius, we obtain an estimate of \$445 ($p < 0.01$), 22\% larger and 65\% more precise than when we adopt the radius used by EHMNW. We then simulate IPW and OLS performance holding geography and observed features of the outcome distribution fixed, but under a linear causal model and two different experimental designs: the one actually used by EHMNW, and a counterfactual one with clusters 4-5 times larger. Under the actual design the MSE of IPW is several multiples that of OLS, while with larger clusters the two estimators perform comparably. We also confirm that OLS confidence intervals provide appropriate coverage at their nominal level.


The primary theoretical contribution of the paper is to show that, broadly speaking, estimating the GATE is more realistic than had previously been understood. Theorists have cast the GATE as an important estimand because it is interpretable even when the researcher is uncertain about the exact channels over which spillovers propagate. For example, in response to \citeapos{SavjeMisspec21} \ characterization of estimands available when exposure mappings are somewhat misspecified, \cite{auerbachcomment} point out that these estimands need not have causal interpretations, to which \cite{michaelcomment} responds that the GATE is one of the few that does, but that it is difficult to estimate in practice. Our results show that the GATE can be rate-optimally estimated using standard IPW estimators under weaker conditions than those in \cite{Leung22}, and that alternative OLS-based estimators converge faster in the presence of common constraints on design. 

This approach complements recent theoretical work that has focused on consistency or on optimal experimental design, leaving open the question of how to jointly choose a design and estimator. \cite{auerbach2023localapproachcausalinference} provide consistency results for a class of estimands including the GATE under very general assumptions, for example, but do not address how choices of design and estimator map into the rate of convergence and what rates are possible. \cite{leung2025crosscluster} and \cite{viviano2024causalclusteringdesigncluster} find optimal experimental designs to estimate the GATE assuming an IPW or difference-in-means estimator will be used. We emphasize the joint choice of estimator and design to achieve the best rate, the ways this is affected by the linearity assumptions on potential outcomes that are implicit in applied work, and the design constraints researchers often face.

An adjacent line of work has focused on estimands other than the GATE, such as the average causal effect of receiving some specific level of neighborhood exposure to treatment \citep{Aronow17,savjearonow, Leung21,auerbach2024regressiondiscontinuitydesignspillovers}, which can be consistently estimated and interpreted even when the exposure mapping is mildly misspecified \citep{Zhang21, SavjeMisspec21}.\footnote{Work in this genre grew out of an earlier literature on peer effects in which outcomes are linear in exogenous and endogenous variables and either the coefficients are known up to scale \citep{Manski93, Lee07, GPandI13, Brom14} or exposure mappings rule out interference from faraway units \citep{volf,Sussman}. Our results under linearity show how one can consistently and rate-optimally estimate global average treatment effects even without such assumptions.} It targets only estimands that can be written in terms of local exposures (and thus not including the GATE). More generally, \cite{Airoldi18} show that there is no consistent estimator of the GATE within such a framework.\footnote{See Section 3.3 of \cite{savjearonow}, for example, for discussion of this point within their framework.}

For applied work we provide a new theoretical foundation and practical guidance for program evaluation in the presence of spillovers. At least since \citet{worms} it has been understood that capturing spillovers can be crucial to getting policy inferences right, and that the decay of spillovers with distance can be exploited to construct estimators that capture them \citep[e.g.][]{worms_at_work_16,EggeretalGE,Muralidharanetal2021GE}. But most studies have limited themselves to estimating the total effect of the experiment actually conducted, rather than the policy-relevant GATE, and done so using estimators and designs for which consistency and optimality results were not available. Our results allow for consistent, rate-optimal estimation of the parameter arguably most relevant to program evaluation, and provide guidance for tuning and inference. In doing so our broader aim is to provide a bridge between the experimental approach to economics---which has rested on the assumption that a ``pure control group'' exists---and economic theory, which typically implies that one does not.

\section{Setup}
\label{sec:setup}

Notational conventions are as follows. For sequences $A_n,B_n$ we say $A_n \lesssim B_n$ if $|A_n/B_n| =\mathcal{O}_p(1)$. $A_n\sim B_n$ means that both $A_n\lesssim B_n$ and $B_n\lesssim A_n$. A subscripted vector ${b}_i$ denotes the $i$th element while a subscripted matrix $G_i$ denotes the $i$th row. 

A researcher observes a finite population $\mathcal{N}_n$ of $n$ units and randomly assigns  each unit $i$ a binary treatment $D_i$. Let $\mathbf{D}\equiv (D_i)_{i\in\mathcal{N}_n}$ denote the full $n\times 1$ vector of treatments. We call the probability distribution of $\mathbf{D}$ the {\it experimental design}. The potential outcome for unit $i$ is the scalar-valued function $Y_i(\mathbf{d})$ where the argument $\mathbf{d}$ is a particular $n\times 1$ vector of counterfactual treatment assignments. The observed outcome is the random variable $Y_i(\mathbf{D})$. Since our estimand is the causal effect of treating an entire population and not a random sample, we model potential outcomes $Y_i(\cdot)$ as deterministic functions and the treatment assignment $\mathbf{D}$ as random (and correspondingly will study asymptotics with respect to a sequence of growing finite populations).\footnote{Fixing potential outcomes is standard in this literature, see for example \citet{Aronow17, Sussman, Leung21, Leung22}.} We remain largely agnostic about the origins of the potential outcomes $Y_i(\cdot)$, which could be the realization of some unspecified spatial process, and in particular we impose no exposure mappings.

Our first assumption states that potential outcomes are uniformly bounded. While less restrictive assumptions could be sufficient, Assumption \ref{assum:boundedoutcomes} eases exposition and is standard \citep[e.g.][]{Aronow17,MANTA2022109331,Leung22,leung2025crosscluster}.

\begin{assumption}{\bf Bounded Outcomes}\label{assum:boundedoutcomes}

There exists a constant $\overline{Y}<\infty$ such that: $$\sup_{n\in\mathbb{N}}\max_{i\in\mathcal{N}_n}\max_{\mathbf{d}\in \{0,1\}^n}|Y_i(\mathbf{d})|<\overline{Y}$$
\end{assumption}

Our estimand of interest is the average causal effect of assigning {\it every} unit in the entire population to treatment vs assigning {\it no} units at all to treatment.  This is called the {\it global average treatment effect}  (GATE) because it sums both the own-effects and spillover effects of treating all units (within some relevant class).
\begin{definition}{\bf Global Average Treatment Effect  (GATE)}\label{def:AGE}
$$\theta_n\equiv \frac{1}{n}\sum_{i=1}^n\left(Y_{i}(\mathbf{1})-Y_{i}(\mathbf{0})\right) $$
\end{definition}
The GATE is often the estimand most relevant for policy decisions; a policymaker using RCT data to decide whether to scale up a program to the control group, for example, should ideally base this decision on the GATE rather than the direct treatment-on-treated effect. Note that we assume the researcher observes outcomes for the entire population for whom she wishes to estimate the GATE.\footnote{Appendix \ref{sec:subset} extends to settings where only a subset of the entire population is eligible for treatment.} Results concerning the limiting properties of our estimators will require a notion of a growing population, and our estimand $\theta_n$ can change as the population grows. Later assumptions will characterize the conditions this sequence of populations must follow.

To solve the core identification challenge, we consider restrictions on the magnitude of spillovers, as opposed to less economically plausible restrictions on their sparsity. Specifically, we consider a restriction on the (maximum) impact that faraway treatments have on each unit. To that end, we define $\rho(i,j)$ as the distance metric between units $i$ and $j$, which in turn induces a natural notion of neighborhoods as $s$-balls:
\begin{equation}\label{eq:neighborhoods}
    \mathcal{N}(i,s) \equiv \{j \in \mathcal{N}_n : \rho(i,j) \leq s\}
\end{equation}
Here $s > 0$ parameterizes the radius of the neighborhood. This will allow us to study estimators that make use of increasingly large neighborhoods as $n$ grows. 

For expositional clarity our focus here will be on the purely spatial case where units are located in $\mathbb{R}^d$ and $\rho$ is the Euclidean distance. This captures many economic applications: economic interactions decay geometrically with spatial distance in the gravity \citep{AllenArkolakisTakahashiUniversal} and market access \citep{DH16} traditions, for example. Space need not be the only factor governing interactions between units, provided that a spatial bound on the rate of decay holds. In an earlier version of the paper \citep{arxivprevious} we showed more explicitly that the main results still hold if we relax the triangle inequality or symmetry property of distance,\footnote{Another case we discuss in that draft in which the more general results are useful is when the researcher is uncertain about which of several potential channels for spillovers matter. For example, training small business owners might lead to knowledge spillovers for which the social network defines the relevant notion of distance, businesses stealing for which physical distance and product substitutability between firms define distance, input demand effects for which the buyer-supplier network defines distance, and so on. We thank David McKenzie for suggesting this example. See \citet{SavjeMisspec21}, \citet{auerbachcomment}, and \citet{michaelcomment} for further discussion of the desirability of robustness to misspecification.} while in ongoing work we are applying similar methods to a ``small-world'' network setting in which results change more fundamentally.\footnote{In this setting neighborhoods instead have the ``small-world'' feature that their size grows exponentially rather than polynomially with their radius. Our proof techniques extend naturally to this setting and yield an analogous (but slower) bound on the rate of convergence.}

In addition to locating units in Euclidean space, we require that the population is not too densely or sparsely distributed. Assumption \ref{assum:euclidean_space} says that (a) the population can fit inside a ball with volume proportional to $n$ and (b) every pair of units is at least distance $\rho_0>0$ apart.

\begin{assumption}{\bf Spatial Density}\label{assum:euclidean_space}
      
    \begin{enumerate}[label=(\alph*)]
        \item  There is a universal constant $R>0$ such that: $\max_{i,j\in \mathcal{N}_n} \rho(i,j) \leq Rn^{1/d}$ for all $n\in\mathbb{N}$
        \item  There is a universal constant $\rho_0>0$ such that: $\min_{i\neq j} \rho(i,j)  \geq\rho_0$ for all $n\in \mathbb{N}$
    \end{enumerate}
\end{assumption}

\begin{example}\label{ex:space_egger}\normalfont
    A common example of Assumption \ref{assum:euclidean_space} in practice is when units are households and $\rho(\cdot,\cdot)$ is the geographic distance between them in meters. For example, in EHMNW the study is composed of 5,419 households that fit inside a circle of diameter $560$ kilometers and are all at least $0.89$ meters apart from one another. Such a setting satisfies Assumption \ref{assum:euclidean_space} with $d=2$, $\rho_0=0.89\text{m}$, and $R=7618\text{m}$ if we view it as one of a sequence of growing populations which keep the same spatial density---i.e., that are situated on a map that grows larger but not denser.
\end{example}


\begin{rem}\label{ex:space_violation}\normalfont
    Condition (a) in Assumption \ref{assum:euclidean_space} assumes that units all fit inside a  $d$-dimensional ball with volume that scales with $n$. This implies that all $d$ dimensions ``matter" asymptotically. It would be violated by a sequence of populations  in $\mathbb{R}^2$ located along a road, for example, and which therefore grew longer but not wider with $n$. Since only the $x$-axis location matters, these units are really best understood as existing in $\mathbb{R}^1$. A practitioner studying such a situation should model the population in a lower-dimensional space.
\end{rem}

Our next assumption captures the idea of spatial decay in the spillover effects. Specifically, it states that changing the treatment assignments of any subset of units more than distance $s$ away from unit $i$ changes the potential outcome of unit $i$ by an amount upper-bounded by a function that decays with $s^{-\gamma}$. The researcher need not know the exact fashion in which spillovers decay; they need only accept that spillovers are upper-bounded in this way. Assumption \ref{assum:ANI} is identical to Assumption 3 of \cite{Leung22} and \cite{leung2025crosscluster} except that, as we discuss further below, we require only $\gamma > 0$ rather than $\gamma > d$. This also implies that our inference results will relax fast spatial decay requirements common in the spatial central limit theorem (CLT) literature \citep{JENISH_NED}. 

\begin{assumption}\label{assum:ANI}
{\bf Decay of Spillovers}

There exist constants \(c> 0 \) and \( \gamma > 0 \) such that for all \( s > 0 \),
\[
\sup_{n \in \mathbb{N}} \max_{i \in \mathcal{N}_n} \max \left\{\left| Y_i(\mathbf{d}) - Y_i(\mathbf{d}') \right|\: :\: d_j=d'_j \: \forall j \in \mathcal{N}(i,s)\right\} \leq c\min\{s^{-\gamma}, 1\}
\]
\end{assumption}

Assumption \ref{assum:ANI} bounds spillovers deterministically, as is standard in the literature on the GATE and related estimands \citep{Leung22,leung2025crosscluster}.\footnote{Other papers bound long-range spillovers deterministically in a more subtle way via exposure mappings that apply directly to the potential outcome function itself rather than to expectations over it \citep{Manski13,Aronow17}.} For readers familiar with the spatial econometrics literatures it may be helpful to contrast it with the Near Epoch Dependence   used by (for example) \cite{JENISH_NED}, which instead bound spillovers using expectations:
\begin{equation}\label{eq:ANI_expectation}
    \mathbb{E}\left[\left|Y_i(\mathbf{D})-\mathbb{E}[Y_i(\mathbf{D})|\mathcal{F}_{i,s}]\right|\right]\leq  c\min\{s^{-\gamma},1\}
\end{equation}

\noindent where $\mathcal{F}_{i,s}$ is the $\sigma$-algebra generated by all the treatments within distance $s$ of unit $i$ and the expectation is taken over the distribution of treatment assignments $\mathbf{D}$. While closely related to Assumption \ref{assum:ANI}, condition (\ref{eq:ANI_expectation}) is not suitable for GATE estimation, for two reasons. First, (\ref{eq:ANI_expectation}) is not primitive in this setting because it depends on the experimental design chosen by the researcher, over which we want to allow for optimization. Second, and more importantly, (\ref{eq:ANI_expectation}) is not strong enough to identify the GATE, as Remark \ref{rem:gate_identify_ned} below illustrates. This is why we (and others, such as \cite{Leung22,leung2025crosscluster}) bound spillovers deterministically using Assumption \ref{assum:ANI}. 

\begin{rem}\label{rem:gate_identify_ned}\normalfont
    To see why condition (\ref{eq:ANI_expectation}) does not identify the GATE, suppose an experimental design that randomly treats exactly half of all units, and then consider the following two examples of potential outcomes both of which satisfy condition (\ref{eq:ANI_expectation}). First, consider potential outcomes where $Y_i(\mathbf{D})=1$ regardless of $\mathbf{D}$, i.e. where treatment is irrelevant and the GATE is 0. These produce realized outcomes $Y_i = 1 \:\forall i$ for any design. Next, consider alternative potential outcomes $Y_i'(\mathbf{D}) = \max_{j\in \mathcal{N}_n} {D}_j$. These might result, for example, from a social learning process where information reaches the entire population provided one member receives it. In this case the GATE is 1, but realized outcomes are again $Y_i = 1 \:\forall i$ for any design that treats at least one unit. The realized outcomes thus do not uniquely identify the GATE. This happens because sensitivity to the treatment status of a small number of units breaks the connection between $\mathbb{E}[Y_i(\mathbf{D})|\mathcal{F}_{i,s}]$ and $Y_i(\mathbf{1}),Y_i({\mathbf{0}})$ even when $s$ is large. Equation (\ref{eq:ANI_expectation}) does not rule this out because for both sets of potential outcomes, $\mathbb{E}[Y_i(\mathbf{D})|\mathcal{F}_{i,s}]=Y_i(\mathbf{D})=1 \:\forall i$ almost surely. Assumption \ref{assum:ANI} does rule this out by connecting $Y_i(\mathbf{D})$ directly to $Y_i(\mathbf{1}),Y_i({\mathbf{0}})$.
    \end{rem}

\section{Rate-Optimal Linear Estimation}
\label{sec:minimax}

We study estimators that are linear, i.e. that take the form of weighted averages of observed outcomes $Y_i$ where the weights $\omega_{in}(\mathbf{D})$ are arbitrary functions of the full vector of treatment assignments $\mathbf{D}$. Weights can be specific to units $i$ and may change with $n$, though we will sometimes suppress the latter subscript for brevity. Such estimators can thus be written as:
\begin{equation}\label{eq:linear_estimators}
 \widehat{\theta}_n \equiv \frac{1}{n}\sum_{i=1}^n Y_i\omega_{in}\left(\mathbf{D}\right)  
\end{equation}
The class of linear estimators includes the IPW estimators common in theoretical analysis \citep{Aronow17,Leung22} as well as (after conditioning on covariates) the regression-based estimators common in applied work.

\begin{theorem}{\bf Optimal Rate}
\label{thm:optimalrate}

  Let $\mathcal{N}_n$ be any sequence of populations satisfying Assumption \ref{assum:euclidean_space}.  Let $\mathcal{Y}_n$ be the set of all potential outcomes $\mathbf{Y}$ that satisfy Assumptions \ref{assum:boundedoutcomes} and \ref{assum:ANI}.  Let $\widehat{\theta}_n$ be any estimator of the form of Equation (\ref{eq:linear_estimators}). Fix any sequence of experimental designs. Then for any $\delta >0$ and any sequence $b_n\to\infty$,
 $$\limsup_{n\to\infty}\sup_{\mathbf{Y}\in\mathcal{Y}_n} \mathbb{P}\left[b_nn^{\frac{1}{2+d/\gamma}}|\widehat{\theta}_n-\theta_n|>\delta\right] >0$$ 

Proof: Section \ref{proof:thm:optimalrate}
\end{theorem}

Theorem \ref{thm:optimalrate} states that no estimator taking the form of a weighted mean of outcomes can be guaranteed to converge to the GATE at a rate strictly faster than $n^{-\frac{1}{2+ d /\gamma}}$ for all potential outcomes satisfying Assumptions \ref{assum:boundedoutcomes} and \ref{assum:ANI}. The following sketch illustrates the workings of the proof. Posit a sequence $a_n\to \infty$ such that $a_n(\widehat{\theta}_n-\theta_n)={o}_p(1)$ for any set of potential outcomes satisfying our assumptions. Let $\mathcal{J}_n\subset \mathcal{N}_n$ be a minimal subset of the population such that every unit $i\in \mathcal{N}_n$ is within distance $a_n^{1/\gamma}$ of some $j\in \mathcal{J}_n$ called $j(i)$; call these the ``focal units.'' Now let the potential outcomes be $Y_i(\mathbf{D})=a_n^{-1}D_{j(i)}$, so that they are entirely determined by the treatment statuses of the focal units. (Note that these satisfy Assumption \ref{assum:ANI}, as no focal unit $j(i)$ is too far from $i$.\footnote{Although the GATE here is decreasing, it is not $o\left(a_n^{-1}\right)$ and therefore must be meaningfully estimated by hypothesis.}) The effective sample size is consequently $|\mathcal{J}_n|$. And this quantity is bounded: by Assumption \ref{assum:euclidean_space} the population is contained by a ball of radius $n^{1/d}$ in $\mathbb{R}^d$, and it can be shown that the number of balls of radius $a_n^{1/\gamma}$ required to cover a larger ball of radius $n^{1/d}$ is just $\mathcal{O}(na_n^{-d/\gamma})$, i.e. the volume of the whole space divided by the volume of the balls used to cover it. We thus have: $a_n \lesssim |\mathcal{J}_n|^{1/2}\lesssim {n^{1/2}a_n^{-d/(2\gamma)}}$. This relation is solved for $a_n$ to yield the final result $a_n\lesssim n^{\frac{1}{2+d/\gamma}}$.\footnote{This sketch is partial; another set of potential outcomes is needed to resolve special cases, e.g. estimators where the weights $\omega_{i,n}(\mathbf{D})$ converge rapidly to zero.}

The rates allowed by Theorem \ref{thm:optimalrate} are slower than the parametric rate $n^{-\frac{1}{2}}$ because $d/\gamma >0$. The result thus illustrates a sense in which GATE estimation in the presence of global spillovers is a more difficult problem than estimation with localized spillovers, where the parametric rate can be guaranteed \citep{Aronow17}. Theorem \ref{thm:optimalrate} also links the spatial rate of decay of the spillover effects with the rate of convergence of linear estimators. The economic principle this suggests is that estimators based on Assumption \ref{assum:ANI} will tend to converge faster in less integrated economies, where distance is an important constraint on economic interactions, than in highly integrated ones where all units are ``close'' to each other.

Theorem \ref{thm:optimalrate} extends the main results of \cite{Leung22} and \cite{leung2025crosscluster} in two ways. First, it extends the rate limit result to the case $\gamma < d$. We will show below that this ``speed limit" is still achievable in that case. Small values of $\gamma$ will also play a key role in our approach to bandwidth selection and inference because they can serve as conservative lower bounds which facilitate undersmoothing. Second, Theorem \ref{thm:optimalrate} finds the optimal rate jointly over all experimental designs as well as all linear estimators. This is relevant given that applied work has used linear estimators other than IPW.

\subsection{Achieving the optimal rate}

We next show that the optimal rate is achievable using an experimental design from a class of ``Scaling Clusters'' designs we define, and a suitable estimator.

Consider \emph{cluster-randomized} experimental designs in which the researcher partitions the population $\mathcal{N}_n$ into $m_n$ mutually exclusive clusters $\{\mathcal{P}_c\}_{c=1}^{m_n}$ and assigns treatment cluster-by-cluster, treating each cluster independently of the rest with probability $p$, so that all units within a cluster share the same treatment status, denoted $W_c \in \{0,1\}$. 
\begin{equation}\label{eq:design}
    \{D_i\}_{i\in \mathcal{P}_c} = W_c \overset{\text{i.i.d.}}{\sim}\text{Bernoulli}(p) 
\end{equation}
We focus on such designs throughout the main text for expositional clarity, and following the literature \citep{Leung22,leung2025crosscluster}. In practice researchers often use variants of this design where \emph{exactly} fraction $p$ of clusters are treated within predefined strata (``stratified randomization''). Our empirical application (EHMNW) is one prominent example, among many others.\footnote{For instance, \citet{worms} (which randomized schools stratified by administrative zone), \citet{CaseyGlennersterMiguel2012reshaping} (villages stratified by ward), or \citet{Alatasetal2016selftargeting} (villages stratified by groups of one or more subdistricts).} We defer analysis of such cases to Appendix \ref{sec:correlated_designs} as the arguments are substantially more involved but the results are substantively the same. In particular we show that the same rates of convergence can be achieved when the design is spatially stratified provided clusters still grow, and provide consistency and CLT results for the OLS estimator applied to designs such as that in our empirical application.\footnote{Notice the absence of an overlap (or ``positivity'') assumption bounding away from zero the probability that any given unit experiences full exposure to treatment or to control conditions in some relevant neighborhood. Such restrictions on the design are often used when the estimand compares the effects of ``local" exposures. In these settings it (helpfully) ensures that the probability that the researcher observes both units whose relevant neighborhoods are fully treated and units whose relevant neighborhoods are fully untreated goes to one, and that the researcher can construct consistent estimators by comparing these units. In our setting, however, the relevant neighborhood is the entire population, so we can \emph{never} observe both such units at once: if one unit is exposed to universal treatment, then there cannot be a comparison unit that was exposed to universal non-treatment. As a result a positivity assumption would hurt rather than help us.}

\begin{figure}[tp]
    \caption{ \label{fig:ScalingClusters_temp} Scaling Clusters (left) vs Counterexample (right)}

    \begin{subfigure}[t]{0.48\textwidth}
        \centering
        \includegraphics[width=\textwidth]{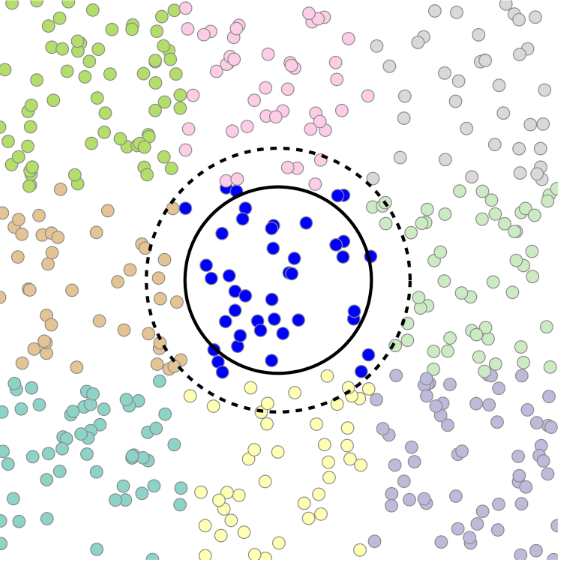}
       
    \end{subfigure}
    \hfill
    \begin{subfigure}[t]{0.48\textwidth}
        \centering
        \includegraphics[width=\textwidth]{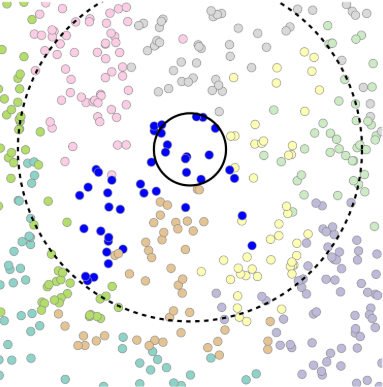}
    \end{subfigure}

    {\footnotesize This figure illustrates the distinction between an experimental design in which clusters are similar to neighborhoods (left), so that circles bounded by and bounding each cluster have similar radii, and one in which they are not (right), so that the two radii are very different. Each point represents a distinct unit, and common colors indicate units assigned to common clusters.} 
\end{figure}

To consistently estimate the GATE we will need clusters to be of a shape and size such that the number of clusters within any $s$-neighborhood does not grow too rapidly with $s$. To achieve this, consider a class of {\it Scaling Clusters} designs where the clusters are shaped like neighborhoods and grow with $n$. The rate of cluster growth is governed by a non-decreasing sequence of positive numbers $\{g_n\}_{n=1}^\infty \subseteq \mathbb{R}^+$ as follows:

\begin{definition}{\bf Scaling Clusters Designs}\label{def:scaling_clusters}

Consider any non-decreasing sequence $\{g_n\}_{n=1}^\infty \subseteq \mathbb{R}^+$. A cluster-randomized design is called a ``Scaling Clusters" design when there is a $K_{S} > 1$ such that  for every cluster $\mathcal{P}_c$ there is a corresponding unit $q_c \in \mathcal{N}_n$ such that $ \mathcal{N}(q_c,g_n) \subseteq \mathcal{P}_c \subseteq \mathcal{N}(q_c, g_nK_{S}) $.
\end{definition}
A Scaling Clusters design is thus one where treatment clusters both contain and are contained by neighborhoods that grow in proportion to $g_n$. Assumption \ref{assum:euclidean_space} guarantees that the number of clusters $m_n$ satisfies $m_n\sim\frac{n}{g_n^d}$, i.e. it scales with the ratio of volume of the whole space to the volumes of the individual clusters. This requirement is loose, ruling out only clusters that become progressively more ``gerrymandered" as the population grows. One simple example of a Scaling Clusters design is a grid of expanding squares in $\mathbb{R}^2$. Another is provided by \citet{leung2025crosscluster}, whose Algorithm SA.1.1 constructs clusters using unsupervised learning to optimize the rate at which MSE shrinks; these satisfy Definition \ref{def:scaling_clusters} by his Theorem 1.\footnote{While \citet{leung2025crosscluster} optimizes the rate at which MSE shrinks and not its finite-sample level, we view his proposal as intuitively appealing as it aims to minimize the number of units near cluster boundaries which should lower estimator bias. Alternatively, we conjecture that it may be possible to extend the minimax risk procedure for radius selection we develop in Section \ref{sec:minimax_radius} to also select among a small finite set of candidate designs.} Given a specific experimental design in a fixed population, one can think of it as satisfying Scaling Clusters if the largest circle contained within each cluster is close to the smallest circle containing it. Figure \ref{fig:ScalingClusters_temp} illustrates this: in the left panel the solid (contained) and dashed (containing) circles are close together because the clusters are ``convex enough," while in the right panel the solid and dashed circles are far apart because the clusters are very ``gerrymandered." 
 
The researcher can achieve the optimal rate without further assumptions using an inverse probability weighting (IPW) estimator. This estimator is conveniently tractable and standard in theoretical work on network effects \citep[see for example][]{Aronow17, Sussman, Leung21, Leung22,leung2025crosscluster}. To define it, fix any sequence of neighborhood radii $\{\kappa_n\}_{n\in \mathbb{N}} \subset \mathbb{R}^+$ chosen by the researcher which grow to infinity with the sample size. Define the random variables $T_{i1},T_{i0}$ as indicators that all of $i$'s neighbors within distance $\kappa_n$ are treated or untreated: 
\begin{equation}
    \label{eq:define_Ti1Ti0}
    T_{i1} \equiv \prod_{j \in \mathcal{N}(i,\kappa_n) } D_j,\qquad
T_{i0} \equiv \prod_{j \in \mathcal{N}(i,\kappa_n) } (1-D_j)
\end{equation}

The IPW estimator compares units with saturated $\kappa_n$-neighborhoods to units with dissaturated neighborhoods, following the recommendation in \cite{HiranoImbensIPW} to take a difference between means rather than the simple Horvitz-Thompson procedure.
\begin{equation}
       \widehat{\theta}_{n,\kappa_n}^{IPW} \equiv \frac{1}{n}\sum_{i=1}^n Y_i\left(\frac{T_{i1}}{\mathbb{P}\left[T_{i1}=1\right]\frac{1}{n}\sum_{k=1}^n \frac{T_{k1}}{\mathbb{P}\left[T_{k1}=1\right]}} - \frac{T_{i0}}{\mathbb{P}\left[T_{i0}=1\right]\frac{1}{n}\sum_{k=1}^n \frac{T_{k0}}{\mathbb{P}\left[T_{k0}=1\right]}}\right)
\end{equation}

For the IPW estimator to converge at a polynomial rate it is necessary to control the probabilities in the denominators of the weights, which is equivalent to upper-bounding the number of clusters $\phi(i,\kappa_n)$ that each neighborhood $\mathcal{N}(i,\kappa_n)$ intersects. Lemma \ref{lem:phi} in Appendix \ref{app:lemmas} shows that a Scaling Clusters design  makes $\phi(i,\kappa_n)$ scale with $(\kappa_n^d+g_n^d)/g_n^d$. In order for IPW to be consistent, the radius of the estimator thus cannot outpace the cluster size. Theorem \ref{thm:rateipw} then shows that the IPW estimator paired with a Scaling Clusters design achieves the optimal rate from Theorem \ref{thm:optimalrate} when the radius and cluster size grow at the appropriate rate.

\begin{theorem}{\bf Inverse Probability Weighting}\label{thm:rateipw}

If Assumptions \ref{assum:boundedoutcomes}--\ref{assum:ANI} hold and the researcher uses a Scaling Clusters Design with $g_n\sim \kappa_n\to \infty$, then
$$ \left|\widehat{\theta}_{n,\kappa_n}^{IPW}-\theta_n\right| \lesssim \sqrt{\frac{\kappa_n^d}{n}}+\kappa_n^{-\gamma}$$

When $g_n\sim\kappa_n \sim n^{\frac{1}{2\gamma+d}}$, the optimal rate is achieved:
$$\sqrt{\frac{\kappa_n^d}{n}}+\kappa_n^{-\gamma}\lesssim  n^{-\frac{1}{2+d/\gamma}}$$

Proof: Section \ref{proof:thm:rateipw}.
\end{theorem}
\noindent In contrast to the IPW estimators for local effects studied in \cite{Aronow17}, the IPW estimator of the GATE here converges at a subparametric rate. Extending results for the Horvitz-Thompson estimator and square cluster design in $\mathbb{R}^2$ studied by \cite{Leung22}, Theorem \ref{thm:rateipw} applies to $\mathbb{R}^d$ for any $d$ and (in conjunction with Theorem \ref{thm:optimalrate}) shows that pairing any Scaling Clusters design with an IPW estimator is rate-optimal over all pairs of experimental designs and linear estimators when the clusters and radius grow at the right rates.

Most importantly, Theorem \ref{thm:rateipw} allows for any $\gamma > 0$, relaxing the requirement that $\gamma > d$ from \cite{Leung21,leung2025crosscluster}. A phase change does occur at $\gamma = d$, adding a possibly non-normal stochastic term to the estimation error arising from sums of long-range spillovers. However, we show that even though the ``bad" stochastic term can converge more slowly than $n^{-1/2}$, it is controlled by the radius of the estimator and in fact shrinks with $\kappa_n^{-\gamma}$ (see Step 2 of the proof in Section \ref{proof:thm:rateipw}). Since this is the same rate, for any $\gamma > 0$, as the bias, the overall rate of convergence is unaffected by the phase change. Inference will also be unaffected when the researcher undersmooths, i.e. selects a $\kappa_n$ large enough that variance dominates bias.\footnote{On the other hand, $\gamma>d$ would be required for many estimands other than the GATE. For example, \cite{JENISH_NED} study inference for a population average estimated by a simple sample mean in a non-experimental setting. When $\gamma < d$ in this setup, the ``bad" stochastic term converges slower than $n^{-1/2}$. Since there is neither a ``radius" nor ``clusters" to grow, there is no way to ensure that this term is dominated by the variance of the sample mean, posing a major problem for inference.}

This positive result regarding IPW is sensitive, however, to the experimental design. Theorem \ref{thm:smallclusters} shows that when treatment clusters grow more slowly than the optimal rate, and barring implausible topologies,\footnote{Specifically, the theorem requires that units are not spatially arranged into dense and well-separated clumps that all fly apart as the population grows. A study that grows across a map covered by evenly-spaced villages meets this requirement, for example.} IPW will converge slowly. Growing $\kappa_n$ faster will not prevent this. Appendix \ref{sec:ipw_clt} shows additionally that in this case IPW's bias will usually dominate its standard error, posing challenges for inference.

\begin{theorem}{\bf IPW when Clusters are Small}\label{thm:smallclusters} 

    Let $\gamma > 0$ and suppose that Assumption \ref{assum:euclidean_space} holds and that the researcher uses a Scaling Clusters design with $g_n \sim n^{q_1}$, where $q_1 < \frac{1}{2\gamma+d}$, and sets $\kappa_n\sim n^{q_2}$ for some $q_2\geq 0$. Assume also that for every sequence $s_n\to \infty$ such that $s_n/n^{1/d} \to 0$, $\liminf_{n\to \infty}\frac{1}{n}\sum_{i=1}^n \mathbf{1}\left\{|\mathcal{N}(i,s_n)\setminus \mathcal{N}(i,s_n/2)|\geq 1\right\}>0$. Then there exist potential outcomes (which, additionally, will satisfy Assumptions \ref{assum:boundedoutcomes}-\ref{assum:linearity} with parameter $\gamma >0$) and some $\delta > 0$ such that:
    $$\limsup_{n\to\infty}\mathbb{P}\left[ n^{q_1\gamma }\left|\widehat{\theta}_{n,\kappa_n}^{IPW}-\theta_n\right|>\delta\right] >0 $$

    Proof: Section \ref{proof:thm:smallclusters}.
\end{theorem}

\begin{rem}\normalfont
\label{rem:random_saturation}
The use here of randomization by clusters, which we define as spatial groupings of units that are all assigned together to treatment or to control (Equation \ref{eq:design}), is also consequential. An alternative approach, sometimes called a ``random saturation design,'' is to group units spatially and then treat each unit within group $g$ with group-specific probability $p_g \in (0,1)$ i.i.d.  where $p_g$ may be random \citep[c.f.][]{Baird}. In Appendix \ref{subsec:random_saturation} we show that with such a design IPW cannot be guaranteed to consistently estimate the GATE at any polynomial rate, let alone the optimal one. This is because the probability that a neighborhood around any unit contains only treated or only control units vanishes as $\kappa_n$ grows.
\end{rem}

\section{Estimation under linearity}
\label{sec:linearity}

We have seen that GATE estimation with IPW requires that randomization clusters be ``large,'' in the sense that their radii grow rapidly. But this contrasts with practice, where many field experiments use small clusters. In their review, for instance, \citet{MuralidharanNiehaus2017} documented a median cluster size of just 26 units, with 34\% of designs not clustered at all. Where the primary purpose of the study is to inform a decision about scale-up, so that the GATE is the relevant estimand, such designs are unlikely to be optimal. But researchers may face constraints: equity considerations or the logistics of treatment delivery may require that the experimental design be based on pre-existing administrative units. This was true, for example, of the large-scale experiments referenced above: schools in \citet{worms}, villages in \citet{EggeretalGE}, and sub-districts in \citet{Muralidharanetal2021GE}. Or they may choose to randomize in small clusters---or even at the individual level---due to a scientific interest in studying direct and indirect effects as well as the GATE.

How should researchers approach GATE estimation in such scenarios? Existing work---while targeting other, related estimands---rarely uses IPW: none of the studies covered by \citet{MuralidharanNiehaus2017}, nor any of the other examples we have provided, do so. Instead, studies that take a spatial approach typically report results from OLS regressions on measures of the average treatment intensity in a surrounding neighborhood.

We next study an analogous approach to GATE estimation. The core idea is that, to learn about the GATE from an experiment with small clusters, we must learn about the outcomes a unit would experience if its neighborhood were fully (dis)saturated by extrapolating from the outcomes of units whose neighborhoods are less than fully (dis)saturated. Assumption \ref{assum:linearity}, which we introduce next, allows for such extrapolation.

\begin{assumption}\label{assum:linearity}

There exist a scalar $\beta_0$, an $n\times 1$ vector of numbers $\mathbf{\epsilon}$, and an $n\times n$ matrix of numbers $A$ such that potential outcomes follow:
    $$ Y_i(\mathbf{D}) = \beta_0 +\sum_{j\in\mathcal{N}_n}A_{ij}D_j +\epsilon_i,\qquad \sum_{i=1}^n \epsilon_i=0$$
\end{assumption}

Regression-based estimators are an intuitive way to take advantage of this posited linear structure, and are common practice in applied work. One could consider various regression-based estimators; we focus here on the simplest, which is to regress outcomes on a single measure of neighborhood treatment intensity.\footnote{Applied papers often regress outcomes on an indicator for own-cluster treatment status as well as a measure of treatment intensity in a neighborhood beyond one's own cluster; more generally, one could regress outcomes on measures of treatment intensity in multiple bands around each unit. In our empirical application we will see that separating out the own-cluster treatment status in the regression does not substantially change the results (see appendix Table \ref{table:rings}).} To that end, define the scalar-valued regressor $X_i$, which is the centered sum of treated clusters within distance $\kappa_n$ of unit $i$ divided by the average number of clusters $\overline{\phi}$ within distance $\kappa_n$ over all units. 
\begin{equation}
\label{eqn:X}
    X_{i} \equiv \frac{1}{\overline{\phi}} \sum_{c=1 }^{m_n}\mathbbm{1}\left\{\mathcal{N}(i,\kappa_n) \cap \mathcal{P}_c\neq \emptyset\right\}(W_c-p)
\end{equation}
Here $\phi(i,\kappa_n)\equiv \sum_{c=1}^{m_n} \mathbbm{1}\left\{\mathcal{P}_{c} \cap \mathcal{N}(i,\kappa_n) \neq \emptyset\right\}$ denotes the number of clusters within $\kappa_n$ of unit $i$, and $\overline{\phi}\equiv \frac{1}{n}\sum_{i=1}^n\phi(i,\kappa_n)$ denotes its mean. Then define the OLS estimator as the following simple linear regression of $Y$ on $X$.
\begin{equation}\label{eq:ols_recommended}
   \widehat{\theta}_{n,\kappa_n}^{OLS} \equiv \frac{\frac{1}{n}\sum_{i=1}^n Y_iX_i - \left(\frac{1}{n}\sum_{i=1}^n Y_i\right)\left(\frac{1}{n}\sum_{i=1}^n X_i\right) }{\frac{1}{n}\sum_{i=1}^n X_i^2 - \left(\frac{1}{n}\sum_{i=1}^n X_i\right)^2}
\end{equation}

\begin{rem}
\label{rem:intensity_measure}
\normalfont
    One difference between the regression in (\ref{eq:ols_recommended}) and those used in recent applications is that it measures neighborhood treatment intensity by the number of neighboring clusters that are treated, not the number of neighboring units. The latter regression is problematic because the number of treated nearby units can contain information about which clusters were treated, generating a nonlinear relationship between outcomes and neighborhood treatment intensity even when Assumption \ref{assum:linearity} holds. Consider the following example. Suppose that unit $i$ has 10 neighbors within distance $\kappa_n$, 3 of which are in its own cluster and the other 7 of which are in a distinct cluster.  Suppose that the causal effect of treating $i$'s own cluster on $i$ is 2, the effect of treating the neighboring cluster on $i$ is $-1$, and that in the absence of either treatment $Y_i=0$. Then the expectation of the outcome conditional on the realized fraction $U_i$ of nearby neighbors who are treated is $\mathbb{E}[Y_i|U_i=1]=1$, $\mathbb{E}[Y_i|U_i=0.7]=-1$, $\mathbb{E}[Y_i|U_i=0.3]=2$, or $\mathbb{E}[Y_i|U_i=0]=0$. Overall, the expectation of $Y_i$ conditional on $U_i$ is thus nonlinear in $U_i$! Worse, the best linear approximation to this conditional expectation function has negative slope even though the GATE equals positive one. This phenomenon has consequences for the point estimate: we show formally in Appendix \ref{sec:ols_unit_regression} that regressing on the share of treated neighboring units yields a weighted average of spillover effects, as opposed to the unweighted GATE.\footnote{\cite{VAZQUEZBARE2022} study a related weighting phenomenon for OLS in a local setting where spillovers are confined to non-overlapping groups of units and estimands are defined in terms of exposure mappings.}
\end{rem}

Our next pair of results demonstrates that imposing Assumption \ref{assum:linearity} does not change the optimal rate, but does allow the OLS estimator to estimate the GATE consistently and to achieve that optimal rate. For the latter result we require that $\frac{\kappa_n^d+g_n^d}{n}\to 0$, which simply says that the number of units in any cluster or $\kappa_n$-neighborhood grows more slowly than $n$. We also impose the mild condition $\kappa_n \to \infty$ simply to shorten the proofs.

\newenvironment{thmbis}[1]
  {\renewcommand{\thetheorem}{\ref{#1}$'$}%
   \addtocounter{theorem}{-1}%
   \begin{theorem}}
  {\end{theorem}}

\begin{thmbis}{thm:optimalrate}{\bf Optimal Rate for Linear DGPs}
\label{cor:minimaxlin}

Requiring that potential outcomes also satisfy Assumption \ref{assum:linearity} does not change the conclusions of Theorem \ref{thm:optimalrate}. Proof: Section \ref{proof:thm:optimalrate}
\end{thmbis}

\begin{theorem}{\bf Consistency of OLS}\label{thm:ols_bstar}

If Assumptions \ref{assum:boundedoutcomes}-\ref{assum:linearity} hold and the researcher uses a Scaling Clusters design with scaling sequence $g_n$ and sets $\frac{\kappa_n^d+g_n^d}{n}\to 0$ and $\kappa_n\to \infty$, then:
\begin{align*}
\left|\widehat{\theta}_{n,\kappa_n}^{OLS}  -{\theta}_n\right|\lesssim \frac{\kappa_n^{d}+g_n^{d}}{g_n^{d/2}n^{1/2}} +\kappa_n^{-\gamma}
\end{align*}

When $g_n\sim \kappa_n \sim n^{\frac{1}{2\gamma +d}}$, then the optimal rate is achieved:
$$ \frac{\kappa_n^{d}+g_n^{d}}{g_n^{d/2}n^{1/2}} +\kappa_n^{-\gamma}\sim n^{-\frac{1}{2+d/\gamma}}$$

If instead $g_n \sim n^{q_1}$ where $0\leq q_1 < \frac{1}{2\gamma+d}$, then setting $\kappa_n\sim n^{\frac{dq_1+1}{2(\gamma+d)}}$ yields a faster rate than IPW can achieve for this design:

$$ \frac{\kappa_n^{d}+g_n^{d}}{g_n^{d/2}n^{1/2}} +\kappa_n^{-\gamma}\sim  n^{-\frac{dq_1+1}{2(1+d/\gamma)}}=o\left(n^{-q_1\gamma}\right) $$

Proof: Section \ref{proof:thm:ols_bstar}.
\end{theorem}

The last statement of Theorem \ref{thm:ols_bstar} is particularly significant in that, unlike Theorem \ref{thm:rateipw}, it requires no lower bound on the rate $g_n$ at which clusters grow. It thus shows that OLS converges faster than IPW when clusters grow slower than the optimal rate, and in fact that OLS provides a consistent estimator even when the design is unclustered, i.e. randomization is at the individual level. (To see this, consider Theorem \ref{thm:ols_bstar} with $g_n = \rho_0/2$, or half the minimum distance between two units from Assumption \ref{assum:euclidean_space}.b, which ensures that each cluster contains just one unit.) Intuitively, OLS is able to solve the small-clusters problem by using the linearity assumption to extrapolate from nearly-(dis)saturated neighborhoods to fully-(dis)saturated ones. (Our variance estimator, on the other hand, will require that clusters grow, but this can be at an arbitrarily slow rate.)

 \begin{rem}\normalfont
     The upper bound on the rate of convergence in the first line of Theorem \ref{thm:ols_bstar} is sharp. It holds exactly if, for example, $Y_i(\mathbf{D})=\sum_{j\neq i} \rho(i,j)^{-(\gamma+d)}D_j\mathbf{1}\left\{\rho(i,j)>\kappa_n\right\}+\epsilon_i$, where $\epsilon_i$ equals one if unit $i$ is in the northern half of the population and negative one otherwise, and units are located on an evenly spaced grid in $\mathbb{R}^d$. See the proof of Theorem \ref{thm:minimax_ols_selection}. 
 \end{rem}


\begin{rem}\normalfont
    If Assumption \ref{assum:linearity} does not hold and potential outcomes are nonlinear functions of treatment, OLS loses its advantage over IPW, incurring a bias term that converges slower than $\kappa_n^{-\gamma}$ for the optimized choice of $\kappa_n$. However, it turns out to still be consistent provided clusters are growing (even if slowly) and mostly convexly shaped. The intuition is that when clusters become large and have ``thin crusts" with low conductance, nonlinearities contribute diminishing bias since most units are surrounded by neighbors of the same treatment status as themselves. Appendix \ref{sec:nonlinear_outcomes} shows and discusses this robustness property. 
\end{rem}

Taken together, these results suggest the following tradeoff for practitioners. If nonlinearities are expected to be economically unimportant, and the design of an experiment requires small clusters for reasons other than GATE estimation, then the linear approach developed here will be attractive. If, on the other hand economic interactions between treatment effects are expected to be important, and large clusters are feasible, then the IPW-based approach developed in the previous section is preferable from the point of view of the GATE.

\section{Radius Selection for OLS}
\label{sec:minimax_radius}


We turn next to the problem of selecting a radius $\kappa_n$ for the OLS estimator given a specific, finite study population and experimental design. Radius selection for IPW is largely determined by the experimental design, since saturated neighborhoods will not occur with reasonable probability if the IPW radius much exceeds the cluster size. But this logic does not apply to OLS, since it does not depend on saturated neighborhoods; nor do the convergence results above select a radius, as they hold provided that $\kappa_n=C_{\kappa} n^{\frac{1}{2\gamma+d}}$  for an arbitrary constant $C_{\kappa}$. In fact, without further structure nearly any choice of radius and clustering can be justified by some DGP satisfying Assumptions  \ref{assum:boundedoutcomes}, \ref{assum:ANI}, and \ref{assum:linearity}. To see this, notice that for any $\kappa_n>0$ and any experimental design, if we set potential outcomes to be $Y_i = \frac{\overline{\phi}c\kappa_n^{-\gamma}}{\max_j \phi(j,\kappa_n)}X_i$, then $\widehat{\theta}_{n,\kappa}^{OLS}=\theta_n$ with probability one. Thus, any choice of radius (and design) is optimal under some circumstances. 

We aim to provide a default selection method that requires minimal contextual knowledge and does not depend on units of measurement. Data from a pilot experiment could of course be used to guide radius selection if available, but typically they are not. Alternatively, a researcher willing to commit to specific values of $c,\gamma$ and to accept bias of up to $B$ could simply choose $\kappa_n=(B/c)^{-1/\gamma}$, but this is demanding; choosing $c$ in particular amounts to choosing the scale of the spillover effects overall---which is essentially what the experimenter wishes to learn in the first place. Instead we will develop a rule that requires only that the researcher place a conservative lower bound $\widetilde{\gamma}$ on the rate of spatial decay $\gamma$. This approach takes advantage of the fact that our convergence results hold for any $\gamma>0$, allowing the bound to be arbitrarily conservative provided only that it is positive.

To define and motivate the approach, we first isolate the leading bias and variance terms in OLS. Proposition \ref{prop:ols_asymptotic_decomp} decomposes OLS into an expectation-zero stochastic sum of regression residuals $[\mathcal{A}]$, a deterministic bias term $[\mathcal{B}]$, and a dominated term.
\begin{proposition}\label{prop:ols_asymptotic_decomp}
Define: $\bar{\theta}_{n,\kappa_n}\equiv  \frac{\sum_{i=1}^n \mathbb{E}[X_iY_i]}{\sum_{i=1}^n\mathbb{E}[X_i^2]}$ and $r_i\equiv X_i\left(\mathbb{E}\left[Y_i|\mathcal{F}_{i,\kappa_n}\right]- \frac{1}{n}\sum_{j=1}^n \mathbb{E}[Y_j] -X_i\bar{\theta}_{n,\kappa_n}\right)$.  Let the conditions of Theorem \ref{thm:ols_bstar} hold.  Then:
\begin{align*}
    \widehat{\theta}_{n,\kappa_n}^{OLS}-\theta_n =    \underbrace{\frac{1}{p(1-p)}\frac{\overline{\phi}}{n}\sum_{i=1}^nr_i }_{[\mathcal{A}]}\quad+\quad\underbrace{\bar{\theta}_{n,\kappa_n}-\theta_n}_{[\mathcal{B}]}\quad +\quad  o_p\left(\frac{\kappa_n^d+g_n^d}{g_n^{d/2}n^{1/2}}\right)
\end{align*}

    Proof: Section \ref{proof:prop:ols_asymptotic_decomp}
\end{proposition}
\noindent To help interpret these terms, recall that $\mathcal{F}_{i,\kappa_n}$ is the $\sigma$-algebra generated by all the treatments within distance $\kappa_n$ of unit $i$. The random variables $\mathbb{E}\left[Y_i|\mathcal{F}_{i,\kappa_n}\right]$ can thus be interpreted as the outcomes after ignoring ``long-range'' spillovers, and so the proposition shows (among other things) that these may affect bias but do not affect the leading variance term, which will be useful for radius selection as well as obtaining a CLT in the next section. 

We would ideally like to directly minimize the sum of the variance of $[\mathcal{A}]$ and the square of the bias $[\mathcal{B}]$, to which we refer henceforth as the risk.\footnote{To be precise, it is equal to the expected MSE modulo the vanishingly rare event in which OLS's denominator is near zero. This approach is standard; see for example the term  $\sigma_n^2$ defined prior to Theorem 4 in \cite{leung2025crosscluster}.} For an OLS estimator $\widehat{\theta}_{n,\kappa_n}^{OLS}$ that uses radius $\kappa_n$, given potential outcomes determined by $A$ and $\epsilon$ as defined in Assumption \ref{assum:linearity}, we write this as 
\begin{definition} For potential outcomes $\mathbf{Y}$ satisfying Assumptions \ref{assum:boundedoutcomes}-\ref{assum:linearity}, define risk as:
    \label{eq:risk}
$$\mathcal{R}^*\left(\kappa_n,\mathbf{Y}\right) \equiv \mathbb{V}([\mathcal{A}])+[\mathcal{B}]^2$$
\end{definition}

\noindent where the variance is taken with respect to random assignments induced by the design $\mathcal{D}$. A larger radius will tend to increase variance and decrease bias. Finding the exact minimizer that balances this tradeoff is not feasible, however, since $A,\epsilon$ are unknown. Instead we will consider the radius that minimizes the worst-case risk over a set $\mathcal{G}_n$ of plausible DGPs. To motivate that set, we first need some insight into how the potential outcomes map into risk asymptotically. Proposition \ref{prop:risk_simplification} isolates the leading terms of the risk:

\begin{proposition}\label{prop:risk_simplification}
If the conditions of Proposition \ref{prop:ols_asymptotic_decomp} hold and $||A||_1\lesssim 1$, then:
\begin{align*}
    \mathcal{R}^*\left(\kappa_n,\mathbf{Y}\right)= &\underbrace{\frac{1}{n^2p(1-p)} \sum_{c=1}^{m_n}\left(\sum_{i=1}^n(p(A_i\mathbf{1}-\mathbf{1}'A\mathbf{1}/n)+\epsilon_i)\mathbbm{1}\left\{\mathcal{N}(i,\kappa_n) \cap \mathcal{P}_{c}\neq \emptyset\right\}\right)^2}_{\text{ Leading term of $\mathbb{V}([\mathcal{A}])$}}\\
    &\quad +\underbrace{\left(\frac{1}{n}\sum_{i=1}^n\sum_{j=1}^nA_{ij}\mathbbm{1}\left\{\mathcal{N}(i,\kappa_n) \cap \mathcal{P}_{c(j)}= \emptyset\right\}\right)^2}_{\text{$[\mathcal{B}]^2$}}+\mathcal{O}\left({\frac{\kappa_n^d+g_n^d}{n}}\right)
\end{align*}

Proof: Section \ref{proof:prop:risk_simplification}.
\end{proposition}

Proposition \ref{prop:risk_simplification} reveals a helpful insight: provided that no single unit's treatment has an unbounded effect on the population ($||A||_1\lesssim 1$), the leading variance term does not depend on all features of the spillover matrix $A$. Instead, it depends asymptotically only on the global treatment effect heterogeneity, i.e. variation in the row sums $A_i\mathbf{1}$. Similarly, the bias depends only on the tail behavior of the rows of $A$. Exploiting these facts, we define below a large set of DGPs $\mathcal{G}_n$ over which it will be straightforward to maximize risk:

\begin{definition}\label{def:Gn}
    Given scalar constants $\sigma_e \geq 0$, and $\tau,\widetilde{\gamma},\widetilde{c},\overline{Y}>0$, define $\mathcal{G}_n$ to  be the set of all potential outcomes $\mathbf{Y}$ such that:
\begin{enumerate}
\item $\mathbf{Y}$ satisfies Assumption \ref{assum:boundedoutcomes} with constant $\overline{Y}$, Assumption \ref{assum:ANI} with $c=\widetilde{c}$ and $\gamma =\widetilde{\gamma}$,  and Assumption with \ref{assum:linearity} for some $n\times n$ matrix $A$ and $n\times 1$ vector $\epsilon$
    \item $|A_{ij}|\leq \widetilde{c} \rho(i,j)^{-d-\widetilde{\gamma}} $ for all $i\neq j$\footnote{In practice, GPS error can meaningfully affect this term via a small number of implausibly tiny values of $\rho(i,j)$. To avoid this, we recommend using the upper bound $\widetilde{c} \max\{\rho(i,j),\overline{m}\}^{-d-\widetilde{\gamma}} $ where $\overline{m}$ is the median distance to the closest neighbor} 
\item $||\epsilon||_\infty \leq \sigma_e$
        \item $\max_{i\in \mathcal{N}_n}|A_i\mathbf{1}- \mathbf{1}'A\mathbf{1}/n|\leq \tau$ 
\end{enumerate}
\end{definition}

\noindent Condition 1 simply says that all DGPs must satisfy the previous assumptions. Condition 2 strengthens Assumption \ref{assum:ANI} from a condition on sums of spillovers to a condition on individual spillovers. Specifically, it stipulates that individual spillovers decay with distance no slower than a classic spatial moving average. Conditions 3 and 4 impose uniform bounds on the magnitudes of the error term and treatment effect heterogeneity, respectively.  This broad class of linear DGPs includes, e.g. the spatial moving average models from \cite{Leung22}.  

 Define the minimax radius ${\kappa}^*_n(\mathcal{G}_n)$ as the  minimizer of the worst-case risk over these DGPs.\footnote{Since increasing $\kappa_n$ by less than the minimum distance between units $\rho_0$ does not necessarily change the risk, minimizers cannot be unique. Let the $\argmin$ here denote the largest minimizer.}
\begin{equation}
    {\kappa}^*_n(\mathcal{G}_n)\equiv \argmin_{\kappa_n \geq 0} \sup_{\mathbf{Y}\in \mathcal{G}_n} \mathcal{R}^*\left(\kappa_n,\mathbf{Y}\right)
\end{equation}

To solve this minimax problem, we will show that the maximized risk can be asymptotically approximated by the deterministic function $ R(\kappa_n,\mathcal{G}_n)$ defined by:
\begin{align*}
    R(\kappa_n,\mathcal{G}_n) \equiv &  \underbrace{\frac{(p\tau+\sigma_e)^2}{n^2p(1-p)}\sum_{c=1}^{m_n}\left(\sum_{i=1}^n\mathbbm{1}\left\{\mathcal{N}(i,\kappa_n) \cap \mathcal{P}_{c}\neq \emptyset\right\}\right)^2}_{\text{ Leading term of worst-case variance}}
    \\&+ \underbrace{\left(\frac{\widetilde{c}}{n}\sum_{i=1}^n\sum_{j=1}^n  \rho(i,j)^{-d-\widetilde{\gamma}}\mathbf{1}\left\{\mathcal{P}_{c(j)} \cap \mathcal{N}(i,\kappa_n)= \emptyset\right\}\right)^2}_{\text{Worst-case bias squared}} 
\end{align*}

Theorem \ref{thm:minimax_ols_selection} establishes that minimaxing $   R(\kappa_n,\mathcal{G}_n)$ is asymptotically equivalent to minimaxing the true risk $\mathcal{R}^*$. This result requires that clusters grow slower than their optimal rate (which is the relevant case since, if they grew optimally or faster, the researcher should use IPW instead).

\begin{theorem}{\bf Minimax Risk of OLS}\label{thm:minimax_ols_selection}

 Let the population satisfy Assumption \ref{assum:euclidean_space}  and let the experimental design be a Scaling Clusters design with $g_n =o\left(n^{\frac{1}{2\widetilde{\gamma}+d}}\right)$ and $\overline{Y} > \tau+\sigma_e$. Then:
 $$ \frac{  \sup_{\mathbf{Y}\in \mathcal{G}_n}\mathcal{R}^*\left(\widehat{\kappa}_n(\mathcal{G}_n) ,\mathbf{Y}\right)}{\sup_{\mathbf{Y}\in \mathcal{G}_n}\mathcal{R}^*\left({\kappa}^*_n(\mathcal{G}_n) ,\mathbf{Y}\right)} \to 1$$

\noindent where $ \widehat{\kappa}_n(\mathcal{G}_n) \equiv \argmin_{\kappa_n\geq 0}R(\kappa_n,\mathcal{G}_n)$.
\vskip 0.1in
Proof: Appendix \ref{proof:thm:minimax_ols_selection}
\end{theorem} 

To use this rule, the researcher must select the five scalar parameters that Definition \ref{def:Gn} uses to define $\mathcal{G}_n$. Since $\overline{Y}$ does not affect the minimax radius so long as it is sufficiently large, the researcher can simply set it to an arbitrary large number and ignore it. For the others we suggest a simple default: setting them to maximize the resulting minimax radius. This is conservative from the point of view of inference, since large $\kappa_n$ leads to less bias, more variance, and thus better coverage of the confidence intervals. 

Doing so involves setting $\sigma_e$ and $\tau$ as small as possible, as this reduces the effect of increasing $\kappa_n$ on the worst-case variance and these parameters do not affect worst-case bias. For $\sigma_e$ this means setting $\sigma_e = 0$, as it affects only Condition 3 of Definition \ref{def:Gn}. Note that this does not cause the suggested radius to diverge because treatment effect heterogeneity can still drive estimator variance. In fact, since the variance term of $R(\kappa_n,\mathcal{G}_n)$ is now proportional to $\tau^2$ and the bias term to $\widetilde{c}^2$, only the relative magnitude of these two parameters (and not their individual values) will determine the minimizing radius of $R(\kappa_n,\mathcal{G}_n)$. Structurally, $\tau$ ought to scale with $\widetilde{c}$ since the latter governs the overall scale of the spillover effects and the former controls their heterogeneity. We suggest setting $\tau= \max_{i\in \mathcal{N}_n}\left|\sum_{j\neq i} \widetilde{c}\rho(i,j)^{-d-\widetilde{\gamma}} - \frac{1}{n}\sum_{k=1}^n\sum_{j=1}^n\widetilde{c}\rho(k,j)^{-d-\widetilde{\gamma}}\right|$, which is the smallest value that does not rule out the spatial moving average model from \cite{Leung22} with spillovers as large as Condition 2 of Definition \ref{def:Gn} allows.  Given these defaults, the value of $\widetilde{c}$ no longer matters, as $R(\kappa_n,\mathcal{G}_n)$ is now proportional to $\widetilde{c}^2$ for all $\kappa_n$. 

This leaves the choice of $\widetilde{\gamma}$. Lower values will tend to yield wider minimax radii. Any value $\widetilde{\gamma}$ below the true value of $\gamma$ will cause the estimator to be undersmoothed, ensuring that confidence intervals cover at nominal rates asymptotically.\footnote{Undersmoothing is guaranteed because minimizing $ R(\kappa_n,\mathcal{G}_n)$ will set the standard deviation to scale with a bias term $\mathcal{O}\left(\kappa_n^{-\widetilde{\gamma}}\right)$ that dominates actual bias $\mathcal{O}\left(\kappa_n^{-\gamma}\right)$.} While in principle this might also yield undesirably large radii, the effect is moderated by the fact that when $\widetilde{\gamma}$ is small the worst-case bias is not only large but also insensitive to small changes in the radius.

Taken as a whole, this approach trades off between bias and variance with relatively little calibration; does not depend on any random quantities or on the units in which distance is measured; and does not interfere with inference. Provided the researcher can commit to a lower bound on $\gamma$, it will undersmooth by picking a radius that grows fast enough to make bias dominated, ensuring asymptotic coverage of confidence intervals in large experiments. And we will see shortly in our empirical application that this radius can be surprisingly small in practice: even when we set $\widetilde{\gamma}=0.01$, the resulting radius is substantially smaller than that originally used by EHMNW.

\section{Inference for OLS}
\label{sec:inference}

This section presents a central limit theorem and variance estimator in order to construct confidence intervals, focusing on the OLS case (which is more involved). Appendix \ref{sec:ipw_clt} provides a CLT for IPW when $\gamma < d$, and we recommend the standard errors from \cite{Leung22,leung2025crosscluster}.  

Proposition \ref{prop:ols_asymptotic_decomp} above decomposed OLS's estimation error into a variance term $[\mathcal{A}]$ with expectation zero, a nonstochastic bias term $[\mathcal{B}]$, and a dominated term. Assumption \ref{assum:ANI} guarantees that the bias is of order: $[\mathcal{B}]\lesssim \kappa_n^{-\gamma}$. We control the asymptotic distribution of the variance term $[\mathcal{A}]$  using the following CLT. 

\begin{theorem}{\bf Central Limit Theorem}\label{thm:clt}

    Suppose the researcher uses a Scaling Clusters design such that  $\frac{\kappa_n^{2d}+g_n^{2d}}{ng_n^d}\to 0$, Assumptions \ref{assum:boundedoutcomes}-\ref{assum:linearity} hold, and $ \liminf_{n\to \infty}\mathbb{V}([\mathcal{A}])\frac{n g_n^{d}}{\kappa_n^{2d}+g_n^{2d}} >0$, then:
    \begin{align*} 
        [\mathcal{A}]/\sqrt{\mathbb{V}\left([\mathcal{A}]\right)} \to_d N(0,1)
    \end{align*}

    Proof: Section \ref{proof:thm:clt}.
\end{theorem}

Theorem \ref{thm:clt} shows that, provided the estimator does not exhibit degenerate variance, any choice of $\kappa_n$ that delivers consistency and grows relative to cluster size will ensure that $[\mathcal{A}]$ is asymptotically normal, no matter how small $\gamma$ or the clusters are. The proof is similar to the CLT in \cite{Leung22}, but permits any $\gamma >0$. 

If we knew  $\mathbb{V}\left([\mathcal{A}]\right)$ then Theorem \ref{thm:clt} would allow us to conduct inference. The challenge is that $\mathbb{V}\left([\mathcal{A}]\right)$ is unidentified. Specifically, the variance of OLS is equal to the sum of an identified component and the unidentified term
\begin{align*}
    H_n &\equiv \frac{1}{n^2}\sum_{i=1}^n\sum_{j=1}^n \Lambda_{ij} \left( \widetilde{A}_i\mathbf{1}-\frac{1}{n}\sum_{k=1}^n\widetilde{A}_k\mathbf{1} \right)\left( \widetilde{A}_j\mathbf{1}-\frac{1}{n}\sum_{k=1}^n\widetilde{A}_k\mathbf{1} \right)\ 
\end{align*}
where $\Lambda$ denotes the adjacency matrix that connects two units if and only if their $\kappa_n$-neighborhoods intersect a common cluster and $\widetilde{A}_{ij}\equiv A_{ij}\mathbf{1}\left\{\mathcal{P}_{c(j)}\cap \mathcal{N}(i,\kappa_n) \neq \emptyset\right\}$ and $c(j)$ denotes the cluster containing unit $j$. $H_n$ is unidentified because we cannot identify the treatment effect of any unit individually---a well-known consequence of unobserved heterogeneous treatment effects in a fixed-population setting (and not of spillovers, or of the choice of estimator, per se). Moreover, $H_n$'s sign is not known, as $\Lambda$ may have both positive and negative eigenvalues.

A few approaches have been proposed to solving the analogous problem in the IPW case. One can always bound the unidentified term using Young's Inequality \citep{Aronow17}, but this can lead to very conservative confidence sets when clusters are large. Alternatively, if potential outcomes are allowed to be random and treatment effect heterogeneity is not too strongly correlated over large enough neighborhoods, the unidentified term vanishes entirely in large samples \citep{Leung22}.\footnote{\cite{leung2025crosscluster} shows that in fixed populations $H_n$ is asymptotically non-negative for IPW when $\kappa_n\sim g_n$. An analogous approach for OLS would not be useful, however, since OLS has a rate advantage only when $\kappa_n/g_n\to \infty$.} To provide a viable alternative for the OLS case, and holding potential outcomes fixed, we provide a sharper bound on $H_n$ than Young's Inequality affords by leveraging Assumption \ref{assum:ANI} and (slow) treatment cluster growth. In the following variance estimator the first term has a form analogous to a traditional heteroskedasticity and autocorrelation consistent (HAC) variance estimator, while the second term bounds $H_n$.
\begin{equation}
\label{eqn:variance_estimator}
    \widehat{\mathbb{V}}\left([\mathcal{A}]\right)=\underbrace{\frac{\frac{1}{n^2}\sum_{i=1}^n\sum_{j=1}^n\Lambda_{ij}\widehat{r}_i\widehat{r}_j}{\left(\frac{1}{n}\sum_{i=1}^n X_i^2 - \left(\frac{1}{n}\sum_{i=1}^n X_i\right)^2\right)^2}}_{\text{``HAC"}}   -  \underbrace{\frac{ \frac{1}{n^2}\sum_{i=1}^n\sum_{j=1}^n (\Lambda_{ij}-Q_{ij})\widehat{q}_i\widehat{q}_j}{\left(\frac{1}{n}\sum_{i=1}^n X_i^2 - \left(\frac{1}{n}\sum_{i=1}^n X_i\right)^2\right)^2}}_{\text{Bound on $H_n$}} 
\end{equation}

\noindent Here $Q$ denotes the block-diagonal matrix that connects two units if and only if they are in the same cluster, and
\begin{align*}
\widehat{r}_i &\equiv X_i(Y_i-\frac{1}{n}\sum_{j=1}^nY_j-\widehat{\theta}_{n,\kappa_n}X_i)\\
\widehat{q}_i &\equiv p(1-p)\frac{\phi_i}{\overline{\phi}^2}\left(\frac{\overline{\phi}}{\phi_i}\left(\frac{D_i}{p}-\frac{1-D_i}{1-p}\right)\left(Y_i-\frac{1}{n}\sum_{j=1}^n Y_j\right)-\widehat{\theta}_{n,\kappa_n}^{OLS}\right)
\end{align*}
Theorem \ref{thm:varest} shows that this is asymptotically conservative for the variance of $[\mathcal{A}]$, the key step being to show that the second term in (\ref{eqn:variance_estimator}) is an asymptotically valid bound on $H_n$. We require several additional mild conditions for this to hold. First, there is the standard requirement that the variance of the estimator not converge too fast. Second, we require $\frac{\kappa_n^{2d}+g_n^{2d}}{n g_n^{d}} \to 0$, which is satisfied whenever Theorem \ref{thm:ols_bstar} guarantees that OLS is consistent. Third, $\kappa_n$ must grow faster than $g_n$. If this were not the case, then IPW would be preferred to OLS because the two estimators would converge at the same rate. Finally, in order for the upper bound on $H_n$ to hold asymptotically, we need the fraction of the population lying in the ``outer crusts" of the clusters to shrink and the clusters to grow in size, but these rates can be arbitrarily slow. 

\begin{theorem}{\bf Variance Estimation}\label{thm:varest}

If Assumptions \ref{assum:boundedoutcomes}-\ref{assum:linearity} hold, $\liminf_{n\to\infty} \frac{n g_n^{d}}{\kappa_n^{2d}+g_n^{2d}}\mathbb{V}\left([\mathcal{A}]\right)>0$ and $\frac{\kappa_n^{d}+g_n^{d}}{n^{1/2}g_n^{d/2}} +\frac{g_n}{\kappa_n}\to 0$, then under a Scaling Clusters Design that satisfies  $\frac{1}{n}\sum_{c=1}^{m_n}|\mathcal{P}_c \setminus \mathcal{N}(q_c,g_n)|\to 0$ and $g_n\to \infty$,:
$$ \frac{\widehat{\mathbb{V}}\left([\mathcal{A}]\right)}{\mathbb{V}\left([\mathcal{A}]\right)} \geq 1+o_p(1)$$

Proof: Section \ref{proof:thm:varest}.
\end{theorem}

When implementing this approach in small samples we recommend two conservative adjustments. First, we follow \cite{leung2025crosscluster} in using the maximum of the ``HAC" term and the usual cluster-robust standard error where $\Lambda$ is replaced with $Q$. This stabilizes the variance estimator, whose variance can otherwise become nontrivial, when the radius is large. Second, we set the second term to zero if it is numerically positive. This prevents rare cases in which the entire variance can become negative due to volatility in the second term when the radius is large. These adjustments preserve the guarantee from Theorem \ref{thm:varest} that standard errors are asymptotically conservative, while improving coverage in simulation.\footnote{Applying these adjustments, the variance estimator we use in our application is
\begin{equation*}
    \widehat{\mathbb{V}}^*\left([\mathcal{A}]\right)=\frac{\max\left\{\frac{1}{n^2}\sum_{i=1}^n\sum_{j=1}^n{\Lambda}_{ij}\widehat{r}_i\widehat{r}_j,\frac{1}{n^2}\sum_{i=1}^n\sum_{j=1}^n{Q}_{ij}\widehat{r}_i\widehat{r}_j\right\}}{\left(\frac{1}{n}\sum_{i=1}^n X_i^2 - \left(\frac{1}{n}\sum_{i=1}^n X_i\right)^2\right)^2}-\min\left\{0, \frac{ \frac{1}{n^2}\sum_{i=1}^n\sum_{j=1}^n ({\Lambda}_{ij}-Q_{ij})\widehat{q}_i\widehat{q}_j}{\left(\frac{1}{n}\sum_{i=1}^n X_i^2 - \left(\frac{1}{n}\sum_{i=1}^n X_i\right)^2\right)^2}\right\}
\end{equation*}}

\section{Application: choosing the OLS radius}\label{sec:application}

A central theme in the theoretical results has been that large clusters are beneficial for GATE estimation, while an OLS-based approach can be appealing when the original design was \emph{not} optimized for this. We turn now to an empirical examination of these issues, re-analyzing data from EHMNW. We first cover here the prerequisite task of selecting a radius for the OLS estimator given the design actually implemented. In Section \ref{sec:simulations} we then compare the performance of IPW and OLS estimators under a counterfactual design with larger clusters.

EHMNW studied a cluster RCT of cash transfers worth USD 1,871 PPP allocated among 5,419 treatment-eligible households in 653 villages situated within 68 geographic strata formed from 84 sublocations in rural Kenya. The design was as follows: first, half of the strata were randomly assigned to ``high saturation" and half to ``low saturation." Second, in high (low) saturation groups, 2/3 (1/3) of villages were randomly assigned to treatment. All eligible households in treated villages were then treated. Given this design we interpret households as units,  villages as clusters, and (groups of) sublocations as strata.\footnote{We account for this stratified structure using the modified regression described in Appendix \ref{sec:correlated_designs} although row 3 of Table \ref{table:rings} shows that this does not change any point estimate by a meaningful amount.} EHMNW's design can be interpreted as satisfying our Scaling Clusters condition (Definition \ref{def:scaling_clusters}) if we imagine that, as the map expanded, they would have expanded strata and assigned treatment within each stratum in growing groups of 2 villages, then 3 villages, and so on. This design was not selected with GATE estimation in mind, however---EHMNW report estimates of the total effect of the intervention as implemented, but not of the GATE---and is in this sense a useful test case for the OLS-based approach.

We use the radius selection procedure from Section \ref{sec:minimax_radius} under the default values of the parameters there.  We set $\widetilde{\gamma} = 0.01$, which is available (since Theorem \ref{thm:ols_bstar} allows $\gamma <d$) but which we view as very conservative. Assuming that some bound $\gamma > \widetilde{\gamma}$ on the rate of spatial decay exists, this ensures our estimator is undersmoothed and that bias adjustment is asymptotically unnecessary for inference. Given this, the minimax method selects a radius of 500m.\footnote{Increasing $\widetilde{\gamma}$ to 0.25, 0.50, 1.50, and 2.00 leads to suggested radii of 500m, 250m,  250m, and 250m respectively.}


\begin{table}[tp]
\begin{center}
\caption{OLS estimates, by estimator radius\label{table:application_results}}
{
\renewcommand{\arraystretch}{1.3}
\newcolumntype{Y}{>{\centering\arraybackslash}X}
\begin{tabularx}{\textwidth}{p{3cm}YYYYYYY}
  \toprule
  & \multicolumn{7}{c}{Radius ($\kappa_n$)} \\
  \cmidrule(lr){2-8}
 & 0m & 250m & 500m & 750m & 1000m & 1500m & 2000m \\  
  \midrule
  \addlinespace[6pt]

  $\widehat{\theta}^{\text{OLS,STRAT}}$ & 306 & 365 & 445 & 391 & 446 & 389 & 366 \\
  & (58) & (74) & (90) & (122) & (131) & (170) & (260) \\ 

\addlinespace[6pt]
\midrule
\addlinespace[6pt]
  $\widehat{\theta}^{\text{OLS,UNITS}}$ & 306 & 340 & 398 & 444 & 488 & 365 & 317 \\ 

\addlinespace[6pt]
\midrule
  $\overline{\phi}$ & 1.00 & 1.70 & 2.90 & 4.50 & 6.30 & 10.60 & 15.70 \\ 
   \bottomrule
\end{tabularx}
}
\end{center}
{\small Estimates of the GATE on annual household consumption expenditure in PPP US dollars using data from \cite{EggeretalGE}. Units are households and clusters are villages. Columns indicate the estimator radius $\kappa_n$ in meters. $\widehat{\theta}^{\text{OLS,STRAT}}$ indicates estimates from our recommended linear regression on the fraction of nearby treated clusters (Equation \ref{eq:ols_recommended}), with an adjustment for the stratified experimental design detailed in Appendix \ref{sec:correlated_designs}. $\widehat{\theta}^{\text{OLS, UNITS}}$ indicates estimates using an alternative, biased regression on the fraction of nearby treated units (see Remark \ref{rem:intensity_measure}). The bottom row reports, as a diagnostic, the mean number $\overline{\phi}$ of villages within the radius of a given unit.}


\end{table}

Table \ref{table:application_results} presents results, focusing for simplicity on one outcome---annualized household per-capita consumption---from among those that EHMNW consider. It reports estimates from the OLS estimator using the selected 500m radius along with others for comparison, including the 2,000m radius used by EHMNW in constructing their (related but distinct) regressors. It also reports for comparison estimates from an OLS estimator $\widehat{\theta}^{OLS,UNITS}$ constructed along the lines discussed in Remark \ref{rem:intensity_measure}, where the regressor is not the treated share of nearby clusters but the treated share of nearby units (the method that EHMNW used to construct their measures of neighborhood exposure).  All estimates use the same inverse-probability sampling weights as in the original study.\footnote{Table \ref{table:rings} provides additional comparisons to regression-based estimators that ignore stratification, e.g. one that includes in the regression a separate indicator for own-cluster treatment status. These estimates differ only modestly from our preferred estimates.}

We draw two main conclusions. First, radius selection matters. While all the estimates point to substantial effects, the largest point estimate within the range we consider (at $\kappa_n = 1000$m) is 46\% larger than the smallest (at $\kappa_n = 0$m). Of particular interest, the estimates at our selected 500m radius is 22\% larger and 65\% more precise than that at the 2,000m radius which EHMNW used based on their priors.\footnote{To be precise, EHMNW prespecified that they would include a 2,000m-neighborhood regressor and possibly additional ``rings'' if this minimized an empirical information criterion.} Second, the consistent estimator $\widehat{\theta}^{OLS,STRAT}$ and the inconsistent estimator $\widehat{\theta}^{OLS,UNITS}$ differ, but only modestly, by at most 14\% of the former.




\section{Simulation: cluster size and estimator performance}
\label{sec:simulations}

We now turn to questions of cluster size and estimator performance. To that end, this section continues to use the EHMNW data but replaces the outcomes $Y_i$ and treatment assignments $D_i$ with simulated values. Knowing these, we know the true GATE and can compare the performance of IPW and OLS, both under the randomization design that was actually carried out and under a counterfactual one with larger clusters.

The data-generating process for the simulation is as follows. The population again consists of the 5,419 eligible households with their geographic locations, village membership, and sublocation membership unchanged. We impose Assumption \ref{assum:linearity} by setting $Y_i(\mathbf{D}) = \beta_0+\sum_{j=1}^n A_{ij}(D_{j}-p)+\epsilon_i$. We impose Assumption \ref{assum:ANI} by giving the $n\times n$ treatment effect matrix $A$ the spatial moving average structure $A_{ij} \propto \max\{\rho(i,j),\overline{m}\}^{-d-\widetilde{\gamma}}$ for $i \neq j$.\footnote{Here $\overline{m}$ is the median distance to the closest neighbor and the maximization is to prevent GPS measurement error from generating extremely large spillover values for close-together units.} We set $\gamma = 1.0$, consistent with the lower bound $\widetilde{\gamma} = 0.01$ used above to select a radius but still a very slow rate of decay, in order to challenge the performance of the estimators in a slower-decay environment and generate clearly contrasting performance across radii. We set the diagonal of $A$ such that $A_{ii} = \frac{1}{2}\mathbf{1}'A\mathbf{1}/n$ to generate a multiplier of 2, slightly larger than the fiscal multipliers found by \cite{Ramey_multiplier,Gabriel_multiplier}. Together these parameters pin down the relative sizes of $A$'s entries. We then calibrate its scale so that the average effect of having one's own village treatment matches the value (\$300) we actually obtain from a regression of the outcomes on own (village) treatment status (i.e., a regression with $\kappa_n = 0$). Finally, we set the deterministic quantities $\beta_0 + \epsilon_i = Y_i^{Factual} - A(\mathbf{D}^{factual}-p)$ where $Y_i^{factual}$ is the outcome and $\mathbf{D}^{factual}$ the vector of treatment assignments observed in the real data. This aims to ensure that they mimic the spatial patterns and magnitudes in the real-world data. Together these parameter choices imply a true value of the GATE of \$410.

For each iteration of the simulation, we randomly redraw the treatment assignment vector $\mathbf{D}$. We randomize using two different designs for comparison. For the factual design, we use the original STATA code from EHMNW to re-draw treatment from the exact same distribution as the actual experiment. For a counterfactual design with larger clusters, we divide each of the 68 EHMNW strata in half (specifically, northern and southern halves) and treat all members of each of the resulting 136 clusters with probability $0.5$ i.i.d. We view this as a natural way to scale clusters while keeping to the spirit of the original design, which was based on pre-existing administrative units, though of course custom clusters are also feasible. It yields a substantial increase in cluster size compared to the factual design, as the 136 clusters consist of 4--5 villages each. For each iteration of the simulation we draw two independent treatment vectors, one for each design; compute outcomes using each of the two treatment vectors; and then estimate the GATE using IPW and OLS over radii $\kappa$ ranging from 1m radius to 2000m.  

\begin{figure}[tp]
    \caption{Simulation Results for $\gamma = 1.0$}
    \label{fig:simulations_gamma_1.0}
    \vspace{-1em}
    \begin{subfigure}[t]{0.48\textwidth}
        \centering
        \includegraphics[width=\textwidth]{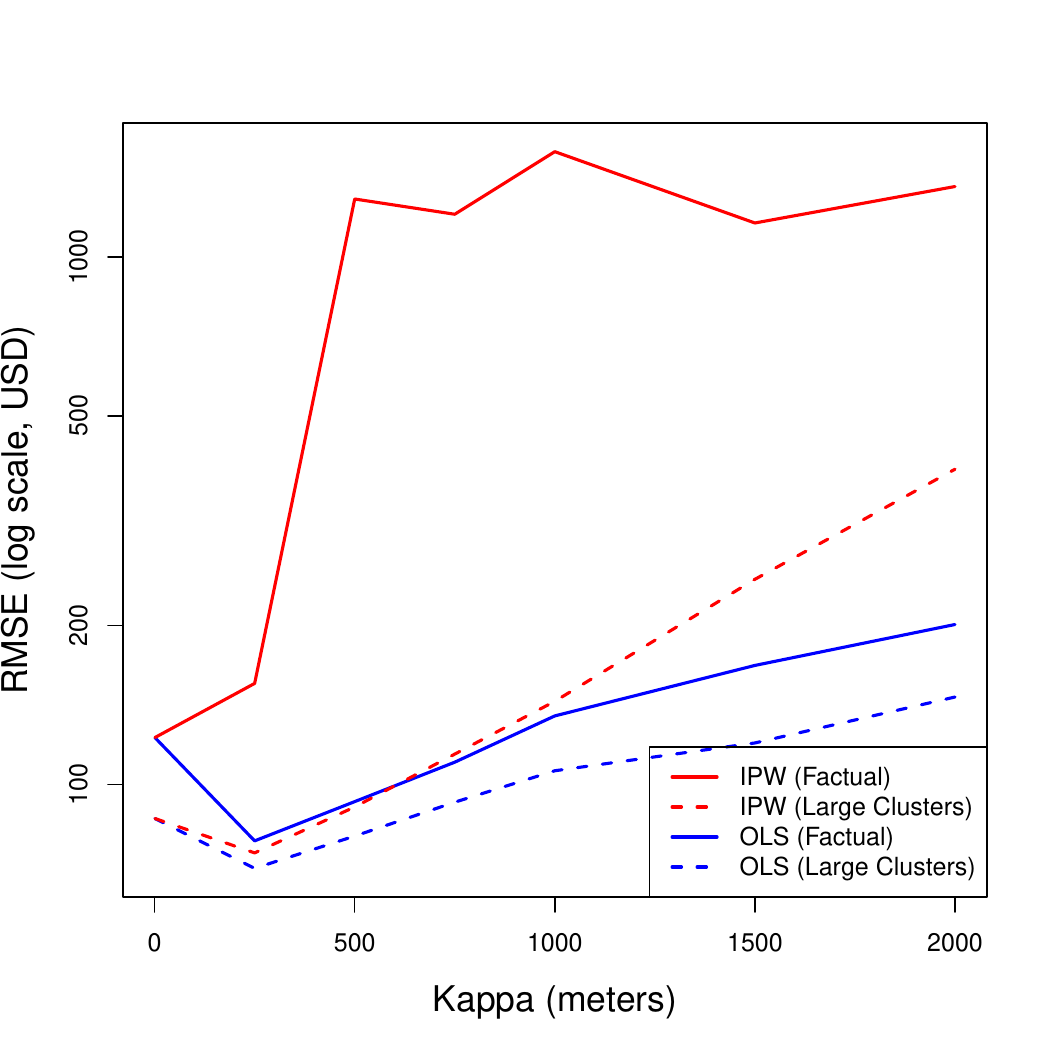}
        
    \end{subfigure}
    \hfill
    \begin{subfigure}[t]{0.48\textwidth}
        \centering
        \includegraphics[width=\textwidth]{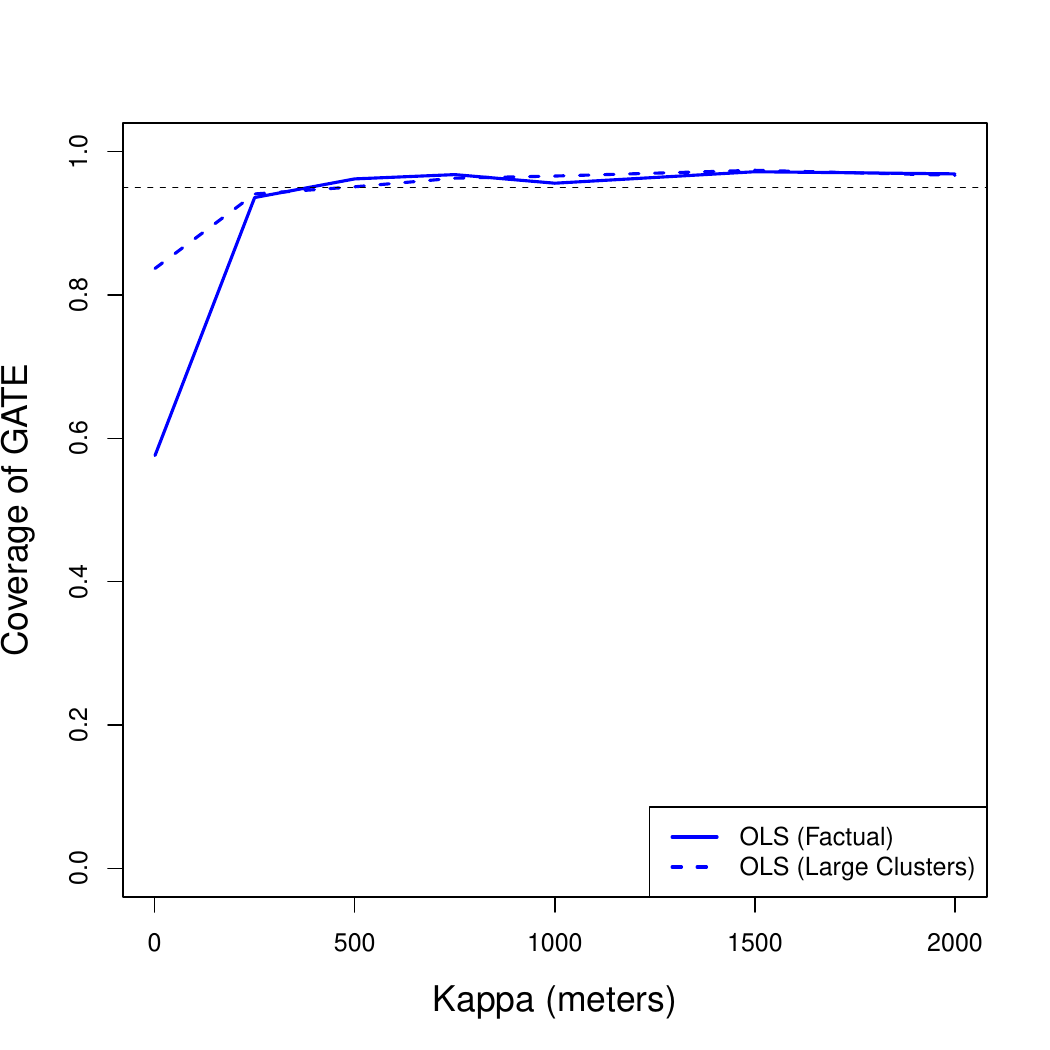}
    \end{subfigure}
    \vspace{-1em}

    
    {\footnotesize This figure plots the root mean squared error (left-hand panel) and coverage of OLS confidence intervals (right-hand panel) in simulation. Solid lines use the factual, small-cluster design and dashed lines use a counterfactual, large-cluster design. Blue lines are OLS and red lines are IPW. The horizontal line in the right-hand panel indicates the nominal coverage level, 95\%. We set $\gamma = 1.00$ given which the true GATE is $\$410$; for comparison, the outcome standard deviation is $\$1,802$.} 
\end{figure}

Figure \ref{fig:simulations_gamma_1.0} displays the results. The left-hand panel shows root mean square error (RMSE), on a log scale, over different radii for the two designs and two estimators. To interpret magnitudes, recall that the true GATE is \$410 and note that the factual outcome variable has standard deviation \$1,802. We draw three main conclusions. First, the IPW estimator has RMSE far higher than that of OLS under the factual, small-cluster design. Second, this performance gap narrows substantially under the counterfactual, large-cluster design. Together these points illustrate the interdependence of design and estimator: IPW can perform well with a design suited for GATE estimation, while OLS performs far better with the small-cluster design EHMNW actually implemented. Third, we note that the RMSE-minimizing OLS radius is 250m, which is shorter than the worst-case-minimizing radius we obtained above (500m) which is itself smaller than that which EHMNW selected based on prior intuition (2,000m). This illustrates the importance of radius selection: using the worst-case-minimizing radius under the factual design, for example, reduces RMSE by 54\% compared to the 2000m radius used by EHMNW. 


The right-hand panel then shows the coverage of 95\% confidence intervals computed using the OLS standard errors defined in Section \ref{sec:inference}, again over different radii. Coverage improves as the radius increases, which is as expected since this reduces bias. Using the factual, small-cluster design and the recommended 500m radius we achieve a coverage level of 96.2\%. The large-cluster counterfactual design yields coverage that is often closer to the nominal 95\% level for all radii. 

Appendix \ref{app:additional_exhibits} shows analogous figures for alternative values of the spatial decay rate $\gamma$. When $\gamma = 1.5$ (Figure \ref{fig:simulations_gamma_1_5}), performance is better across the board: the estimators yield lower RMSE, and coverage of the OLS confidence intervals improves for small radii. This confirms empirically one of the broad and intuitive themes in the theory, that GATE estimation becomes less difficult in populations that are economically less interconnected. When $\gamma = 0.5$ (Figure \ref{fig:simulations_gamma_0.50}), an extremely slow rate, coverage falters for the tightest radii but remains strong for 500m onward.

Considering these results together, a takeaway for practitioners is that the cluster sizes used in EHMNW---while ``large,'' by contemporary standards---were nevertheless too small for reliable IPW estimation. At a 1,000m radius, for instance, IPW's RMSE exceeds the true GATE. Assuming that linearity holds (as EHMNW implicitly did) and estimating the GATE via OLS might then be the best option. Alternatively, a design with larger clusters would have mitigated the need for this assumption.

\section{Conclusion}

Our analysis of the problem of GATE estimation has found that it is feasible---i.e., that consistent and rate-optimal estimator-design pairs exist---under more general conditions than was previously understood; that jointly optimizing over designs as well as linear estimators does not fundamentally change the rate bound; that there are reasonable justifications for using estimators based on OLS regressions like those used in practice to study large-scale field experiments, particularly when clusters must be ``small,'' albeit with the tradeoff that convergence may slow down when the underlying DGP is non-linear; and that disciplined radius selection for and valid inference on such procedures is feasible. Applying these methods to data from one such large-scale experiment, we obtain estimates 22\% larger and 65\% more precise than those obtained using the original authors' choice of radius. Overall, the results demonstrate the viability of GATE estimation without reliance on assumptions about strictly local spillovers, which are often in uncomfortable tension with general equilibrium economic reasoning.

\singlespacing
\typeout{}
\bibliographystyle{aer}
\bibliography{biblio}

\appendix
\onehalfspacing
\renewcommand{\thetable}{\thesection.\arabic{table}}
\renewcommand{\thefigure}{\thesection.\arabic{figure}}
\numberwithin{table}{section}
\numberwithin{figure}{section}

\section{Proofs}
\label{app:proofs}


\subsection{Proof of Theorem \ref{thm:optimalrate}}\label{proof:thm:optimalrate}  

To prove Theorem \ref{thm:optimalrate}, we need Lemma \ref{lem:optimal_rate} and the definition of the covering number below.

\begin{definition}\label{def:covering_number}
    The Covering Number $C\left(s,\mathcal{N}_n\right)$ is defined as the number of units  in the smallest set $\mathcal{J}\subseteq \mathcal{N}_n$ such that $\mathcal{N}_n \subseteq \bigcup_{j\in \mathcal{J}}\mathcal{N}(j,s)$.
\end{definition}

\begin{lemma}\label{lem:optimal_rate}
     Let $\mathcal{N}_n$ be a sequence of populations and let $\widehat{\theta}_n$ be any linear estimator that satisfies Equation (\ref{eq:linear_estimators}). For any sequence $s_n>0$, let the covering number of the population satisfy: $C\left(s_n,\mathcal{N}_n\right)\lesssim f(s_n,n)$   for some function $f \: : \: \mathbb{R}^+\times \mathbb{N}\to \mathbb{R}^+$. Fix any sequence of positive numbers $a_n\to \infty$. Suppose that there is a sequence of experimental designs such that for all $\delta>0$,
\begin{align*}
   \sup_{\mathbf{Y}\in\mathcal{Y}_n}\mathbb{P}\left[a_n\left|\widehat{\theta}_n-\theta_n\right|> \delta\right]\to 0
\end{align*}

Then $a_n$ must satisfy: $$a_n \lesssim \sqrt{f\left(a_n^{1/\gamma},n\right)}$$

Proof: Section \ref{proof:lem:optimal_rate}.
\end{lemma}

To use Lemma \ref{lem:optimal_rate}, we must compute an upper bound on the covering number implied by the Euclidean spatial setup. Assumption \ref{assum:euclidean_space}.a specifies that all units lie within a $d$-ball of radius $n^{1/d}R/2$. For any sequence $s_n$, the $s_n$-covering number of a set of points that all lie inside a $d$-ball of radius $n^{1/d}R/2$ with $d$-balls of radius $s_n$ grows no faster than the volume ratio. So plugging in $a_n$ yields: $ C(a_n,\mathcal{N}_n)\lesssim f(a_n,n)= \max\{\frac{n}{a_n^d},1\}$. So by Lemma \ref{lem:optimal_rate},  $a_n \lesssim \sqrt{ \max\{(\frac{n}{a_n^{d/\gamma}}),1\}}\lesssim  \max\{n^{1/2}a_n^{-d/2\gamma},1 \}$. Rearranging yields: $a_n \lesssim n^{\frac{1}{2+d/\gamma}}$. 

To get the final claim of the theorem, assume for sake of contradiction that for some $b_n\to \infty$ we have:  $\limsup_{n\to\infty}\sup_{\mathbf{Y}\in\mathcal{Y}_n} \mathbb{P}\left[b_nn^{\frac{1}{2+d/\gamma}}\left|\widehat{\theta}_n-\theta_n\right|>\delta\right] =0$. The result of the previous paragraph implies that:  $b_nn^{\frac{1}{2+d/\gamma}} \lesssim n^{\frac{1}{2+d/\gamma}}$. But this cannot be true because by hypothesis any $b_n\to \infty$. So by contradiction we must have:
$$\limsup_{n\to\infty}\sup_{\mathbf{Y}\in\mathcal{Y}_n} \mathbb{P}\left[b_nn^{\frac{1}{2+d/\gamma}}\left|\widehat{\theta}_n-\theta_n\right|>\delta\right] >0$$

\subsection{Proof of Theorem \ref{thm:rateipw}}\label{proof:thm:rateipw}  

Define the (unobserved) random variable $\widetilde{\theta}_n$ below, where factual outcomes are replaced with counterfactuals.  $$\widetilde{\theta}_{n} \equiv \frac{1}{n}\sum_{i=1}^n \left(\frac{Y_i(\mathbf{1})T_{i1}}{\mathbb{P}\left[T_{i1}=1\right]}-\frac{Y_i(\mathbf{0})T_{i0}}{\mathbb{P}\left[T_{i0}=1\right]}\right)$$
\noindent We will first show in Step 1 that  $\widetilde{\theta}_n-\theta_n$ is of the desired order. Then show in Step 2 that the traditional Horvitz-Thompson estimator differs from $\widetilde{\theta}_n$ by that same order. In Step 3, we show that $\widehat{\theta}^{IPW}_{n,\kappa_n}$  differs from the Horvitz-Thompson estimator by that same order. Step 4 completes the proof with the triangle inequality.

{\bf\noindent  Step 1: Convergence of   $\widetilde{\theta}_n-\theta_n$} 

Notice that  $\mathbb{E}[\widetilde{\theta}_{n}]=\theta_n$ by construction so long as the probabilities in the denominators are nonzero. Next we control $\mathbb{V}\left[\widetilde{\theta}_{n}\right]$ and bound the probability weights.  For ease of notation, define $   \widetilde{Z}_{i,\kappa_n}\equiv \frac{Y_i(\mathbf{1})T_{i1}}{\mathbb{P}\left[T_{i1}=1\right]}-\frac{Y_i(\mathbf{0})T_{i0}}{\mathbb{P}\left[T_{i0}=1\right]}$. Recall that $\phi(i,\kappa_n)$ denotes the number of clusters within distance $\kappa_n$ of unit $i$. If clusters are treated iid, then $\mathbb{P}\left[T_{i1}=1\right]\mathbb{P}\left[T_{i0}=1\right]=(p-p^2)^{\phi(i,\kappa_n)}$. By Lemma \ref{lem:phi}, $\phi(i,\kappa_n)$ is uniformly bounded above when $\kappa_n \sim g_n$, so $\mathbb{P}\left[T_{i1}=1\right]\mathbb{P}\left[T_{i0}=1\right]$ is uniformly bounded away from zero. Assumption \ref{assum:boundedoutcomes} guarantees that outcomes are uniformly bounded. So the $\widetilde{Z}_{i,\kappa_n}$ are uniformly bounded and $\sup_{i,n}\mathbb{V}\left[\widetilde{Z}_{i,\kappa_n}\right]<\infty$.  If clusters are treated independently and  $\mathcal{N}\left(i,\kappa_n\right),\mathcal{N}\left(j,\kappa_n\right)$ do not intersect any clusters in common, then $\text{Cov}\left(\widetilde{Z}_{i,\kappa_n},\widetilde{Z}_{j,\kappa_n}\right)=0$. In order for $\mathcal{N}\left(i,\kappa_n\right),\mathcal{N}\left(j,\kappa_n\right)$ to intersect a common cluster, $\rho(i,j)\leq 3\kappa_n+3K_Sg_n$, which by Assumption \ref{assum:euclidean_space}.b can occur for at most $\mathcal{O}\left(\kappa_n^d+g_n^d\right)$ $j$'s for each $i$. Since by hypothesis $g_n\sim \kappa_n$, the number of pairs of neighborhoods which can intersect a common cluster is $\mathcal{O}\left(n\kappa_n^d\right)$.   So $\mathbb{V}\left[\widetilde{\theta}_{n}\right]=\mathcal{O}\left(\frac{\kappa_n^d}{n}\right)$. Therefore: $\mathbb{E}\left[\left|\widetilde{\theta}_n-\theta_n\right|\right]\lesssim \sqrt{\frac{\kappa_n^d}{n}}$. \vskip 0.1in

{\bf\noindent  Step 2:   Convergence of $\widetilde{\theta}_n-\widehat{\theta}_n^{HT}$} 

By Assumption \ref{assum:ANI} and the lower bounds on the probabilities from Step 1,  $\widetilde{\theta}_n$ is close to the Horvitz-Thompson estimator almost surely:
\begin{align*}
    \left|\widetilde{\theta}_n -\frac{1}{n}\sum_{i=1}^n \left(\frac{Y_iT_{i1}}{\mathbb{P}[T_{i1}=1]}-\frac{Y_iT_{i0}}{\mathbb{P}[T_{i0}=1]}\right)\right| &=  \left|\frac{1}{n}\sum_{i=1}^n \left(\frac{(Y_i-Y_i(\mathbf{1}))T_{i1}}{\mathbb{P}[T_{i1}=1]}-\frac{(Y_i-Y_i(\mathbf{0}))T_{i0}}{\mathbb{P}[T_{i0}=1]}\right)\right|\\
    &\leq \frac{1}{n}\sum_{i=1}^n\left|  \frac{(Y_i-Y_i(\mathbf{1}))T_{i1}}{\mathbb{P}[T_{i1}=1]}-\frac{(Y_i-Y_i(\mathbf{0}))T_{i0}}{\mathbb{P}[T_{i0}=1]} \right|\\
    &\leq \max_{i\in \mathcal{N}_n}\max\left\{\frac{T_{i1}\left|Y_i-Y_i(\mathbf{1})\right| }{\mathbb{P}[T_{i1}=1]},\frac{T_{i0}\left|Y_i-Y_i(\mathbf{0})\right| }{\mathbb{P}[T_{i0}=1]}\right\}\\
      &\leq \max_{i\in \mathcal{N}_n}\max\left\{\frac{c\kappa_n^{-\gamma} }{\mathbb{P}[T_{i1}=1]},\frac{c\kappa_n^{-\gamma} }{\mathbb{P}[T_{i0}=1]}\right\}\\
    &\leq  \max_{i\in \mathcal{N}_n}\frac{c\kappa_n^{-\gamma}}{\mathbb{P}[T_{i1}=1]\mathbb{P}[T_{i0}=1]}  \lesssim \kappa_n^{-\gamma}
\end{align*}

{\bf\noindent  Step 3: Convergence of   $\widehat{\theta}^{HT}_{n,\kappa_n}-\widehat{\theta}^{IPW}_{n,\kappa_n}$} 

Next we will show that the IPW estimator converges no slower than the Horvitz-Thompson: $ \left|\widehat{\theta}^{IPW}_{n,\kappa_n}-\frac{1}{n} \sum_{i=1}^n\left(\frac{Y_iT_{i1}}{\mathbb{P}[T_{i1}=1]}-\frac{Y_iT_{i0}}{\mathbb{P}[T_{i0}=1]}\right)\right| \lesssim \sqrt{\frac{\kappa_n^d}{n}}$.  To prove this, it is sufficient to show that $ \left|\frac{1}{\frac{1}{n}\sum_{i=1}^n \frac{T_{i1}}{\mathbb{P}\left[T_{i1}=1\right]}}-1\right|  \lesssim \sqrt{\frac{\kappa_n^d}{n}}$. To do this, first take a first-order Taylor expansion: $ \frac{1}{\frac{1}{n}\sum_{i=1}^n \frac{T_{i1}}{\mathbb{P}\left[T_{i1}=1\right]}}=1-\left(\frac{1}{n}\sum_{i=1}^n \frac{T_{i1}}{\mathbb{P}\left[T_{i1}=1\right]}-1\right)+\mathcal{O}_p\left(\left(\frac{1}{n}\sum_{i=1}^n \frac{T_{i1}}{\mathbb{P}\left[T_{i1}=1\right]}-1\right)^2\right)$. 

\noindent We already know that $\mathbb{E}\left[\left(\frac{1}{n}\sum_{i=1}^n \frac{T_{i1}}{\mathbb{P}\left[T_{i1}=1\right]}-1\right)^2\right]\lesssim {\frac{\kappa_n^d}{n}}$ because this is a special case of $\widetilde{\theta}_n$ when $Y_i(\mathbf{1})=1$ and $Y_i(\mathbf{0})=0$. So the Taylor residual  is $ \mathcal{O}_p\left({\frac{\kappa_n^d}{n}}\right)$. A symmetric argument can be made for $T_{i0}$. So the IPW and Horvitz-Thompson estimators converge at the same rate:\begin{align*}
   \left|\widehat{\theta}_{n,\kappa_n}^{IPW} - \frac{1}{n} \sum_{i=1}^n\left(\frac{Y_iT_{i1}}{\mathbb{P}[T_{i1}=1]}-\frac{Y_iT_{i0}}{\mathbb{P}[T_{i0}=1]}\right)\right| \lesssim \sqrt{\frac{\kappa_n^d}{n}}
\end{align*}

{\bf Step 4:} Combining these inequalities using the triangle inequality completes the proof: 
\begin{align*}
   \left|\widehat{\theta}^{IPW}_{n,\kappa_n}-\theta_n\right| &\lesssim   \left|\widetilde{\theta}_n-\theta_n\right|+\left|\widetilde{\theta}_n -\widehat{\theta}^{HT}_{n,\kappa_n}\right|+   \left|\widehat{\theta}_{n,\kappa_n}^{IPW} - \widehat{\theta}^{HT}_{n,\kappa_n}\right| \lesssim \sqrt{\frac{\kappa_n^d}{n}}+\kappa_n^{-\gamma}
\end{align*}

\subsection{Proof of Theorem \ref{thm:smallclusters}}\label{proof:thm:smallclusters}  

Consider two cases: $q_2 >  q_1$ and $q_2\leq q_1$. In Step 1 we will show that if $q_2 > q_1$, then IPW can be inconsistent for the GATE under some potential outcomes. In Step 2 we will show that if $q_2\leq q_1$, then the convergence is slow due to high bias. 

{\bf Step 1:} If $q_2 > q_1$, then $ \kappa_n^d/g_n^d\sim n^{d(q_2-q_1)}\to \infty$. Since this is a Scaling Clusters design, by Lemma \ref{lem:phi} there exist constants $c_1,c_2>0$ such that for large enough $n$: $  c_1 n^{d(q_2-q_1)}\leq \frac{1}{n}\sum_{i=1}^n \phi(i,\kappa_n)\leq\max_{i\in \mathcal{N}_n}\phi(i,\kappa_n) \leq c_2 n^{d(q_2-q_1)} $. So there exists some $s,t>0$ such that for large enough $n$ there is at least fraction $t$ of units $i$ such that $\phi(i,\kappa_n)>sn^{d(q_2-q_1)}$. By Boole's Inequality, the probability that even just one of these units has $T_{i1}=1$ is:  $$\mathbb{P}\left[  \exists_i \text{ s.t. } \phi(i,\kappa_n) > sn^{d(q_2-q_1)}\text{ and } T_{i1}=1\right]\leq np^{ sn^{d(q_2-q_1)}}=o(1)$$  Consider a set of potential outcomes where: $Y_i(\mathbf{D})=D_i\mathbf{1}\left\{\phi(i,\kappa_n)\geq sn^{d(q_2-q_1)}\right\}$. For these potential outcomes, the GATE is lower-bounded by $\theta_n\geq t$, but IPW equals zero with probability approaching one and IPW is therefore inconsistent. These potential outcomes satisfy Assumptions \ref{assum:boundedoutcomes}-\ref{assum:linearity}. So in order for IPW to be consistent, $q_2 \leq q_1$.

{\bf Step 2:} Now consider the case where $q_2\leq q_1$. Consider the sequence of DGPs where $Y_i(\mathbf{D}) = c\rho(i,j(i))^{-\gamma}(D_{j(i)}-p)$ where $j(i)$ is the closest unit at least distance $\kappa_n+2K_Sg_n$ from unit $i$. So the treatment statuses of all units within $\kappa_n$ of each unit $i$ are independent of $Y_i$ and $\mathbb{E}[Y_{i}|T_{i1}=1] -\mathbb{E}[Y_{i}|T_{i1}=0] = 0$. So the expectation of the Horvitz-Thompson estimator $\widehat{\theta}^{HT}_{n,\kappa_n} \equiv \frac{1}{n}\sum_{i=1}^n Y_i\left(\frac{T_{i1}}{\mathbb{P}[T_{i1}=1]}-\frac{T_{i0}}{\mathbb{P}[T_{i0}=1]}\right)$ is zero.  But, the GATE for this DGP is not zero. Because we assumed a nonvanishing fraction of units must have a neighbor between $\kappa_n+2K_Sg_n$ and $2(\kappa_n+2K_Sg_n)\sim n^{q_1}$ away, $\liminf_{n\to \infty}n^{q_1\gamma }\theta_n>0$. Thus $\liminf_{n\to \infty}n^{q_1\gamma}|\mathbb{E}[\widehat{\theta}^{HT}_{n,\kappa_n}]-\theta_n|>0$.

Next we bound $\mathbb{V}\left[\widehat{\theta}^{HT}_{n,\kappa_n}\right]$. Notice that each summand is bounded because $\kappa_n \lesssim g_n$ controls the probabilities in the denominators. Moreover, each summand $i$ can only be correlated with summands within distance $4(\kappa_n+2K_Sg_n)\lesssim n^{q_1}$. Therefore each summand is correlated with at most $\mathcal{O}\left(n^{dq_1}\right)$ other summands.  So  $\mathbb{V}\left[n^{q_1\gamma}\widehat{\theta}^{HT}_{n,\kappa_n}\right]\lesssim \frac{n^{(d+2\gamma)q_1}}{n}$. Since $q_1 < \frac{1}{d+2\gamma}$ by assumption, $\mathbb{V}\left[n^{q_1\gamma}\widehat{\theta}^{HT}_{n,\kappa_n}\right] \to 0$. Combining this with the lower bound on the bias from the paragraph above, we conclude that $n^{q_1\gamma }\left|\widehat{\theta}^{HT}_{n,\kappa_n}-\theta_n\right|$ cannot converge in probability to zero. By Step 3 of the proof in Section \ref{proof:thm:rateipw}, $|\widehat{\theta}^{HT}_{n,\kappa_n}-\widehat{\theta}^{IPW}_n| \lesssim \sqrt{\frac{\kappa_n^d}{n}}\lesssim n^{\frac{dq_1-1}{2}}$. Since by hypothesis $q_1 < \frac{1}{2\gamma+d}$, we have $n^{\frac{dq_1-1}{2}} n^{q_1\gamma}\to 0$. Thus, $n^{q_1\gamma}|\widehat{\theta}^{HT}_{n,\kappa_n}-\widehat{\theta}^{IPW}_n| \to_p 0$.  So $n^{q_1\gamma}|\widehat{\theta}^{IPW}_n-\theta_n|$ cannot converge in probability to zero.

\subsection{Proof of Theorem \ref{thm:ols_bstar}}\label{proof:thm:ols_bstar}  

Proposition \ref{prop:ols_asymptotic_decomp} decomposes the OLS estimation error into two pieces:  $[\mathcal{A}]$  and  $[\mathcal{B}]$. Since $[\mathcal{A}]$ has expectation zero and $[\mathcal{B}]$ is deterministic, the Triangle Inequality and Chebyshev yield:
 \begin{align*}
      \left|\widehat{\theta}_{n,\kappa_n}^{OLS}-\theta_n\right| \lesssim   \sqrt{ \mathbb{V}\left[[\mathcal{A}]\right] } +\left|[\mathcal{B}]\right| +  \frac{\kappa_n^{d}+g_n^d}{g_n^{d/2}n^{1/2}} 
 \end{align*}
 
We now control the variance of $[\mathcal{A}]$. Notice that the summands of $[\mathcal{A}]$ have a dependency graph with maximum degree $\mathcal{O}\left(\kappa_n^d+g_n^d\right)$. By Proposition \ref{prop:ols_asymptotic_decomp} and Assumption \ref{assum:boundedoutcomes}, $|\bar{\theta}_{n,\kappa_n}|= \left|\frac{1}{n}\sum_{i,j}\widetilde{A}_{ij}\right|\leq 2\overline{Y}$. Using Assumption \ref{assum:boundedoutcomes} again, $\max_i\left|\mathbb{E}[Y_i|\mathcal{F}_{i,\kappa_n}] - \frac{1}{n}\sum_{j=1}^n \mathbb{E}[Y_j]-X_i\bar{\theta}_{n,\kappa_n}\right| \leq 2\overline{Y}(1+\frac{\phi(i,\kappa_n)}{\overline{\phi}})\lesssim 1$. Moreover $\mathbb{E}[X_i]=0$. So $\max_i\mathbb{V}\left[X_i\left(\mathbb{E}[Y_i|\mathcal{F}_{i,\kappa_n}] - \frac{1}{n}\sum_{j=1}^n \mathbb{E}[Y_j]-X_i\bar{\theta}_{n,\kappa_n}\right)\right] \lesssim \max_i\mathbb{E}[X_i^2] = \max_i\mathbb{V}[X_i]$. Since $X_i$ is the sum of $\phi(i,\kappa_n)$ iid bounded random variables divided by $\overline{\phi}$ and Lemma \ref{lem:phi} guarantees that $\max_i\phi(i,\kappa_n) \lesssim \overline{\phi}$, $\max_i\mathbb{V}[X_i] \lesssim \overline{\phi}^{-1}$. So again using Lemma \ref{lem:phi}: $\mathbb{V}\left[[\mathcal{A}]\right] \lesssim \overline{\phi}^2n^{-2}n\overline{\phi}^{-1}\left(\kappa_n^d+g_n^d\right) \lesssim \frac{\kappa_n^{2d}+g_n^{2d}}{g_n^dn} $

The bias term is bounded by Assumption \ref{assum:ANI}: $[\mathcal{B}]=\frac{1}{n}\sum_{i=1}^n\sum_{j=1}^n (A_{ij}-\widetilde{A}_{ij}) \lesssim \kappa_n^{-\gamma}$.

Putting the bias and variance bounds together:
\begin{align*}
     \left|\widehat{\theta}_{n,\kappa_n}^{OLS}-\theta_n\right|  \lesssim \sqrt{ \mathbb{V}\left[[\mathcal{A}]\right] } +\left|[\mathcal{B}]\right| +   \frac{\kappa_n^{d}+g_n^d}{g_n^{d/2}n^{1/2}}  \lesssim \frac{\kappa_n^d+g_n^d}{g_n^{d/2}n^{1/2}}+\kappa_n^{-\gamma}
\end{align*}

\noindent { \bf Sharpness of rate:} See  the proof in Section \ref{proof:thm:minimax_ols_selection} and in particular Equation (\ref{eq:lower_bound_risk}).

\subsection{Proof of Proposition \ref{prop:ols_asymptotic_decomp}}\label{proof:prop:ols_asymptotic_decomp}  

{\bf Notation:} Define $C$ as the $n\times m_n$ matrix where $C_{ic}$ indicates whether unit $i$ is a member of cluster $c$; Define $B$ as the $n\times m_n$ matrix where $B_{ic}= \mathbf{1}\left\{\mathcal{N}(i,\kappa_n) \cap \mathcal{P}_c\neq \emptyset\right\}$ indicates whether unit $i$ is within distance $\kappa_n$ of at least one member of cluster $c$. Define $V_c=W_c-p$. So $X_i = \overline{\phi}^{-1}B_i\mathbf{V}$. Define $\widetilde{A}_{ij}\equiv A_{ij}\mathbf{1}\left\{\mathcal{P}_{c(j)}\cap \mathcal{N}(i,\kappa_n) \neq \emptyset\right\}$. 
\vskip 0.1in

{\noindent \bf Bias term:} First we show that $\frac{\sum_{i=1}^n\mathbb{E}[X_iY_i]}{\sum_{i=1}^n\mathbb{E}[X_i^2]} = \frac{1}{n}\sum_{i=1}^n\sum_{j=1}^n \widetilde{A}_{ij}$. By Lemma \ref{lem:control_x_sharp},  $\sum_{i=1}^n\mathbb{E}[X_i^2]= np(1-p)\overline{\phi}^{-1}$. Since $\sigma(X_i)\subseteq \mathcal{F}_{i,\kappa_n}$, $\mathbb{E}[X_iY_i] = \mathbb{E}[X_i\mathbb{E}[Y_i|\mathcal{F}_{i,\kappa_n}]]$. By Lemma \ref{lem:pot_outcomes_linearity}: $ \mathbb{E}[X_i\mathbb{E}[Y_i|\mathcal{F}_{i,\kappa_n}]]= \overline{\phi}^{-1}\mathbb{E}[\mathbf{V}'B_i'(\widetilde{A}C)_i\mathbf{V}] =\overline{\phi}^{-1}p(1-p)\sum_{j=1}^n\widetilde{A}_{ij}$. So  $\frac{\sum_{i=1}^n\mathbb{E}[X_iY_i]}{\sum_{i=1}^n\mathbb{E}[X_i^2]} = \frac{1}{n}\sum_{i=1}^n\sum_{j=1}^n \widetilde{A}_{ij} = \bar{\theta}_{n,\kappa_n}$.

\vskip 0.1in
{\noindent \bf Stochastic Term:} Starting with the definition of the OLS estimator:
$$ \widehat{\theta}_{n,\kappa_n}^{OLS} \equiv \frac{\frac{1}{n}\sum_{i=1}^n Y_iX_i - \left(\frac{1}{n}\sum_{i=1}^n Y_i\right)\left(\frac{1}{n}\sum_{i=1}^n X_i\right) }{\frac{1}{n}\sum_{i=1}^n X_i^2 - \left(\frac{1}{n}\sum_{i=1}^n X_i\right)^2} = \frac{\overline{\phi}\left(\frac{1}{n}\sum_{i=1}^n Y_iX_i - \left(\frac{1}{n}\sum_{i=1}^n Y_i\right)\left(\frac{1}{n}\sum_{i=1}^n X_i\right)\right) }{\overline{\phi}\left(\frac{1}{n}\sum_{i=1}^n X_i^2 - \left(\frac{1}{n}\sum_{i=1}^n X_i\right)^2\right)}$$

First we address the  denominator and then the numerator. By Lemma \ref{lem:control_x_sharp}, $\left(\frac{1}{n}\sum_{i=1}^n X_i\right)^2 \lesssim \frac{\kappa_n^d+g_n^d}{\overline{\phi}n} $. So $ \overline{\phi} \left(\frac{1}{n}\sum_{i=1}^n X_i^2 - \left(\frac{1}{n}\sum_{i=1}^n X_i\right)^2 \right)= \frac{\overline{\phi}}{n}\sum_{i=1}^n X_i^2 + \mathcal{O}_p\left(\frac{\kappa_n^d+g_n^d}{n}\right)$. 

Next we simplify the numerator below. The last equation is true because by Lemmas \ref{lem:ybar} and then \ref{lem:control_x_sharp}, $\frac{\overline{\phi}}{n}\sum_{i=1}^n X_i \left(\frac{1}{n}\sum_{j=1}^nY_j- \frac{1}{n}\sum_{j=1}^n \mathbb{E}\left[Y_j\right]\right) =o_p\left(\frac{\overline{\phi}}{n}\sum_{i=1}^n X_i \right)=o_p\left(\sqrt{\overline{\phi}\frac{\kappa_n^d+g_n^d}{n}}\right)$. Thus:
\begin{align*}
    \overline{\phi}\left(\frac{1}{n}\sum_{i=1}^n Y_iX_i - \left(\frac{1}{n}\sum_{i=1}^n Y_i\right)\left(\frac{1}{n}\sum_{i=1}^n X_i\right)\right) &= \frac{\overline{\phi}}{n}\sum_{i=1}^n \left(Y_i- \frac{1}{n}\sum_{j=1}^n Y_j\right)\left(X_i-\frac{1}{n}\sum_{j=1}^n X_j\right) \\
    &=\frac{\overline{\phi}}{n}\sum_{i=1}^n \left(Y_i- \frac{1}{n}\sum_{j=1}^n \mathbb{E}\left[Y_j\right]\right)\left(X_i-\frac{1}{n}\sum_{j=1}^n X_j\right) \\
    &= \frac{\overline{\phi}}{n}\sum_{i=1}^n X_i\left(Y_i- \frac{1}{n}\sum_{j=1}^n \mathbb{E}\left[Y_j\right]\right) +o_p\left(\sqrt{\overline{\phi}\frac{\kappa_n^d+g_n^d}{n}}\right)
\end{align*}

By Lemma \ref{lem:phi}: $\sqrt{\overline{\phi}\frac{\kappa_n^d+g_n^d}{n}} \sim \frac{\kappa_n^d+g_n^d}{g_n^{d/2}n^{1/2}}$.  Putting the numerator and denominator together:

$$ \widehat{\theta}_{n,\kappa_n}^{OLS}  = \frac{\frac{\overline{\phi}}{n}\sum_{i=1}^n X_i\left(Y_i- \frac{1}{n}\sum_{j=1}^n \mathbb{E}\left[Y_j\right]\right)  +o_p\left(\frac{\kappa_n^d+g_n^d}{g_n^{d/2}n^{1/2}}\right)}{\frac{\overline{\phi}}{n}\sum_{i=1}^n X_i^2+\mathcal{O}_p\left(\frac{\kappa_n^d+g_n^d}{n}\right)}$$

 Lemmas \ref{lem:phi} and \ref{lem:control_x_sharp} guarantee that 
$\frac{\overline{\phi}}{n}\sum_{i=1}^n X_i^2\to_p p(1-p)$. Since $g_n^d/n\to 0$, we have that $\frac{\kappa_n^d+g_n^d}{n} = o\left(\frac{\kappa_n^d+g_n^d}{g_n^{d/2}n^{1/2}}\right)$. Therefore:
$$ \widehat{\theta}_{n,\kappa_n}^{OLS}  = \frac{\frac{\overline{\phi}}{n}\sum_{i=1}^n X_i\left( Y_i- \frac{1}{n}\sum_{j=1}^n \mathbb{E}\left[Y_j\right]\right)  }{\frac{\overline{\phi}}{n}\sum_{i=1}^n X_i^2}+o_p\left(\frac{\kappa_n^d+g_n^d}{g_n^{d/2}n^{1/2}}\right)$$

{\noindent \bf Long-Range Spillovers:} Now we show that we can replace the $Y_i$ with $\mathbb{E}[Y_i|\mathcal{F}_{i,\kappa_n}]$. Intuitively, this means showing that  the long-range spillovers contribute negligibly to the variance. The first step is to show that their expectation is zero using the fact that $\mathbb{E}[X_i|\mathcal{F}_{i,\kappa_n}]=X_i$ and the law of iterated expectations:
\begin{align*}
     \mathbb{E}\left[\frac{\overline{\phi}}{n}\sum_{i=1}^n X_i\left(Y_i-\mathbb{E}[Y_i|\mathcal{F}_{i,\kappa_n}]\right)\right]&=\frac{\overline{\phi}}{n}\sum_{i=1}^n \mathbb{E}\left[X_i\mathbb{E}\left[\left(Y_i-\mathbb{E}[Y_i|\mathcal{F}_{i,\kappa_n}]\right)|\mathcal{F}_{i,\kappa_n}\right]\right]=0
\end{align*}

Now we control the variance of the long-range spillovers. By Lemma \ref{lem:pot_outcomes_linearity}: $Y_i-\mathbb{E}[Y_i|\mathcal{F}_{i,\kappa_n}] = ((A-\widetilde{A})C)_i\mathbf{V}$. Now we bound the variance with Lemma \ref{lem:covar2}:
\begin{align*}
      \frac{\overline{\phi}}{n}\sum_{i=1}^n X_i\left(Y_i-\mathbb{E}[Y_i|\mathcal{F}_{i,\kappa_n}]\right) &=   \frac{\overline{\phi}}{n}\sum_{i=1}^n X_i\left(((A-\widetilde{A})C)_i\mathbf{V}\right)=\frac{1}{n}\mathbf{V}'B'(A-\widetilde{A})C\mathbf{V}\\
       &\lesssim \sqrt{\mathbb{V}\left[\frac{1}{n}\mathbf{V}'B'(A-\widetilde{A})C\mathbf{V}\right]}  \lesssim   \sqrt{\frac{1}{n^2} \sum_{c=1}^{m_n}\sum_{d=1}^{m_n} \left(B'(A-\widetilde{A})C\right)_{cd}^2 }
\end{align*}

  Each cluster intersects the $\kappa_n$-neighborhoods of at most $\mathcal{O}\left(\kappa_n^d+g_n^d\right)$ units. Combining this fact with Assumption \ref{assum:ANI}  yields $  \sum_{d=1}^{m_n}\left|(B'(\widetilde{A}-A)C)_{cd}\right|\lesssim \kappa_n^{-\gamma} (B')_c\mathbf{1}\lesssim  \kappa_n^{-\gamma}(\kappa_n^d+g_n^d)$ and thus:  
  \begin{align*}
   \frac{1}{n^2} \sum_{c=1}^{m_n}\sum_{d=1}^{m_n} \left(B'(A-\widetilde{A})C\right)_{cd}^2   &\leq  \frac{1}{n^2} \sum_{c=1}^{m_n}\left(\sum_{d=1}^{m_n} |\left(B'(A-\widetilde{A})C\right)_{cd}|\right)^2 \lesssim \frac{1}{n^2}\frac{n}{g_n^d}\kappa_n^{-2\gamma}(\kappa_n^{2d}+g_n^{2d})  \lesssim \kappa_n^{-2\gamma}\frac{\kappa_n^{2d}+g_n^{2d}}{ng_n^d}
\end{align*}

 Since $\kappa_n\to \infty$, the contribution of  the long-range spillovers is dominated.
 \begin{equation}\label{eq:longrange_ols}
    \frac{\overline{\phi}}{n}\sum_{i=1}^n X_i\left(Y_i-\mathbb{E}[Y_i|\mathcal{F}_{i,\kappa_n}]\right)\lesssim \kappa_n^{-\gamma}\frac{\kappa_n^d+g_n^d}{g_n^{d/2}n^{1/2}}=o\left(\frac{\kappa_n^d+g_n^d}{g_n^{d/2}n^{1/2}}\right)
\end{equation}

Substituting this in
$$ \widehat{\theta}_{n,\kappa_n}^{OLS}  = \frac{\frac{\overline{\phi}}{n}\sum_{i=1}^n X_i\left(\mathbb{E}\left[Y_i|\mathcal{F}_{i,\kappa_n}\right]- \frac{1}{n}\sum_{j=1}^n \mathbb{E}\left[Y_j\right]\right) }{\frac{\overline{\phi}}{n}\sum_{i=1}^n X_i^2} +o_p\left(\frac{\kappa_n^d+g_n^d}{g_n^{d/2}n^{1/2}}\right)$$

{\noindent \bf Taylor Residualization:} We would like to use  Lemma \ref{lem:ols_decomp_generic} with:
\begin{align*}
    A_n &= \frac{\overline{\phi}}{n}\sum_{i=1}^n X_i\left(\mathbb{E}\left[Y_i|\mathcal{F}_{i,\kappa_n}\right]- \frac{1}{n}\sum_{j=1}^n \mathbb{E}\left[Y_j\right]\right), \qquad B_n =\frac{\overline{\phi}}{n}\sum_{i=1}^n X_i^2
\end{align*}

We now check the conditions of Lemma \ref{lem:ols_decomp_generic}. By Lemma \ref{lem:control_x_sharp}: $\mathbb{E}[B_n]= p(1-p)$ and $B_n-\mathbb{E}[B_n]\to_p 0 $. We showed above that $\frac{\sum_{i=1}^n\mathbb{E}[X_iY_i]}{\sum_{i=1}^n\mathbb{E}[X_i^2]}=\frac{\mathbb{E}[A_n]}{\mathbb{E}[B_n]} \lesssim 1$. By Assumption \ref{assum:euclidean_space} and the Scaling Clusters design, the summands of $A_n$ have dependency graph with degree at most $\mathcal{O}\left(\kappa_n^d+g_n^d\right)$. By Assumption \ref{assum:boundedoutcomes}, $\left|\mathbb{E}\left[Y_i|\mathcal{F}_{i,\kappa_n}\right]- \frac{1}{n}\sum_{j=1}^n \mathbb{E}\left[Y_j\right]\right|\leq 2\overline{Y}$. Since $\mathbb{E}[X_i]=0$, $\mathbb{V}[X_i\left(\mathbb{E}\left[Y_i|\mathcal{F}_{i,\kappa_n}\right]- \frac{1}{n}\sum_{j=1}^n \mathbb{E}\left[Y_j\right]\right)]\leq 4\overline{Y}^2\mathbb{V}[X_i] \lesssim \overline{\phi}^{-1} $. So $\mathbb{V}[A_n]\lesssim \overline{\phi}\frac{\kappa_n^d+g_n^d}{n}\to 0$. So the conditions of Lemma \ref{lem:ols_decomp_generic} are satisfied and it implies:
\begin{align*}
    \widehat{\theta}_{n,\kappa_n}^{OLS} - \frac{\sum_{i=1}^n\mathbb{E}[X_iY_i]}{\sum_{i=1}^n\mathbb{E}[X_i^2]}  &= \frac{1}{p(1-p)}\frac{\overline{\phi}}{n}\sum_{i=1}^n X_i\left(\mathbb{E}\left[Y_i|\mathcal{F}_{i,\kappa_n}\right]- \frac{1}{n}\sum_{j=1}^n \mathbb{E}\left[Y_j\right]- \frac{\sum_{i=1}^n\mathbb{E}[X_iY_i]}{\sum_{i=1}^n\mathbb{E}[X_i^2]} X_i\right) +o_p\left(\frac{\kappa_n^d+g_n^d}{g_n^{d/2}n^{1/2}}\right)
\end{align*}

\subsection{Proof of Proposition \ref{prop:risk_simplification}}\label{proof:prop:risk_simplification}

 First we isolate the dominant terms of the stochastic part of risk $ [\mathcal{A}]$. To do this we first use Lemma \ref{lem:pot_outcomes_linearity} and then exploit the fact that $\mathbb{E}[[\mathcal{A}]]=0$ and $\mathbb{E}[X_i]=0$:
\begin{align*}
    [\mathcal{A}]&= \frac{1}{p(1-p)}\frac{\overline{\phi}}{n}\sum_{i=1}^n X_i\left(\mathbb{E}\left[Y_i|\mathcal{F}_{i,\kappa_n}\right]- \frac{1}{n}\sum_{j=1}^n \mathbb{E}[Y_j] -X_i\bar{\theta}_{n,\kappa_n} \right)\\
    &= \frac{1}{p(1-p)}\frac{\overline{\phi}}{n}\sum_{i=1}^n X_i\left(\widetilde{A}C\mathbf{V}+p(A_i\mathbf{1}-\theta_n)+\epsilon_i\right) -\frac{1}{p(1-p)}\frac{\overline{\phi}}{n}\sum_{i=1}^n X_i^2\bar{\theta}_{n,\kappa_n} \\
    &=\frac{\frac{\overline{\phi}}{n}\sum_{i=1}^n \left(X_i(\widetilde{A}C)_i\mathbf{V}-\mathbb{E}[X_i(\widetilde{A}C)_i\mathbf{V}] \right)}{p(1-p)}+ \frac{\frac{\overline{\phi}}{n}\sum_{i=1}^nX_i (p(A_i\mathbf{1} -\theta_n)+\epsilon_i)}{p(1-p)}+\frac{\frac{\overline{\phi}}{n}\sum_{i=1}^n (X_i^2 - \mathbb{E}[X_i^2])\bar{\theta}_{n,\kappa_n} }{p(1-p)}
\end{align*}
By Lemma \ref{lem:control_x_sharp}, the third term is $\mathcal{O}_p\left(\sqrt{\frac{\kappa_n^d+g_n^d}{n}}\right)$. 

We now bound the variance of the first term. This is a quadratic form that can be bounded with Lemma \ref{lem:covar2}: 
\begin{equation}\label{eq:VBAC}
   \mathbb{V}\left[\frac{\frac{\overline{\phi}}{n}\sum_{i=1}^n X_i(\widetilde{A}C)_i\mathbf{V} }{p(1-p)}\right] \lesssim   \mathbb{V}\left[ \frac{1}{n} \mathbf{V}'B'\widetilde{A}C\mathbf{V}\right]  \lesssim     \frac{1}{n^2} \sum_{c=1}^{m_n}\sum_{d=1}^{m_n} \left(B'\widetilde{A}C\right)_{cd}^2 
\end{equation}

Now we use the assumption that $||\widetilde{A}||_1\lesssim 1$. So:  $\max_{a,b}\left|\left(B'\widetilde{A}C\right)_{ab}\right| \leq \max_{b}\sum_{i=1}^{n}|(\widetilde{A}C)_{ib}| \leq ||\widetilde{A}||_1  ||C||_1 \lesssim g_n^d$ because the total effect on any subset of the population of treating cluster $b$ grows no faster than the number of units in cluster $b$. Since Assumption \ref{assum:boundedoutcomes} implies $||A||_\infty\lesssim 1$, the sum of absolute causal effects of treating clusters $\{1,\cdots, m_n\}$ on all units within $\kappa_n$ of cluster $c$ is bounded by the number of such units: $\sum_{d=1}^{m_n} \left|\left(B'\widetilde{A}C\right)_{cd}\right| \leq ||B'||_\infty||AC||_\infty \lesssim \kappa_n^d+g_n^d $. Applying H\"{o}lder's Inequality and then this bound to (\ref{eq:VBAC}):
 \begin{align*}
    \frac{1}{n^2} \sum_{d=1}^{m_n} \sum_{c=1}^{m_n} \left(B'\widetilde{A}C\right)_{cd}^2   \lesssim    \frac{1}{n g_n^d}\max_{a,b}\left|\left(B'\widetilde{A}C\right)_{ab}\right|\max_c\sum_{d=1}^{m_n} \left|B'\widetilde{A}C\right|_{cd}\lesssim \frac{1}{n g_n^d} g_n^d (\kappa_n^d+g_n^d)= \frac{\kappa_n^d+g_n^d}{n}
 \end{align*}

So the quadratic form is $\mathcal{O}_p\left(\sqrt{\frac{\kappa_n^d+g_n^d}{n}}\right)  $. Define $\widetilde{\epsilon}_i \equiv p(A_i\mathbf{1}-\theta_n)+\epsilon_i$. Switching to matrix notation defined in \ref{proof:prop:ols_asymptotic_decomp} where $\overline{\phi}X_i=B_i\mathbf{V}$, we have isolated the leading term of the variance:
$$    [\mathcal{A}]= \frac{\frac{\overline{\phi}}{n}\sum_{i=1}^nX_i \widetilde{\epsilon}_i}{p(1-p)} +\mathcal{O}_p\left(\sqrt{\frac{\kappa_n^d+g_n^d}{n}}\right) =\frac{\mathbf{V}'B'\widetilde{\epsilon}}{np(1-p)}+\mathcal{O}_p\left(\sqrt{\frac{\kappa_n^d+g_n^d}{n}}\right)  $$

Calculating variance with Lemma \ref{lem:covar2} and using the fact that $\mathbb{E}[V_c]=0$ we compute:
$\mathbb{V}\left[ \mathbf{V}'B'\widetilde{\epsilon} \right] =  \mathbb{E}\left[\left( \mathbf{V}'B'\widetilde{\epsilon} \right)^2\right]  = p(1-p)||B'\widetilde{\epsilon}||_2^2=p(1-p)\sum_{c=1}^{m_n}\left(\sum_{i=1}^n\widetilde{\epsilon}_iB_{ic}\right)^2$. Recall that $\mu_2\equiv p(1-p)$. We can now express variance as:
\begin{align}\label{eq:leading_variance}
    \mathbb{V}\left[[\mathcal{A}]\right] &=    \frac{1}{n^2\mu_2} \sum_{c=1}^{m_n}\left(\sum_{i=1}^n\widetilde{\epsilon}_iB_{ic}\right)^2+\mathcal{O}\left({\frac{\kappa_n^d+g_n^d}{n}}\right)
\end{align}

The bound on the bias term was shown in step 1 of the proof in Section \ref{proof:prop:ols_asymptotic_decomp}: $\theta_n-\bar{\theta}_{n,\kappa_n}=\frac{1}{n}\sum_{i=1}^n\sum_{j=1}^n (A_{ij}-\widetilde{A}_{ij})$. 

\subsection{Proof of Theorem \ref{thm:minimax_ols_selection}}\label{proof:thm:minimax_ols_selection}  

 {\bf Step 1: Risk-Maximizing DGP:} Recall that $\mu_2\equiv p(1-p)$ and the matrix definitions from Section \ref{proof:prop:ols_asymptotic_decomp}. We must have $||A||_1\lesssim 1$ because $A \in \mathcal{G}_n$ are uniformly upper-bounded by a symmetric matrix with bounded row sums. The leading variance term in Proposition \ref{prop:risk_simplification} is upper bounded over $\mathcal{G}_n$ by:
\begin{align}\label{eq:bound_leading_variance}
    \sup_{\mathbf{Y}\in \mathcal{G}_n}   \frac{1}{n^2\mu_2} \sum_{c=1}^{m_n}\left(\sum_{i=1}^n(p(A_i\mathbf{1}-\mathbf{1}'A\mathbf{1}/n)+\epsilon_i)B_{ic}\right)^2 \leq \frac{1}{n^2\mu_2} \sum_{c=1}^{m_n}\left(\sum_{i=1}^n(p\tau+\sigma_e)B_{ic}\right)^2 
\end{align}

 The upper bound (\ref{eq:bound_leading_variance}) on the leading variance term can be achieved asymptotically while also maximizing bias. First, set $A_{ij}^* = \widetilde{c}\rho(i,j)^{-d-\widetilde{\gamma}}\mathbf{1}\left\{\mathcal{P}_{c(j)}\cap \mathcal{N}(i,\kappa_n) = \emptyset\right\}$ for all pairs $i\neq j$. This maximizes the bias.  Bisect the population and set $A_{ii}^*=\tau -\sum_{j=1}^n \widetilde{c}\rho(i,j)^{-d-\widetilde{\gamma}}\mathbbm{1}\left\{\mathcal{P}_{c(j)}\cap \mathcal{N}(i,\kappa_n) = \emptyset\right\}$ and $\epsilon_i^*=\sigma_e$ for units $i$ in the ``northern" half. Set $A_{ii}^*=-\tau -\sum_{j=1}^n \widetilde{c}\rho(i,j)^{-d-\widetilde{\gamma}}\mathbbm{1}\left\{\mathcal{P}_{c(j)}\cap \mathcal{N}(i,\kappa_n) = \emptyset\right\}$ and $\epsilon_i^*=-\sigma_e$ for units in the ``southern" half. Since the two halves are equal, $\sum_{i}\epsilon_i=0$ and $\mathbf{1}'A^*\mathbf{1}=0$. 

{\bf Step 2: Showing Risk Maximization:} To see that this DGP achieves the  variance upper bound up to a dominated residual, notice that the upper bound  is achieved $\left|\sum_{i=1}^n\widetilde{\epsilon}_i^*B_{ic}\right| = \sum_{i=1}^n(p\tau+\sigma_e)B_{ic}$ for all clusters $c$ except for those that have any member units within distance $2\kappa_n+2K_Sg_n$ of the ``border" between the northern and southern halves. By Assumption \ref{assum:euclidean_space}.a, the largest that this near-border region can be is a slab of thickness $\mathcal{O}\left(\kappa_n+g_n\right)$ bisecting a ball of radius $Rn^{1/d}$. This slab has volume scaling with $\mathcal{O}\left((\kappa_n+g_n)n^{(d-1)/d}\right)$.  By Assumption \ref{assum:euclidean_space}.b, the number of units in a region cannot grow faster than its volume. So the fraction of units in this region is: $\mathcal{O}\left( \frac{\kappa_n+g_n}{n^{1/d}}\right)$. By assumption,  $ \frac{g_n^d}{n}\to 0$. If $\kappa_n^d/n$ did not go to zero, then OLS would not always be consistent. Since OLS can be made consistent for some $\kappa_n$ by Theorem \ref{thm:ols_bstar}, the risk minimizer must also guarantee consistency and thus must satisfy $\kappa_n^d/n\to 0$. So $\frac{\kappa_n^d+g_n^d}{n}\to 0$ and the total weight of clusters for which $\sum_{i=1}^n\widetilde{\epsilon}_i^*B_{ic} <  \sum_{i=1}^n(p\tau+\sigma_e)B_{ic}$ is  the volume of the region divided by the maximum cluster size times the contribution of each cluster which is $\mathcal{O}\left(\frac{\kappa_n+g_n}{n^{1/d}g_n^d} \frac{\kappa_n^{2d}+g_n^{2d}}{n}\right)=o(\frac{\kappa_n^{2d}+g_n^{2d}}{ng_n^d})$. Therefore:
\begin{align*}
    \frac{\sum_{c=1}^{m_n}\left(\sum_{i=1}^n(pA_i^*\mathbf{1} -p\mathbf{1}'A\mathbf{1}+\epsilon_i^*)B_{ic}\right)^2 }{n^2\mu_2} &= \frac{\sum_{c=1}^{m_n}\left(\sum_{i=1}^n(p\tau+\sigma_e)B_{ic}\right)^2 }{n^2\mu_2}+ {o}\left(\frac{\kappa_n^{2d}+g_n^{2d}}{ng_n^d}\right)
\end{align*}

 So asymptotically $A^*,\epsilon^*$ maximize risk: $\sup_{\mathbf{Y}\in\mathcal{G}_n}\mathcal{R}^*\left(\kappa_n,\mathbf{Y}\right)=\mathcal{R}^*\left(\kappa_n,\mathbf{Y}^*\right)+\mathcal{O}\left({\frac{\kappa_n^d+g_n^d}{n}}\right)$. So risk is asymptotically well approximated by the deterministic function $ R(\kappa_n,\mathcal{G}_n)$:
\begin{align*}
    \mathcal{R}^*\left(\kappa_n,\mathbf{Y}^*\right) 
   &= R(\kappa_n,\mathcal{G}_n)+\mathcal{O}\left({\frac{\kappa_n^d+g_n^d}{n}}\right)
\end{align*}

 Now we show that the $\mathcal{O}\left({\frac{\kappa_n^d+g_n^d}{n}}\right)$ term is dominated. By Lemma \ref{lem:columnsumsB}: 
\begin{equation}\label{eq:lower_bound_risk}
\frac{\sum_{c=1}^{m_n}\left(\sum_{i=1}^n(p\tau+\sigma_e)B_{ic}\right)^2 }{n^2\mu_2} \sim \frac{\sum_{c=1}^{m_n}\left(\sum_{i=1}^nB_{ic}\right)^2 }{n^2\mu_2}  \sim  \frac{\kappa_n^{2d}+g_n^{2d}}{g_n^dn}
\end{equation}
Thus $\liminf_{n\to \infty}\frac{g_n^dn}{\kappa_n^{2d}+g_n^{2d}}\sup_{\mathbf{Y}\in \mathcal{G}_n}\mathcal{R}^*\left(\kappa_n,\mathbf{Y}\right) > 0$ and $\liminf_{n\to \infty}\frac{g_n^dn}{\kappa_n^{2d}+g_n^{2d}} R(\kappa_n,\mathcal{G}_n)>0$. So whenever $\kappa_n/g_n \to \infty$, the remainder term is dominated:
\begin{equation}\label{eq:ratiotoone}
    \frac{ R(\kappa_n,\mathcal{G}_n)}{\sup_{\mathbf{Y}\in \mathcal{G}_n}\mathcal{R}^*\left(\kappa_n,\mathbf{Y}\right)} = \frac{ R(\kappa_n,\mathcal{G}_n)}{ R(\kappa_n,\mathcal{G}_n)+\mathcal{O}\left({\frac{\kappa_n^d+g_n^d}{n}}\right)}  \to 1
\end{equation}

{\bf Step 3: Rate Behavior of the Maximizer:} Now we show that:  $\widehat{\kappa}_n(\mathcal{G})/g_n \to \infty$. Since the bias scales as $[B]\sim \int_{\kappa_n}^\infty s^{d-1}s^{-d-\widetilde{\gamma}}ds \sim \kappa_n^{-\widetilde{\gamma}}$ and by Equation (\ref{eq:lower_bound_risk}), there are constants $c_1,c_2>0$ such that $c_1\left(\frac{\kappa_n^{2d}+g_n^{2d}}{g_n^dn}+\kappa_n^{-2\widetilde{\gamma}}\right)\leq R(\kappa_n,\mathcal{G}_n)\leq c_2\left(\frac{\kappa_n^{2d}+g_n^{2d}}{g_n^dn}+\kappa_n^{-2\widetilde{\gamma}}\right)$. The minimizer $\tilde{\kappa}_n \equiv \argmin_{\kappa_n>0}\frac{\kappa_n^{2d}+g_n^{2d}}{g_n^dn}+\kappa_n^{-2\widetilde{\gamma}}$ of the upper and lower bounding functions satisfies $\tilde{\kappa}_n \sim n^{\frac{1}{2d+2\widetilde{\gamma}}}g_n^{\frac{d}{2d+2\widetilde{\gamma}}}  $ and  $\tilde{\kappa}_n /g_n \sim n^{\frac{1}{2d+2\widetilde{\gamma}}}g_n^{\frac{-d-2\widetilde{\gamma}}{2d+2\widetilde{\gamma}}}$. Since by assumption $g_n =o\left(n^{\frac{1}{2\widetilde{{\gamma}}+d}}\right)$, $\tilde{\kappa}_n /g_n \to \infty$. If the minimizer of $R(\kappa_n,\mathcal{G}_n)$ did not satisfy the same property, then it would violate either the upper or lower bounds. Thus,  $  \widehat{\kappa}_n(\mathcal{G}_n)/g_n \to \infty$. An identical argument shows $  {\kappa}^*_n(\mathcal{G}_n)/g_n \to \infty$.

{\bf Step 4: Squeeze Argument:}  Since  $\kappa_n^{*}(\mathcal{G}_n)$ is the minimizer of the denominator and $\widehat{\kappa}_n(\mathcal{G}_n)$ is the minimizer of the numerator:
\begin{align*}
    \frac{ R(\widehat{\kappa}_n(\mathcal{G}_n),\mathcal{G}_n)}{\sup_{\mathbf{Y}\in \mathcal{G}_n}\mathcal{R}^*\left(\kappa_n^{*}(\mathcal{G}_n),\mathbf{Y}\right)}  &\geq  \frac{ R(\widehat{\kappa}_n(\mathcal{G}_n),\mathcal{G}_n)}{\sup_{\mathbf{Y}\in \mathcal{G}_n}\mathcal{R}^*\left(\widehat{\kappa}_n(\mathcal{G}_n),\mathbf{Y}\right)} \to 1\\
       \frac{ R(\widehat{\kappa}_n(\mathcal{G}_n),\mathcal{G}_n)}{\sup_{\mathbf{Y}\in \mathcal{G}_n}\mathcal{R}^*\left(\kappa_n^{*}(\mathcal{G}_n),\mathbf{Y}\right)} &\leq    \frac{ R(\kappa_n^{*}(\mathcal{G}_n),\mathcal{G}_n)}{\sup_{\mathbf{Y}\in \mathcal{G}_n}\mathcal{R}^*\left(\kappa_n^{*}(\mathcal{G}_n),\mathbf{Y}\right)}\to 1
\end{align*}
By the Squeeze Theorem:
$$ \frac{ R(\widehat{\kappa}_n(\mathcal{G}_n),\mathcal{G}_n)}{\sup_{\mathbf{Y}\in \mathcal{G}_n}\mathcal{R}^*\left(\kappa_n^{*}(\mathcal{G}_n),\mathbf{Y}\right)} \to 1 $$

\noindent The final result is obtained by multiplying $ \frac{  \sup_{\mathbf{Y}\in \mathcal{G}_n}\mathcal{R}^*\left(\widehat{\kappa}_n(\mathcal{G}_n) ,\mathbf{Y}\right)}{\sup_{\mathbf{Y}\in \mathcal{G}_n}\mathcal{R}^*\left(\kappa_n^{*}(\mathcal{G}_n) ,\mathbf{Y}\right)} $ by $1=\frac{ R(\widehat{\kappa}_n(\mathcal{G}_n),\mathcal{G}_n)}{ R(\widehat{\kappa}_n(\mathcal{G}_n),\mathcal{G}_n)}$ and decomposing into the product of two ratios already shown to be converging to one.

\subsection{Proof of Theorem \ref{thm:clt}}\label{proof:thm:clt}  

This CLT is similar to the proof in \cite{Leung22}. There are two additional issues in the context of this paper: to allow $\gamma < d$ and to take extra care because $|X_i|$ in OLS is not necessarily bounded when $\kappa_n/g_n\to \infty$. We have already addressed the former with Proposition \ref{prop:ols_asymptotic_decomp}. Now we address the latter. Let $U_i\equiv \frac{1}{p(1-p)}\frac{\overline{\phi}}{n} X_i\left(\mathbb{E}\left[Y_i|\mathcal{F}_{i,\kappa_n}\right]- \frac{1}{n}\sum_{j=1}^n \mathbb{E}\left[Y_j\right]- \frac{\sum_{i=1}^n\mathbb{E}[X_iY_i]}{\sum_{i=1}^n\mathbb{E}[X_i^2]} X_i\right) $. These are the summands of $[\mathcal{A}]$. Define $\sigma_{n}^2 \equiv \mathbb{V}\left[[\mathcal{A}]\right]$. Define $\Psi_n$ to be the maximum degree of the dependency graph among the summands $U_i$. The dependency graph has degree $\mathcal{O}\left(\kappa_n^d+g_n^d\right)$ because $U_i$ depends only on treatment statuses of units within $\kappa_n+g_n$ of unit $i$ which has membership bounded by $\mathcal{O}\left(\kappa_n^d+g_n^d\right)$ by Assumption \ref{assum:euclidean_space}.b. There is a constant $C>0$ such that: $|U_i| \leq C|\overline{\phi}X_i/n| $ because Lemma \ref{lem:phi} forces $\max_i|X_i|\leq \max_i\phi(i,\kappa_n)/\overline{\phi}\lesssim 1$ and Assumption \ref{assum:boundedoutcomes} forces $|\mathbb{E}\left[Y_i|\mathcal{F}_{i,\kappa_n}\right]- \frac{1}{n}\sum_{j=1}^n \mathbb{E}\left[Y_j\right]|\leq 2\overline{Y}$. Furthermore by the Marcinkiewicz–Zygmund inequality: $\max_i\mathbb{E}[|U_i|^3]\lesssim \max_i\overline{\phi}^3n^{-3}\mathbb{E}[|X_i|^3] \lesssim \frac{\overline{\phi}^{3/2}}{n^3}\lesssim \frac{\kappa_n^{(3/2)d}+g_n^{(3/2)d}}{g_n^{(3/2)d} n^3} $. Similarly, $\max_i\mathbb{E}[|U_i|^4]\lesssim \frac{\kappa_n^{2d}+g_n^{2d}}{g_n^{2d} n^4}$. So we can use   Theorem 3.6 from \cite{RossSteins} to bound the Wasserstein distance $d\left([\mathcal{A}]/\sigma_{n},\mathcal{Z}\right) $ between a standardized version of the distribution of $[\mathcal{A}]$ and the standard normal distribution. Substituting in the assumed lower-bound on $\sigma_n$ shows the Wasserstein distance goes to zero:
\begin{align*}
    d\left([\mathcal{A}]/\sigma_{n},\mathcal{Z}\right) &\leq \frac{\Psi_n^2}{\sigma_{n}^3}\sum_{i=1}^n \mathbb{E}[|U_i|^3] +\frac{\sqrt{28}\Psi_n^{3/2}}{\sqrt{\pi}\sigma_{n}^2}\sqrt{\sum_{i=1}^n \mathbb{E}[U_i^4]}
    \\
    &\lesssim \frac{(\kappa_n^{2d}+g_n^{2d})n^{3/2}g_n^{(3/2)d}}{\kappa_n^{3d}+g_n^{3d}}\frac{\kappa_n^{(3/2)d}+g_n^{(3/2)d}}{n^2g_n^{(3/2)d}}+\frac{(\kappa_n^{(5/2)d}+g_n^{(5/2)d})g_n^{d}n}{\kappa_n^{2d}+g_n^{2d}}\frac{1}{g_n^{d}n^{3/2}}\lesssim  \sqrt{\frac{\kappa_n^d+g_n^d}{n}} \to 0
\end{align*}

\subsection{Proof of Theorem \ref{thm:varest}}\label{proof:thm:varest}

Let $\Lambda$ be the $n\times n$ matrix where $\Lambda_{ij}$ indicates that unit $j$ belongs to a cluster that intersects $\mathcal{N}(i,\kappa_n)$. The summands $r_i,r_j$ are independent when $\Lambda_{ij}=0$ because each $r_i$ is $\mathcal{F}_{i,\kappa_n}$-measurable. So the variance of $[\mathcal{A}]$ is:
\begin{align*}
   \mathbb{V}\left([\mathcal{A}]\right) &=   \frac{1}{\mu_2^2} \frac{\overline{\phi}^2}{n^2}\sum_{i=1}^n\sum_{j=1}^n\Lambda_{ij}\text{Cov}(r_j,r_i)
   = \frac{1}{\mu_2^2}\frac{\overline{\phi}^2}{n^2}\sum_{i=1}^n\sum_{j=1}^n\Lambda_{ij}\mathbb{E}[r_jr_i]- \frac{1}{\mu_2^2}\frac{\overline{\phi}^2}{n^2}\sum_{i=1}^n\sum_{j=1}^n\Lambda_{ij}\mathbb{E}[r_j]\mathbb{E}[r_i]\\
   &=  \frac{1}{\mu_2^2} \frac{\overline{\phi}^2}{n^2}\sum_{i=1}^n\sum_{j=1}^n\Lambda_{ij}\mathbb{E}[r_jr_i] - \frac{1}{\mu_2^2} \frac{\overline{\phi}^2}{n^2}\sum_{i=1}^n\sum_{j=1}^n(\Lambda_{ij}-Q_{ij})\mathbb{E}[r_j]\mathbb{E}[r_i]- \frac{1}{\mu_2^2} \frac{\overline{\phi}^2}{n^2}\sum_{i=1}^n\sum_{j=1}^nQ_{ij}\mathbb{E}[r_j]\mathbb{E}[r_i]\\
   &\equiv [D]-[E]-[F]
\end{align*}

 Where $Q$ is the $n\times n$ matrix where $Q_{ij}=1$ if $i,j$ are in the same cluster. So $Q$ is block-diagonal and therefore positive semidefinite which implies that $[F]\geq 0$. $[F]$ is unidentified but non-negative so it can be ignored without making inference anticonservative. Here the unidentified variance term from the main text is: $H_n = -[E]-[F]$.

\noindent {\bf Estimating [D]:} $[D]$ can be estimated consistently using a traditional HAC estimator. First we confirm that using the empirical $\widehat{r}_i$ in place of $r_i$ does not add meaningful bias:  $\widehat{r}_i-r_i= X_i\left((Y_i-\mathbb{E}[Y_i|\mathcal{F}_{i,\kappa_n}])+\frac{1}{n}\sum_{j=1}^n (Y_j-\mathbb{E}[Y_j]) -(\widehat{\theta}_{n,\kappa_n}^{OLS}-\bar{\theta}_{n,\kappa_n})X_i\right)$. Bounding the sum::\begin{align}\label{eq:rrhat1}
    \left|\frac{\overline{\phi}^2}{n^2}\sum_{i=1}^n\sum_{j=1}^n\Lambda_{ij}(\widehat{r}_i\widehat{r}_j-r_ir_j)\right| &\leq 2\left|\widehat{\theta}_{n,\kappa_n}^{OLS}-\bar{\theta}_{n,\kappa_n}\right|  \frac{\overline{\phi}^2}{n^2}\sum_{i=1}^n\sum_{j=1}^n\Lambda_{ij}X_i^2X_j^2\\
    &+2\left|\kappa_n^{-\gamma}+\frac{1}{n}\sum_{j=1}^n (Y_j-\mathbb{E}[Y_j]) \right| \frac{\overline{\phi}^2}{n^2}\sum_{i=1}^n\sum_{j=1}^n\Lambda_{ij}|X_iX_j|\label{eq:rrhat2}
\end{align} 

Since $\Lambda$ has maximum degree $\mathcal{O}\left(\kappa_n^d+g_n^d\right)$, each sum has $\mathcal{O}\left(n(\kappa_n^d+g_n^d)\right)$ nonzero terms, each of which is correlated with at most $\mathcal{O}\left((\kappa_n^d+g_n^d)^2\right)$ other terms. Since each $X_i$ is the sum of $\phi(i,\kappa_n)$ iid random variables, the Marcinkiewicz–Zygmund inequality bounds each covariance: $Cov(X_i,X_j)\lesssim \max_i\mathbb{V}[X_i]\lesssim \overline{\phi}^{-1}$ and $Cov(X_i^2,X_j^2)\leq \max_i \mathbb{E}[X_i^4]\lesssim \overline{\phi}^{-2}$. By Lemma \ref{lem:phi}, $\overline{\phi}\sim \frac{\kappa_n^d+g_n^d}{g_n^d}$. So:
\begin{align}\label{eq:varxxbound}
     \mathbb{V}\left[\frac{\overline{\phi}^2}{n^2}\sum_{i=1}^n\sum_{j=1}^n\Lambda_{ij}X_i^2X_j^2\right],\mathbb{V}\left[\frac{\overline{\phi}^2}{n^2}\sum_{i=1}^n\sum_{j=1}^n\Lambda_{ij}|X_iX_j|\right] \lesssim \overline{\phi}^3 \frac{n(\kappa_n^d+g_n^d)^3}{n^4} \lesssim \left(\frac{\kappa_n^{d}+g_n^{d}}{n^{1/2}g_n^{d/2}}\right)^6
\end{align}

Moreover $\mathbb{E}[|X_iX_j|]\leq \max_i\mathbb{E}[X_i^2]\lesssim \overline{\phi}^{-1}$. So:
\begin{align*}
  \mathbb{E}\left[\frac{\overline{\phi}^2}{n^2}\sum_{i=1}^n\sum_{j=1}^n\Lambda_{ij}X_i^2X_j^2\right]\lesssim   \mathbb{E}\left[\frac{\overline{\phi}^2}{n^2}\sum_{i=1}^n\sum_{j=1}^n\Lambda_{ij}|X_iX_j|\right] \lesssim \overline{\phi} \frac{(\kappa_n^d+g_n^d)}{n}\lesssim \left(\frac{\kappa_n^{d}+g_n^{d}}{n^{1/2}g_n^{d/2}}\right)^2
\end{align*}

The leading sequences inside the absolute values of (\ref{eq:rrhat1}) and (\ref{eq:rrhat2}) vanish by  Theorem \ref{thm:ols_bstar} for the former and Lemma \ref{lem:ybar} for the latter. So the error from substituting $\widehat{r}_i$ for $r_i$ is negligible:
\begin{align*}
     \left|\frac{\overline{\phi}^2}{n^2}\sum_{i=1}^n\sum_{j=1}^n\Lambda_{ij}(\widehat{r}_i\widehat{r}_j-r_ir_j)\right| = o_p\left(\frac{\kappa_n^{2d}+g_n^{2d}}{ng_n^{d}}\right)
\end{align*}

 Now we control $ \mathbb{V}\left[ \frac{\overline{\phi}^2}{n^2}\sum_{i=1}^n\sum_{j=1}^n\Lambda_{ij}\widehat{r}_i\widehat{r}_j\right]$. Since each covariance  $Cov(\widehat{r}_i,\widehat{r}_j)\lesssim \max_i\mathbb{V}[\widehat{r}_i]\lesssim \max_i\mathbb{V}[X_i]\lesssim \overline{\phi}^{-1}$, we can re-use the argument that justified (\ref{eq:varxxbound}) to bound the variance of the sum over the $\widehat{r}_i$. Thus, the estimator converges faster than the variance it is estimating: 
\begin{align*}
    \mathbb{V}\left[ \frac{\overline{\phi}^2}{n^2}\sum_{i=1}^n\sum_{j=1}^n\Lambda_{ij}\widehat{r}_i\widehat{r}_j\right]\lesssim \overline{\phi}^4 \frac{1}{n^4} n(\kappa_n^d+g_n^d)^3\overline{\phi}^{-1} \lesssim \overline{\phi}^3\frac{(\kappa_n^d+g_n^d)^3}{n^3} =o\left(\frac{\kappa_n^{2d}+g_n^{2d}}{ng_n^{d}}\right)
\end{align*}

Finally,  by Lemma \ref{lem:control_x_sharp}, exchanging the denominator for its expectation incurs an error term of $\frac{\overline{\phi}^2}{n}\sum_{i=1}^n X_i^2 - \overline{\phi}^2\left(\frac{1}{n}\sum_{i=1}^n X_i\right)^2-\mu_2^{2}={o}_p\left(1\right)$ in the denominator. Since the  denominator error is $\mathcal{O}_p\left( \frac{\kappa_n^{2d}+g_n^{2d}}{ng_n^{d}}\right)$ and the numerator is $o_p(1)$:
\begin{align*}
      \frac{1}{\mu_2^2} \frac{\overline{\phi}^2}{n^2}\sum_{i=1}^n\sum_{j=1}^n\Lambda_{ij}\widehat{r}_i\widehat{r}_j - \frac{\frac{1}{n^2}\sum_{i=1}^n\sum_{j=1}^n\Lambda_{ij}\widehat{r}_i\widehat{r}_j}{\left(\frac{1}{n}\sum_{i=1}^n X_i^2 - \left(\frac{1}{n}\sum_{i=1}^n X_i\right)^2\right)^2} =o_p\left(\frac{\kappa_n^{2d}+g_n^{2d}}{ng_n^{d}}\right)
\end{align*}

Thus, $\frac{\frac{1}{n^2}\sum_{i=1}^n\sum_{j=1}^n\Lambda_{ij}\widehat{r}_i\widehat{r}_j}{\left(\frac{1}{n}\sum_{i=1}^n X_i^2 - \left(\frac{1}{n}\sum_{i=1}^n X_i\right)^2\right)^2} = [D]+o_p\left(\frac{\kappa_n^{2d}+g_n^{2d}}{ng_n^{d}}\right)$.

\noindent {\bf Estimating [E]:} Next we exploit the growing clusters to estimate [E] with $\frac{ \frac{1}{n^2}\sum_{i=1}^n\sum_{j=1}^n (\Lambda_{ij}-Q_{ij})\widehat{q}_i\widehat{q}_j}{\left(\frac{1}{n}\sum_{i=1}^n X_i^2 - \left(\frac{1}{n}\sum_{i=1}^n X_i\right)^2\right)^2}$. By the proof of Lemma \ref{lem:control_x_sharp}: $\mathbb{E}[X_i^2] = p(1-p) \frac{\phi_i}{\overline{\phi}^2}$. Moreover, $ \frac{1}{n^2}\sum_{i=1}^n\sum_{j=1}^n \Lambda_{ij}\lesssim \frac{\kappa_n^d+g_n^d}{n}$. By Lemma \ref{lem:pot_outcomes_linearity}: $\mathbb{E}[X_i Y_i]= \mathbb{E}[X_i\mathbb{E}[Y_i|\mathcal{F}_{i,\kappa_n}]]= \overline{\phi}^{-1}\mathbb{E}[\mathbf{V}'B_i'(\widetilde{A}C)_i\mathbf{V}] =\overline{\phi}^{-1}p(1-p)\sum_{j=1}^n\widetilde{A}_{ij}$.  So:
\begin{align*}
    \mathbb{E}[r_i] &= p(1-p)\left(\overline{\phi}^{-1}(Y_i(\mathbf{1})-Y_i(\mathbf{0})) - \frac{\phi_i}{\overline{\phi}^2}\bar{\theta}_{n,\kappa_n}\right) + p(1-p) \overline{\phi}^{-1}\sum_{j}(\widetilde{A}_{ij}-A_{ij})\\
    &\equiv \widetilde{r}_i +p(1-p) \overline{\phi}^{-1}\sum_{j}(\widetilde{A}_{ij}-A_{ij})
    \end{align*}Taking the sum and bounding $\sum_{j}(\widetilde{A}_{ij}-A_{ij})$ with Assumption \ref{assum:ANI}:
    \begin{align}\label{eq:Ertilde}
    [E] &= \frac{1}{\mu_2^2}\frac{\overline{\phi}^2}{n^2}\sum_{i=1}^n\sum_{j=1}^n(\Lambda_{ij}-Q_{ij})\widetilde{r}_i\widetilde{r}_j+\mathcal{O}\left(\kappa_n^{-\gamma}\frac{\kappa_n^d+g_n^d}{n}\right)
\end{align}

The $\widetilde{r}_i$ are hard to work with. Define $\widetilde{q}_i$ below. We will now show that the sum over $\widetilde{r}_i$ is asymptotically equivalent to the sum over  $\widetilde{q}_i$.
\begin{align*}
    \widetilde{q}_i &\equiv p(1-p)\frac{\phi_i}{\overline{\phi}^2}\left(\frac{\overline{\phi}}{\phi_i}\left(\frac{D_i}{p}Y_i(\mathbf{1})-\frac{1-D_i}{1-p}Y_i(\mathbf{0})\right)-\bar{\theta}_{n,\kappa_n}\right)
\end{align*}

\noindent Notice that $\mathbb{E}[\widetilde{q}_i]=\widetilde{r}_i$. Because $D_i,D_j$ are assumed to be independent if $\Lambda_{ij}-Q_{ij}\neq 0$ :
\begin{align*}
   (\Lambda_{ij}-Q_{ij})\mathbb{E}\left[\widetilde{q}_i\widetilde{q}_j\right]= (\Lambda_{ij}-Q_{ij})\mathbb{E}\left[\widetilde{q}_i]\mathbb{E}[\widetilde{q}_j\right] =(\Lambda_{ij}-Q_{ij})\widetilde{r}_i\widetilde{r}_j
\end{align*}

\noindent Since there are $\mathcal{O}\left(n(\kappa_n^d+g_n^d)\right)$ nonzero terms of $ \frac{1}{n^2}\sum_{i=1}^n\sum_{j=1}^n(\Lambda_{ij}-Q_{ij})\widetilde{q}_i\widetilde{q}_j$ and each term is correlated with at most $\mathcal{O}\left((\kappa_n^d+g_n^d)^2\right)$ other terms and each term is bounded by $\overline{\phi}^{-2}$:
\begin{align*}
      \mathbb{V}\left[\frac{\overline{\phi}^2}{n^2}\sum_{i=1}^n\sum_{j=1}^n (\Lambda_{ij}-Q_{ij})\widetilde{q}_i\widetilde{q}_j \right] \lesssim \frac{n(\kappa_n^d+g_n^d)^3}{n^4}=\frac{(\kappa_n^d+g_n^d)^3}{n^3}
\end{align*}

\noindent Substituting into (\ref{eq:Ertilde}), summing over $\widetilde{q}_i$ is asymptotically equivalent to summing over $\widetilde{r}_i$: 
\begin{align}\label{eq:Eqtilde}
  [E] =   \frac{1}{\mu_2^2}\frac{\overline{\phi}^2}{n^2}\sum_{i=1}^n\sum_{j=1}^n (\Lambda_{ij}-Q_{ij})\widetilde{q}_i\widetilde{q}_j+o_p\left(\frac{\kappa_n^d+g_n^d}{n}\right)
\end{align}

Since $\widetilde{q}_i$ is unobserved, we will next show that $[E]$ can be estimated by replacing $\widetilde{q}_i$ with $\widehat{q}_i$. By Assumption \ref{assum:ANI}, if $\rho_i$ is the distance from unit $i$ to the nearest cluster not its own:
\begin{equation}
 \frac{1}{n}\sum_{i=1}^n\left|\frac{D_i}{p}(Y_i-Y_i(\mathbf{1}))+\frac{1-D_i}{1-p}(Y_i-Y_i(\mathbf{0}))\right|\leq c  \frac{1}{n}\sum_{i=1}^n\rho_i^{-\gamma}
\end{equation}

We now bound  $\frac{1}{n}\sum_{i=1}^n\rho_i^{-\gamma}$. The theorem assumes that only a vanishing fraction of each cluster lies in its crust, so  if $c(i)$ denotes the cluster of unit $i$: $\frac{1}{n}\sum_{i=1}^n\rho_i^{-\gamma}\mathbf{1}\left\{i\notin \mathcal{N}(q_{c(i)},g_n)\right\} \to 0$. Now we bound $\frac{1}{n}\sum_{i=1}^n\rho_i^{-\gamma}\mathbf{1}\left\{i\in \mathcal{N}(q_{c(i)},g_n)\right\}$. Under a scaling clusters design, any unit $i$ that lies within the neighborhood $\mathcal{N}(q_c,g_n-\sqrt{g_n})$ corresponding to the centerpoint $q_c$ of its own cluster defined in Definition \ref{def:scaling_clusters} has $\rho_i \geq \sqrt{g_n}$. The number of remaining units in each cluster lie inside $\mathcal{N}(q_c,g_n) \setminus \mathcal{N}(q_c,g_n-\sqrt{g_n})$, which is a region with volume $g_n^d -(g_n-\sqrt{g_n})^d $.  The leading terms $g_n^d$ cancel, so the next leading term is $\mathcal{O}\left(g_n^{d-1/2}\right)$. By Assumption \ref{assum:euclidean_space}.b, the number of units in any region cannot grow faster than its volume. So the fraction of the population within $\sqrt{g_n}$ of the boundary of their own cluster is therefore $\mathcal{O}\left(g_n^{d-1/2}/g_n^d\right)=\mathcal{O}\left(g_n^{-1/2}\right)\to 0$. So when clusters grow,  $\frac{1}{n}\sum_{i=1}^n\rho_i^{-\gamma}=\mathcal{O}\left(g_n^{-1/2}+g_n^{-\gamma /2}\right)\to 0$. 

Notice that $ \frac{1}{n^2}\sum_{i=1}^n\sum_{j=1}^n \Lambda_{ij}\lesssim \frac{\kappa_n^d+g_n^d}{n}$ and $\widehat{\theta}_{n,\kappa_n}^{OLS}$ is consistent. $g_n\to\infty$ and $\max_i|\widetilde{q}_i|<\overline{\phi}^{-1}$.  Since $\frac{1}{n}\sum_{i=1}^n\rho_i^{-\gamma} \to 0$ and we replace $\frac{1}{n}\sum_{i=1}^n Y_i$ in $\widehat{q}_i$ with its expectation by Lemma \ref{lem:ybar}:
\begin{align*}
     \frac{\overline{\phi}^2}{n^2}\sum_{i=1}^n\sum_{j=1}^n (\Lambda_{ij}-Q_{ij})\widehat{q}_i\widehat{q}_j =  \frac{\overline{\phi}^2}{n^2}\sum_{i=1}^n\sum_{j=1}^n (\Lambda_{ij}-Q_{ij})\widetilde{q}_i\widetilde{q}_j+o_p\left(\frac{\kappa_n^d+g_n^d}{n}\right)
\end{align*}

Substituting into (\ref{eq:Eqtilde}):
\begin{align}\label{eq:Efinal}
 [E] = \frac{1}{\mu_2^2}\frac{\overline{\phi}^2}{n^2}\sum_{i=1}^n\sum_{j=1}^n (\Lambda_{ij}-Q_{ij})\widehat{q}_i\widehat{q}_j+o_p\left(\frac{\kappa_n^d+g_n^d}{n}\right)
\end{align}

{\bf Concluding the Proof:} Combining the HAC estimator with (\ref{eq:Efinal}) and since [F] is positive:
\begin{align*}
&\frac{\frac{1}{n^2}\sum_{i=1}^n\sum_{j=1}^n\Lambda_{ij}\widehat{r}_i\widehat{r}_j}{\left(\frac{1}{n}\sum_{i=1}^n X_i^2 - \left(\frac{1}{n}\sum_{i=1}^n X_i\right)^2\right)^2}   -  \frac{ \frac{1}{n^2}\sum_{i=1}^n\sum_{j=1}^n (\Lambda_{ij}-Q_{ij})\widehat{q}_i\widehat{q}_j}{\left(\frac{1}{n}\sum_{i=1}^n X_i^2 - \left(\frac{1}{n}\sum_{i=1}^n X_i\right)^2\right)^2} \geq \mathbb{V}([\mathcal{A}]) +o_p\left(\frac{\kappa_n^{2d}+g_n^{2d}}{ng_n^{d}}\right)
\end{align*}

Since we assumed that $\liminf_{n\to\infty} \frac{n g_n^{d}}{\kappa_n^{2d}+g_n^{2d}}\mathbb{V}([\mathcal{A}])>0 $, we have:
$\frac{ \widehat{\mathbb{V}([\mathcal{A}])}}{\mathbb{V}([\mathcal{A}])}\geq 1+o_p(1) $

\section{Lemmas}\label{app:lemmas}

\subsection{Proof of Lemma \ref{lem:optimal_rate}}\label{proof:lem:optimal_rate}

 Fix a sequence of positive numbers $a_n$. Suppose that there is a sequence of experimental designs such that for all $\delta>0$,
\begin{align*}\sup_{\mathbf{Y}\in\mathcal{Y}_n}\mathbb{P}\left[a_n\left|\widehat{\theta}_n-\theta_n\right|> \delta\right]\to 0
\end{align*}
Let $\mathcal{J}_n\subseteq \mathcal{N}_n$ be a minimal $a_n^{1/\gamma}$ cover of $\mathcal{N}_n$. Let $k(i)\equiv \arg\min_{k\in\mathcal{J}_n} \rho(i,k)$ be the closest cover member to $i$. By construction $\rho(i,k(i))\leq a_n^{1/\gamma}$. It will be convenient to sum weights over all units $i$ that share a member of $k(i)\in\mathcal{J}_n$, so define $w_k=\sum_{i=1}^n\mathbf{1}\left\{k(i)=k\right\}\omega_i$.

The proof will proceed in four steps. In Step 1, we show that the ``effective sample size" cannot be greater than $|\mathcal{J}_n|$. In Steps 2-3 we show the estimator cannot always converge faster than the square root of its effective sample size. In Step 4 we conclude the proof and solve for the minimax rate of convergence.

{\bf Step 1:} Without loss of generality, let $\overline{Y}=c=1$. Consider the potential outcomes: $Y_i(\mathbf{d})=d_{k(i)}a_n^{-1}$ which makes $\theta_n=a_n^{-1}$ and $\widehat{\theta}_n = \frac{1}{n}\sum_{k=1}^{|\mathcal{J}_n|} a_n^{-1}D_kw_k$. These potential outcomes satisfy Assumptions \ref{assum:boundedoutcomes}, and \ref{assum:ANI}-\ref{assum:linearity}. By hypothesis, $\mathbb{P}\left[a_n\left|\frac{1}{n}\sum_{k=1}^{|\mathcal{J}_n|} a_n^{-1}D_kw_k - a_n^{-1}\right|>\delta\right]\to 0$. Distributing: $\frac{1}{n}\sum_{k=1}^{|\mathcal{J}_n|} D_kw_k\to_p 1$. By an identical argument, $\frac{1}{n}\sum_{k=1}^{|\mathcal{J}_n|} (1-D_k)w_k\to_p -1$. Combining: $\frac{1}{|\mathcal{J}_n|}\sum_{k=1}^{|\mathcal{J}_n|} \frac{|\mathcal{J}_n|}{n}|w_k| \geq \frac{1}{2}\left|\frac{1}{n}\sum_{k=1}^{|\mathcal{J}_n|} D_kw_k\right|+\frac{1}{2}\left|\frac{1}{n}\sum_{k=1}^{|\mathcal{J}_n|} (1-D_k)w_k\right|\geq 1+o_p(1)$. So:
\begin{equation}
    \mathbb{P}\left[\frac{1}{|\mathcal{J}_n|}\sum_{k=1}^{|\mathcal{J}_n|} \frac{|\mathcal{J}_n|}{n}|w_k| >1-\delta\right]\to 1
\end{equation}

{\bf Step 2:} Define the $|\mathcal{J}_n| \times 1$ random vectors $\mathbf{R}$ where the elements $R_k$ are  iid Rademacher random variables independent of $\mathbf{D}$. Next we will use the Paley–Zygmund Inequality to show:
$$ \liminf_{n\to \infty}\mathbb{P}\left[\frac{\sqrt{|\mathcal{J}_n|}}{|\mathcal{J}_n|}\sum_k \frac{|\mathcal{J}_n|}{n} w_k R_k >\delta\right] >0 $$
First notice that by independence and symmetry of $R_k$, $\frac{1}{|\mathcal{J}_n|}\sum_{k=1}^{|\mathcal{J}_n|} \frac{|\mathcal{J}_n|}{n}w_k R_k$ has the same distribution as $\frac{1}{|\mathcal{J}_n|}\sum_{k=1}^{|\mathcal{J}_n|} \frac{|\mathcal{J}_n|}{n}|w_k|R_k$. So it suffices to work with $|\omega_k|$. Define $c_k \equiv \frac{|\mathcal{J}_n|}{n}|w_k|$ and define $S_n \equiv \sum_{k=1}^{|\mathcal{J}_n|}c_kR_k$. So $\frac{\sqrt{|\mathcal{J}_n|}}{|\mathcal{J}_n|}\sum_{k=1}^{|\mathcal{J}_n|} \frac{|\mathcal{J}_n|}{n} w_k R_k  \stackrel{d}{=} \frac{S_n}{\sqrt{|\mathcal{J}_n|}}$. By Step 1: $\mathbb{P}\left[\frac{1}{|\mathcal{J}_n|}\sum_{k=1}^{|\mathcal{J}_n|} c_k > 1-\delta\right] \to 1$. Conditional on the $c_k$: $\mathbb{E}[S_n|\mathbf{c}]=0$ and $\mathbb{V}\left[S_n|\mathbf{c}\right] = \sum_{k=1}^{|\mathcal{J}_n|} c_k^2$. By Cauchy-Schwarz: $\sum_{k=1}^{|\mathcal{J}_n|} c_k^2 \geq |\mathcal{J}_n| \left(\frac{1}{|\mathcal{J}_n|}\sum_{k=1}^{|\mathcal{J}_n|} c_k\right)^2 $. By the Marcinkiewicz–Zygmund Inequality: $\mathbb{E}\left[S_n^4\mid \mathbf{c}\right] \leq 3\left(\sum_{k=1}^{|\mathcal{J}_n|} c_k^2\right)^2$. By the  Paley–Zygmund Inequality with constant $\frac{1}{2}$:
\begin{align*}
    \mathbb{P}\left[|S_n| \geq \sqrt{\frac{1}{2}\sum_{k=1}^{|\mathcal{J}_n|} c_k^2}\mid \mathbf{c}\right]=\mathbb{P}\left[S_n^2 \geq \frac{1}{2}\mathbb{E}\left[S_n^2 \mid \mathbf{c}\right]\mid \mathbf{c}\right]\geq \left(1-\frac{1}{2}\right)^2 \frac{\mathbb{E}\left[S_n^2 \mid \mathbf{c}\right]^2}{\mathbb{E}\left[S_n^4 \mid \mathbf{c}\right]} \geq \frac{\left(\sum_{k=1}^{|\mathcal{J}_n|} c_k^2\right)^2}{4\times 3\left(\sum_{k=1}^{|\mathcal{J}_n|} c_k^2\right)^2} \geq \frac{1}{12}
\end{align*}
By Cauchy-Schwarz:  $\sqrt{\sum_{k=1}^{|\mathcal{J}_n|} c_k^2} \geq \frac{1}{\sqrt{|\mathcal{J}_n|}}\left(\sum_{k=1}^{|\mathcal{J}_n|} c_k\right)$. By Step 1, for any $\epsilon <1$:
\begin{align*}
    \mathbb{P}\left[\sqrt{\sum_{k=1}^{|\mathcal{J}_n|} c_k^2} >\epsilon\sqrt{|\mathcal{J}_n|}\right] \geq  \mathbb{P}\left[  \frac{1}{{|\mathcal{J}_n|}}\sum_{k=1}^{|\mathcal{J}_n|} c_k  >\epsilon\right]\to 1
\end{align*}

By the symmetry of $S_n$, for any $\delta \in \left(0,\frac{1}{\sqrt{2}}\right)$: $  \mathbb{P}\left[\frac{S_n}{\sqrt{|\mathcal{J}_n|}} \geq  \delta \right] \geq \frac{1}{2} \mathbb{P}\left[|S_n|\geq \delta\sqrt{|\mathcal{J}_n|}\right] $. Then by the two  bounds just proven:
\begin{align*}
    \mathbb{P}\left[|S_n|\geq \delta\sqrt{|\mathcal{J}_n|}\right]&\geq \mathbb{P}\left[|S_n| \geq \sqrt{\frac{1}{2}\sum_{k=1}^{|\mathcal{J}_n|} c_k^2} \cap \sqrt{\sum_{k=1}^{|\mathcal{J}_n|} c_k^2} > \sqrt{2}\delta\sqrt{|\mathcal{J}_n|}\right]  \\&\geq   \mathbb{P}\left[|S_n| \geq \sqrt{\frac{1}{2}\sum_{k=1}^{|\mathcal{J}_n|} c_k^2}\mid \sqrt{\sum_{k=1}^{|\mathcal{J}_n|} c_k^2} > \sqrt{2}\delta\sqrt{|\mathcal{J}_n|}\right]  \mathbb{P}\left[\sqrt{\sum_{k=1}^{|\mathcal{J}_n|} c_k^2} > \sqrt{2}\delta\sqrt{|\mathcal{J}_n|}\right] \\
    &\geq \frac{1}{12}+o(1)
\end{align*}


So $\liminf_{n\to \infty}\mathbb{P}\left[\frac{S_n}{\sqrt{|\mathcal{J}_n|}} \geq  \delta \right] > 0$. Now substitute $\frac{\sqrt{|\mathcal{J}_n|}}{|\mathcal{J}_n|}\sum_{k=1}^{|\mathcal{J}_n|} \frac{|\mathcal{J}_n|}{n} w_k R_k  = \frac{S_n}{\sqrt{|\mathcal{J}_n|}}$. So we have shown:
\begin{equation}\label{eq:rademacher}
    \liminf_{n\to \infty}\mathbb{P}\left[\frac{\sqrt{|\mathcal{J}_n|}}{|\mathcal{J}_n|}\sum_{k=1}^{|\mathcal{J}_n|}  \frac{|\mathcal{J}_n|}{n}w_kR_k >\delta\right]=\liminf_{n\to \infty}\mathbb{P}\left[\frac{\sqrt{|\mathcal{J}_n|}}{|\mathcal{J}_n|}\sum_{k=1}^{|\mathcal{J}_n|}  \frac{|\mathcal{J}_n|}{n}|w_k|R_k >\delta\right] >0
\end{equation}

{\bf Step 3:} We now show that there exists a sequence of potential outcomes such that: $ \liminf_{n\to \infty}\sup_{\mathbf{Y}\in\mathcal{Y}_n}\mathbb{P}\left[\sqrt{|\mathcal{J}_n|}\left|\widehat{\theta}_n-\theta_n\right|> \delta\right] >0$. To do this, notice that since (\ref{eq:rademacher}) holds over randomly selected $\mathbf{R}$, it also holds conditional on some non-stochastic realized vectors $\mathbf{r}\in \{-1,1\}$:
$$  \liminf_{n\to \infty}\max_{\mathbf{r}\in \{-1,1\}^{|\mathcal{J}_n|}}\mathbb{P}\left[\frac{\sqrt{|\mathcal{J}_n|}}{|\mathcal{J}_n|}\sum_k \frac{|\mathcal{J}_n|}{n} w_k r_k >\delta\right] >0 $$

Now consider the potential outcomes $Y_i=r_{k(i)}$ where the vector $\mathbf{r}$ for each $n$ is the maximizer in the limit above. Since these do not depend on treatment at all, $\theta_n=0$. They satisfy Assumptions \ref{assum:boundedoutcomes} and \ref{assum:ANI}-\ref{assum:linearity}. Under these potential outcomes: 
$$ \liminf_{n\to \infty}\mathbb{P}\left[\sqrt{|\mathcal{J}_n|}\left|\widehat{\theta}_n-\theta_n\right|> \delta\right]=\liminf_{n\to \infty}\max_{\mathbf{r}\in \{-1,1\}^{|\mathcal{J}_n|}}\mathbb{P}\left[\frac{\sqrt{|\mathcal{J}_n|}}{|\mathcal{J}_n|}\sum_k \frac{|\mathcal{J}_n|}{n} w_k r_k >\delta\right] >0$$

{\bf Step 4:} Let $\delta  = \frac{1}{2}$. We have two guarantees, one by hypothesis and the other just shown:\begin{align*}
\sup_{\mathbf{Y}\in\mathcal{Y}_n}\mathbb{P}\left[a_n\left|\widehat{\theta}_n-\theta_n\right|> \frac{1}{2}\right]&\to 0 \quad \text{ and}\quad 
      \liminf_{n\to \infty}\sup_{\mathbf{Y}\in\mathcal{Y}_n}\mathbb{P}\left[\sqrt{|\mathcal{J}_n|}\left|\widehat{\theta}_n-\theta_n\right|> \frac{1}{2}\right]>0
\end{align*}
For both to be true, it is necessary that: $a_n/\sqrt{|\mathcal{J}_n|} \to 0$. So $a_n \lesssim \sqrt{|\mathcal{J}_n|} \lesssim \sqrt{f\left(a_n^{1/\gamma},n\right)}$.

\subsection{Lemma \ref{lem:phi} }

\begin{lemma}{\bf Bounding $\phi$}\label{lem:phi}  

Consider any sequence of populations $\mathcal{N}_n$ in $\mathbb{R}^d$. Let Assumption \ref{assum:euclidean_space} hold and let the design be a Scaling Clusters design with sequence of cluster sizes $g_n>\rho_0/2$ and constant $K_{S}$. Let $s_n$ be a sequence of positive numbers. Then:
$$\left(\frac{s_n+g_n}{g_n}\right)^d  \lesssim \overline{\phi} \leq \max_{i\in \mathcal{N}_n}\phi(i,s_n) \lesssim \left(\frac{s_n+g_n}{g_n}\right)^d $$

\end{lemma}

{\bf Proof:}   First we derive the upper bounds and then the lower bounds. Since the design is Scaling Clusters, $\phi(i,s_n) $ is upper bounded by the number of cluster centers $q_c$ with $K_{S}g_n$-neighborhoods that intersect $\mathcal{N}(i,s_n)$. By the triangle inequality,  $\phi(i,s_n) $ is upper bounded by the number of cluster centers inside $\mathcal{N}(i, s_n+K_{S}g_n)$. The $g_n$-neighborhoods of all of these cluster centers are mutually exclusive and must be contained within $\mathcal{N}(i, s_n+(K_{S}+1)g_n)$. Thus $\phi(i,s_n)$ is upper bounded by the number of balls of radius $g_n$ that can fit inside a ball of radius $s_n+(K_{S}+1)g_n$. The number of balls in $\mathbb{R}^d$ of radius $x$ that can fit inside a ball of radius $y>x$ is upper bounded by one plus the volume ratio: $\mathcal{O}\left(\frac{y^d}{x^d}+1\right)$. So $\phi(i,s_n) \lesssim \left(\frac{s_n}{g_n}+1\right)^d$.

Now we derive the lower bound on $\overline{\phi}$. By Assumption \ref{assum:euclidean_space}.b, each cluster can have at most $\mathcal{O}\left(g_n^d\right)$ members. Therefore $\frac{|\mathcal{N}(i,s_n)|}{1+g_n^d}\lesssim \phi(i,s_n)$. Taking the average: $ \frac{1}{n(1+g_n^d)} \sum_{i=1}^n |\mathcal{N}(i,s_n)|\lesssim \overline{\phi} $. To bound the mean neighborhood membership $\frac{1}{n} \sum_{i=1}^n |\mathcal{N}(i,s_n)|$, first notice that by Assumption \ref{assum:euclidean_space}.a, there is a ball in $\mathbb{R}^d$ of diameter $Rn^{1/d}$ that contains the entire population. Any such ball can be covered by a set of $M_n$ mutually exclusive  hypercubes $m_k$ each with diameter $\frac{s_n}{2}$ where $M_n \lesssim \frac{n}{s_n^d}$. So if any two units $i,j$ are inside the same hypercube $m_k$, then $\rho(i,j)\leq s_n$. We can now lower bound the mean neighborhood membership with Cauchy-Schwarz: $\frac{1}{n} \sum_{i=1}^n |\mathcal{N}(i,s_n)|=\frac{1}{n} \sum_{i=1}^n \sum_{j=1}^n\mathbf{1}\{\rho(i,j)\leq s_n\}\geq \frac{1}{n} \sum_{k=1}^{M_n} |m_k|^2 \geq \frac{M_n}{n} \left(\frac{n}{M_n}\right)^2 = \frac{n}{M_n}$. Since $M_n\lesssim  \frac{n}{s_n^d}$, by substitution we have: $s_n^d \lesssim \frac{1}{n} \sum_{i=1}^n |\mathcal{N}(i,s_n)|$. So we have:  $ \frac{s_n^d}{1+g_n^d}\lesssim \overline{\phi} $. Since $g_n > \rho_0/2>0$, $\frac{s_n^d}{1+g_n^d}= \frac{s_n^d}{g_n^d}\frac{g_n^d}{1+g_n^d}>\frac{s_n^d}{g_n^d}\frac{\rho_0/2}{1+\rho_0/2}$. So $\frac{s_n^d}{g_n^d} \lesssim \frac{s_n^d}{1+g_n^d}$. Since $\overline{\phi} \geq 1$, $\overline{\phi}+1 \lesssim \overline{\phi}$. So $ \frac{s_n^d+g_n^d}{g_n^d}\lesssim \overline{\phi}$. Since $ (s_n+g_n)^d \leq 2^{d-1}(s_n^d+g_n^d)$ and $d$ is constant,  $ \left(\frac{s_n+g_n}{g_n}\right)^d\lesssim \overline{\phi}$.

\subsection{ Lemma \ref{lem:covar2}} 

\begin{lemma} {\bf Moments of the empirical covariance}\label{lem:covar2}

Let $\mathbf{v}$ be a $C\times 1$ vector of mean zero i.i.d.  random variables with second moment $\mu_2$ and fourth moment $\mu_4$. Let $M,W$ be $n\times C$ matrices of real numbers. Denote $Q \equiv M'W$. Then,
\begin{align*}
\mathbb{E}\left[\left(\frac{\mathbf{1}'M\mathbf{v}}{n}\right)^2\right]&= \frac{\mu_2||M'\mathbf{1}||_2^2}{n^2},\qquad 
\mathbb{E}\left[\frac{\mathbf{v}'M'W\mathbf{v}}{n}\right] = \frac{\mu_2tr(Q)}{n}\\ 
\mathbb{V}\left[\frac{\mathbf{v}'M'W\mathbf{v}}{n}\right] &= \frac{(\mu_4-3\mu_2^2)\sum_{i=1}^CQ_{i,i}^2 + \mu_2^2tr(Q^2)+\mu_2^2tr(Q'Q)}{n^2}\lesssim \frac{tr(Q'Q)}{n^2} 
\end{align*}

\end{lemma}

   \paragraph{Proof of Claim 1: Expectation of the vector product.}$$\mathbb{E}\left[\left(\frac{\mathbf{1}'M\mathbf{v}}{n}\right)^2\right]=\frac{1}{n^2}\sum_{i=1}^n\sum_{c=1}^C\sum_{j=1}^n\sum_{d=1}^CM_{ic}M_{jd}\mathbb{E}[v_cv_d]= \frac{1}{n^2}\sum_{i=1}^n\sum_{c=1}^C\sum_{j=1}^nM_{ic}M_{jc}\mu_2= \frac{\mu_2}{n^2}||M'\mathbf{1}||_2^2 $$
    
\paragraph{Proof of Claim 2: Expectation of the quadratic form.} Since $\mathbf{v}$ are mean zero and i.i.d: 
\begin{align*}
    \mathbb{E}\left[ \frac{\mathbf{v}'Q\mathbf{v}}{n}\right] &= \frac{\sum_{j=1}^C\sum_{i=1}^C Q_{ij}\mathbf{v}_i\mathbf{v}_j}{n}
    = \frac{\mu_2\sum_{i=1}^C Q_{ii}}{n}
    = \frac{\mu_2tr(Q)}{n}
\end{align*}

\paragraph{Proof of Claim 3: Variance of the quadratic form.}

Now we compute $\mathbb{V}\left(\frac{\mathbf{v}'Q\mathbf{v}}{n}\right)$. Evaluating the expectation of the squared quadratic form: We compute $\mathbb{V}\!\left(\frac{\mathbf{v}'Q\mathbf{v}}{n}\right)$.
Using independence and zero mean of the $v_i$'s we see that: $\mathbb{E}[v_i v_j v_k v_m]= \mu_4$ if  $i=j=k=m$. This expectation equals $\mu_2^2$ if  $(i,j,k,m)$ \text{ form two equal pairs} and equals $0$ otherwise. Therefore:
\begin{align*}
\mathbb{E}\!\left[(\mathbf{v}'Q\mathbf{v})^2\right]
&= \sum_{i,j,k,m} Q_{ij}Q_{km}\,\mathbb{E}[v_i v_j v_k v_m]= (\mu_4-3\mu_2^2)\sum_{i} Q_{ii}^2
   + \mu_2^2\!\left(\operatorname{tr}(Q)^2
   + \operatorname{tr}(Q^2)
   + \operatorname{tr}(Q'Q)\right).
\end{align*}

Since $
\mathbb{E}\!\left[\frac{\mathbf{v}'Q\mathbf{v}}{n}\right]
   = \frac{\mu_2\,\operatorname{tr}(Q)}{n},$, 
its square equals $\mu_2^2\operatorname{tr}(Q)^2/n^2$.  
Therefore:
\[
\mathbb{V}\!\left(\frac{\mathbf{v}'Q\mathbf{v}}{n}\right)
= \frac{(\mu_4-3\mu_2^2)\sum_{i}Q_{ii}^2
+ \mu_2^2\operatorname{tr}(Q^2)
+ \mu_2^2\operatorname{tr}(Q'Q)}{n^2}.
\]

\noindent By Cauchy-Schwartz: $tr(QQ) \leq tr(Q'Q)$ and $\sum_{i=1}^C Q_{ii}^2 \leq tr(Q'Q)$. So the leading variance term is $\frac{tr(Q'Q)}{n^2}$.

\subsection{Lemma \ref{lem:ols_decomp_generic} and proof}  


\begin{lemma}\label{lem:ols_decomp_generic}
Consider sequences of random variables $A_n,B_n$. Assume that: $\frac{A_n-\mathbb{E}[A_n]}{ \mathbb{E}[B_n]} \to_p 0$, $\sup_n\left|\frac{\mathbb{E}[A_n]}{\mathbb{E}[B_n]}\right|< \infty$, and $\frac{B_n-\mathbb{E}[B_n]}{ \mathbb{E}[B_n]} \to_p 0$ and $\liminf_{n\to \infty}\mathbb{E}[B_n] >0$ and $B_n \neq 0$ a.s. Then:

 $$  \frac{A_n}{B_n} -\frac{\mathbb{E}[A_n]}{\mathbb{E}[B_n]}=
\frac{A_n - B_n \frac{\mathbb{E}[A_n]}{\mathbb{E}[B_n]}}{\mathbb{E}[B_n]} + o_p\left(\frac{B_n-\mathbb{E}[B_n]}{\mathbb{E}[B_n]}\right)$$

\end{lemma}
    
{\bf Proof:} This is a typical Taylor result and so this proof is terse. Taking a first-order Taylor Expansion of $\frac{\mathbb{E}[B_n]}{B_n}$ about 1: $\frac{\mathbb{E}[B_n]}{B_n} = 1- \frac{B_n-\mathbb{E}[B_n]}{\mathbb{E}[B_n]}+\mathcal{O}_p\left(\left(\frac{B_n-\mathbb{E}[B_n]}{\mathbb{E}[B_n]}\right)^2\right)$. Therefore: $
    \frac{A_n}{B_n} =  \frac{A_n}{\mathbb{E}[B_n]} \frac{\mathbb{E}[B_n]}{B_n} =  \frac{A_n}{\mathbb{E}[B_n]} \left(1- \frac{B_n-\mathbb{E}[B_n]}{\mathbb{E}[B_n]}+\mathcal{O}_p\left(\left(\frac{B_n-\mathbb{E}[B_n]}{\mathbb{E}[B_n]}\right)^2\right)\right)$ Since $\frac{A_n-\mathbb{E}[A_n]}{\mathbb{E}[B_n]}\to_p 0$, we have: $
  \frac{A_n}{\mathbb{E}[B_n]}\frac{B_n-\mathbb{E}[B_n]}{\mathbb{E}[B_n]} =   \frac{\mathbb{E}[A_n]}{\mathbb{E}[B_n]}\frac{B_n-\mathbb{E}[B_n]}{\mathbb{E}[B_n]}+o_p\left(\frac{B_n-\mathbb{E}[B_n]}{\mathbb{E}[B_n]}\right) $. Substituting this back in (and using the assumption that $\frac{B_n-\mathbb{E}[B_n]}{ \mathbb{E}[B_n]} \to_p 0$ and $\sup_n\left|\frac{\mathbb{E}[A_n]}{\mathbb{E}[B_n]}\right|< \infty$ to combine remainders) yields the desired result: $ \frac{A_n}{B_n} -\frac{\mathbb{E}[A_n]}{\mathbb{E}[B_n]}=   \frac{A_n - B_n \frac{\mathbb{E}[A_n]}{\mathbb{E}[B_n]}}{\mathbb{E}[B_n]} +o_p\left( \frac{B_n-\mathbb{E}[B_n]}{\mathbb{E}[B_n]} \right)$

\subsection{Lemma \ref{lem:control_x_sharp} and proof} 

\begin{lemma}\label{lem:control_x_sharp}
Under the assumptions of Theorem \ref{thm:ols_bstar} and $\frac{\kappa_n^d+g_n^d}{n}\to 0$, the following  statements hold:
$$ \frac{1}{n}\sum_{i=1}^n \mathbb{E}[X_i^2] = p(1-p) \overline{\phi}^{-1},\qquad \mathbb{E}\left[\overline{X}^2\right] \lesssim {\frac{\kappa_n^d+g_n^d}{\overline{\phi}n}}, \quad    \mathbb{V}\left[\frac{1}{n} \sum_{i=1}^n X_i^2\right]\lesssim \frac{\kappa_n^d+g_n^d}{n\overline{\phi}^2} $$
\end{lemma}
{\bf Proof:} Throughout we make use of Lemma \ref{lem:phi}. First we show that the expectation is: $ \mathbb{E}\left[\frac{1}{n} \sum_{i=1}^n X_i^2\right] = p(1-p)\overline{\phi}^{-1}$.  Since clusters are treated iid, we compute: 

\noindent $\mathbb{E}[X_i^2] = \overline{\phi}^{-2}\sum_{c=1 }^{m_n}\mathbbm{1}\left\{\mathcal{N}(i,\kappa_n) \cap \mathcal{P}_c\neq \emptyset\right\}\mathbb{E}\left[(W_c-p)^2\right] = \overline{\phi}^{-2}\phi(i,\kappa_n)p(1-p)$. Taking the mean: $\frac{1}{n}\sum_{i=1}^n \mathbb{E}[X_i^2] =\frac{1}{n}\sum_{i=1}^n \overline{\phi}^{-2}\phi(i,\kappa_n)p(1-p) = p(1-p) \overline{\phi}^{-1} $.  

Now we prove the second claim. By construction $\mathbb{E}[X_i]=0$ and each $X_i$ depends on at most $\mathcal{O}\left(\kappa_n^d+g_n^d\right)$ other $X_j$ and by the first claim just proven $\max_i\mathbb{V}[X_i]\lesssim \overline{\phi}^{-1}$. So  $\mathbb{E}\left[\overline{X}^2\right] \lesssim \frac{\kappa_n^d+g_n^d}{\overline{\phi}n}$. 

To show the third claim, we first bound $\mathbb{V}[X_i^2]$ with the Marcinkiewicz–Zygmund inequality using the clustered randomization. To see the last step below, recall the definition of $B$ from Section \ref{proof:prop:ols_asymptotic_decomp} and notice that $\sum_{c=1}^{m_n}(B_{ic}V_c)^2 \leq \phi(i,\kappa_n)$. Then use Lemma \ref{lem:phi}.
   \begin{align*}
   \mathbb{V}[X_i^2] &= \frac{1}{\overline{\phi}^4}\mathbb{V}\left[\left(\sum_{c=1}^{m_n}B_{ic}V_c\right)^2\right]\leq  \frac{1}{\overline{\phi}^4}\mathbb{E}\left[\left(\sum_{c=1}^{m_n}B_{ic}V_c\right)^4\right]\lesssim   \frac{1}{\overline{\phi}^4}\mathbb{E}\left[\left(\sum_{c=1}^{m_n}(B_{ic}V_c)^2\right)^2\right]\lesssim \frac{1}{\overline{\phi}^4}\overline{\phi}^2
    \end{align*}

The variance of the mean of squares can then be bounded using the fact that each $X_i$ is independent of all but $\mathcal{O}\left(\kappa_n^d+g_n^d\right)$ other $X_j$: $ \mathbb{V}\left[\frac{1}{n} \sum_{i=1}^n X_i^2\right]\lesssim  \frac{\kappa_n^d+g_n^d}{n}\max_i\mathbb{V}[X_i^2]\lesssim \frac{\kappa_n^d+g_n^d}{n\overline{\phi}^2}$.


\subsection{Lemma \ref{lem:ybar} and proof}  

\begin{lemma}\label{lem:ybar}
Let Assumptions \ref{assum:boundedoutcomes}-\ref{assum:ANI} hold and suppose that the researcher uses a Scaling Clusters design with cluster scaling sequence $g_n$. Then:  $\frac{1}{n}\sum_{i=1}^n \left(Y_i - \mathbb{E}[Y_i]\right)\lesssim \sqrt{\frac{g_n^d}{n}}+g_n^{-\gamma} $.
\end{lemma}

{\noindent \bf Proof:} We can always decompose: \begin{align*}
    \frac{1}{n}\sum_{i=1}^n \left(Y_i-\mathbb{E}[Y_i]\right) = \frac{1}{n} \sum_{i=1}^n \left(Y_i-\mathbb{E}\left[Y_i|\mathcal{F}_{i,g_n}\right]\right)+ \frac{1}{n} \sum_{i=1}^n \left(\mathbb{E}\left[Y_i|\mathcal{F}_{i,g_n}\right]-\mathbb{E}[Y_i]\right)
\end{align*}

By Assumption \ref{assum:ANI}, $\left|\frac{1}{n} \sum_{i=1}^n \left(Y_i-\mathbb{E}\left[Y_i|\mathcal{F}_{i,g_n}\right]\right)\right| \leq cg_n^{-\gamma}$ almost surely. 

The sum $\frac{1}{n} \sum_{i=1}^n \left(\mathbb{E}\left[Y_i|\mathcal{F}_{i,g_n}\right]-\mathbb{E}[Y_i]\right)$ has expectation zero by construction. By the clustered randomization and Scaling Clusters design, each summand $\left(\mathbb{E}\left[Y_i|\mathcal{F}_{i,g_n}\right]-\mathbb{E}[Y_i]\right)$  is independent of $\left(\mathbb{E}\left[Y_j|\mathcal{F}_{j,g_n}\right]-\mathbb{E}[Y_j]\right)$ when $\rho(i,j)> Cg_n$ for some constant $C>2(1+K_S)$. So each $\left(\mathbb{E}\left[Y_i|\mathcal{F}_{i,g_n}\right]-\mathbb{E}[Y_i]\right)$  is dependent on at most $\mathcal{O}\left(g_n^d\right)$ others. By Assumption \ref{assum:boundedoutcomes}, $\max_i|\mathbb{E}\left[Y_i|\mathcal{F}_{i,g_n}\right]-\mathbb{E}[Y_i]|<2\overline{Y}$  and therefore $\max_i\mathbb{V}\left[\mathbb{E}\left[Y_i|\mathcal{F}_{i,g_n}\right]-\mathbb{E}[Y_i]\right] \leq 4\overline{Y}^2$. So, $\mathbb{V}\left[\frac{1}{n} \sum_{i=1}^n \left(\mathbb{E}\left[Y_i|\mathcal{F}_{i,g_n}\right]-\mathbb{E}[Y_i]\right)\right] \lesssim \frac{g_n^d}{n}$. By Chebyshev: $ \frac{1}{n}\sum_{i=1}^n \left(Y_i-\mathbb{E}[Y_i]\right) \lesssim \sqrt{\frac{g_n^d}{n}}+g_n^{-\gamma}$.

\subsection{Lemma \ref{lem:columnsumsB} and proof}  

\begin{lemma}\label{lem:columnsumsB}
    Let Assumption \ref{assum:euclidean_space} hold. If the experimental design is a Scaling Clusters design with $\frac{g_n^{d}+\kappa_n^{d}}{n}\to 0$. Then: $\frac{1}{n^2}\sum_{c=1}^{m_n}\left(\sum_{i=1}^n\mathbbm{1}\left\{\mathcal{N}(i,\kappa_n) \cap \mathcal{P}_c\neq \emptyset\right\}\right)^2  \sim \frac{\kappa_n^{2d}+g_n^{2d}}{ng_n^d} $
    
\end{lemma}
\begin{proof}  {\noindent \bf Lower Bound:} Notice that: $\frac{1}{m_n}\sum_{c=1}^{m_n}\sum_{i=1}^n\mathbbm{1}\left\{\mathcal{N}(i,\kappa_n) \cap \mathcal{P}_c\neq \emptyset\right\} = \frac{n\overline{\phi}}{m_n}$. The sum of squares is minimized by setting all summands equal to their mean. So:  $$\frac{1}{n^2}\sum_{c=1}^{m_n}\left(\sum_{i=1}^n\mathbbm{1}\left\{\mathcal{N}(i,\kappa_n) \cap \mathcal{P}_c\neq \emptyset\right\}\right)^2 \geq \frac{1}{n^2}\sum_{c=1}^{m_n}\frac{n^2\overline{\phi}^2}{m_n^2}  \sim  \frac{\overline{\phi}^2}{m_n}$$ 

By Assumption \ref{assum:euclidean_space}.b and the Scaling Clusters Design, $m_n \sim \frac{n}{g_n^d}$. By Lemma \ref{lem:phi}, $\overline{\phi} \sim \frac{\kappa_n^d+g_n^d}{g_n^d}$. So $\frac{\overline{\phi}^2}{m_n} \sim \frac{\kappa_n^{2d}+g_n^{2d}}{g_n^dn}$. So $\frac{\kappa_n^{2d}+g_n^{2d}}{g_n^dn}\lesssim \frac{1}{n^2}\sum_{c=1}^{m_n}\left(\sum_{i=1}^n\mathbbm{1}\left\{\mathcal{N}(i,\kappa_n) \cap \mathcal{P}_c\neq \emptyset\right\}\right)^2$.

{\noindent \bf Upper Bound:} Since the Scaling Clusters design requires that all clusters be contained in balls of radius $K_Sg_n$, then by the triangle inequality: $\sum_{i=1}^n\mathbbm{1}\left\{\mathcal{N}(i,\kappa_n) \cap \mathcal{P}_c\neq \emptyset\right\} \leq |\mathcal{N}(q_c,\kappa_n+K_Sg_n)|$. By Assumption \ref{assum:euclidean_space}.b, $|\mathcal{N}(i,\kappa_n+K_Sg_n)|\lesssim \kappa_n^d+g_n^d$. So $\frac{1}{n^2}\sum_{c=1}^{m_n}\left(\sum_{i=1}^n\mathbbm{1}\left\{\mathcal{N}(i,\kappa_n) \cap \mathcal{P}_c\neq \emptyset\right\}\right)^2\lesssim \frac{m_n}{n^2} (\kappa_n^{2d}+g_n^{2d}) \lesssim \frac{\kappa_n^{2d}+g_n^{2d} }{ng_n^d}$
\end{proof}

\subsection{Lemma \ref{lem:pot_outcomes_linearity} and proof}\label{proof:lem:pot_outcomes_linearity} 

\begin{lemma}\label{lem:pot_outcomes_linearity}
    Recall the matrices from Section \ref{proof:prop:ols_asymptotic_decomp}. If Assumptions \ref{assum:boundedoutcomes}-\ref{assum:linearity} hold, then:
\begin{align*}
    Y_i &=(\beta_0+p\theta_n)+ (AC)_i\mathbf{V} +p(A_i\mathbf{1}-\theta_n)+\epsilon_i\\
    \mathbb{E}[Y_i|\mathcal{F}_{i,\kappa_n}] &= (\beta_0+p\theta_n)+ (\widetilde{A}C)_i\mathbf{V} +p(A_i\mathbf{1}-\theta_n)+\epsilon_i
\end{align*}
\end{lemma}

{\bf Proof:} To prove the first equality, we start with Assumption \ref{assum:linearity}: $Y_i = \beta_0 +\sum_{j\in\mathcal{N}_n}A_{ij}D_j +\epsilon_i$. Since $\mathbf{D} = C\mathbf{W}$, we have $Y_i = \beta_0 +(AC)_i\mathbf{W} +\epsilon_i$. Since $\mathbf{V} = \mathbf{W}-p\mathbf{1}$, we have: $Y_i = \beta_0 +(AC)_i\mathbf{V}+p(AC)_i\mathbf{1} +\epsilon_i$. To make the deterministic terms that vary with $i$ sum to zero, we subtract off their mean:  $Y_i = (\beta_0+p\frac{1}{n}\sum_{j=1}^n(AC)_j\mathbf{1}) +(AC)_i\mathbf{V}+p((AC)_i\mathbf{1}-\frac{1}{n}\sum_{j=1}^n(AC)_j\mathbf{1}) +\epsilon_i$. Since $\frac{1}{n}\sum_{j=1}^n(AC)_j\mathbf{1} = \theta_n$, we have $Y_i = (\beta_0+p\theta_n) +(AC)_i\mathbf{V}+p((AC)_i\mathbf{1}-\theta_n) +\epsilon_i$. Finally, since $C\mathbf{1} = \mathbf{1}$, we have $Y_i = (\beta_0+p\theta_n) +(AC)_i\mathbf{V}+p(A_i\mathbf{1}-\theta_n) +\epsilon_i$. By Assumption \ref{assum:linearity}, $\sum_{i=1}^n \epsilon_i=0$, so  $\sum_{i=1}^n p(A_i\mathbf{1}-\theta_n)+\epsilon_i = 0 $.

To prove the second equality, we take the conditional expectation of the first: $ \mathbb{E}[Y_i|\mathcal{F}_{i,\kappa_n}] = (\beta_0+p\theta_n)+ \sum_{c=1}^{m_n}({A}C)_{ic}\mathbb{E}[V_c|\mathcal{F}_{i,\kappa_n}] +p(A_i\mathbf{1}-\theta_n)+\epsilon_i$. For any cluster $c$ with at least one member unit inside $\mathcal{N}(i,\kappa_n)$, $\sigma(V_c)\subset \mathcal{F}_{i,\kappa_n}$. By the cluster-randomized design, if cluster $c$ does not intersect $\mathcal{N}(i,\kappa_n)$, $V_c \independent  \mathcal{F}(i,\kappa_n)$. So, $\mathbb{E}[V_c|\mathcal{F}_{i,\kappa_n}] = V_c\mathbf{1}\left\{\mathcal{N}(i,\kappa_n)\cap \mathcal{P}_c \neq \emptyset \right\}+\mathbb{E}[V_c]\mathbf{1}\left\{\mathcal{N}(i,\kappa_n)\cap \mathcal{P}_c = \emptyset \right\}=V_c\mathbf{1}\left\{\mathcal{N}(i,\kappa_n)\cap \mathcal{P}_c \neq \emptyset \right\}$. Moreover, $(\widetilde{A}C)_{ic} = (AC)_{ic}\mathbf{1}\left\{\mathcal{N}(i,\kappa_n)\cap \mathcal{P}_c \neq \emptyset \right\}$. So $\sum_{c=1}^{m_n}({A}C)_{ic}\mathbb{E}[V_c|\mathcal{F}_{i,\kappa_n}] = \sum_{c=1}^{m_n}({A}C)_{ic}\mathbf{1}\left\{\mathcal{N}(i,\kappa_n)\cap \mathcal{P}_c \neq \emptyset \right\}V_c =  \sum_{c=1}^{m_n}(\widetilde{A}C)_{ic}V_c = (\widetilde{A}C)_{i}\mathbf{V} $ and the second claim follows.


\clearpage
\begin{center}
    {\huge \centering Online Appendix}
\end{center}

\section{The GATE when a Subset are Ineligible for Treatment}\label{sec:subset}  

This appendix considers a scenario in which, unlike in the main text, a subset $\mathcal{I}_n \subset \mathcal{N}_n$ of units in the population are ``treatment ineligible," meaning that they are never treated in the experiment, but still contribute to the GATE. We show that the results in the main text go through in this scenario after a suitable redefinition of the estimand.

Call the non-ineligible units in $\mathcal{N}_n \setminus \mathcal{I}_n$ the ``treatment eligible" units and assume that they are all members of a scaling clusters design with cluster size parameter $g_n$ and Bernoulli cluster treatment probability $p\in(0,1)$. The researcher now wishes to estimate a variant of the GATE, $\theta_n^{(1)}$, which is the average causal effect of treating all eligible units, as opposed to none of them, on the entire study population $\mathcal{N}_n$. Let $\mathbf{1}_{\mathcal{N}_n\setminus\mathcal{I}_n}$ be the $n\times 1$ vector which indicates membership in $\mathcal{N}_n\setminus\mathcal{I}_n$. Define this variant as:

$$\theta_n^{(1)}\equiv \frac{1}{n}\sum_{i=1}^n(Y_i(\mathbf{1}_{\mathcal{N}_n\setminus\mathcal{I}_n}) - Y_i(\mathbf{0}))$$

The estimand $\theta_n^{(1)}$ is straightforward to estimate: the researcher need only (a) design and run an experiment that treats eligibles, temporarily ignoring the ineligibles, but then (b) calculate estimators including the ineligibles, treating each ineligible unit as if it were a member of its respective nearest cluster. If the clusters thus created are still a scaling clusters design, then all of the results in this paper will apply, but for the estimand $\theta_n^{(1)}$ instead of $\theta_n$.
\vskip 0.1in
{\bf Proof:} Consider an alternative DGP $(\mathcal{D}_n',\mathbf{Y}')$ with the same population and distance function as the factual one but a different experimental design  $\mathcal{D}_n'$ and set of potential outcomes $\mathbf{Y}'$. Let the alternative experimental design $\mathcal{D}_n'$ be the same as the factual one $\mathcal{D}_n$, except that now each ineligible unit is instead a member of the cluster nearest to it. Let the alternative set of potential outcomes be $Y_i'(\mathbf{D})=Y_i(\mathbf{D}\odot \mathbf{1}_{\mathcal{N}_n\setminus\mathcal{I}_n} )$ where $\odot$ is the element-wise product. The GATE under the alternative DGP is now equal to $\theta_n^{(1)}$ since under alternative DGP, the treatment assignment of ineligible units has no effect on any outcome. If $Y_i$ satisfies Assumptions \ref{assum:boundedoutcomes}, \ref{assum:ANI}, and/or \ref{assum:linearity} then so does $Y_i'$. So under the alternative DGP, the conclusions of Theorems \ref{thm:optimalrate}-\ref{thm:varest} would still hold. 

Since the vector of unit treatment assignments $\mathbf{D}$ is a function of the vector of cluster assignments $\mathbf{W}$ we can always write any set of potential outcomes as functions of $\mathbf{W}$ rather than $\mathbf{D}$. Notice also that when we do this, $Y_i'(\mathbf{W})=Y_i(\mathbf{W})$ for all vectors of cluster treatment assignments $\mathbf{W}$. The vector $\mathbf{W}$ is identical under both the factual and alternative DGP. So the factual joint distribution of $\mathbf{Y},\mathbf{W}$ under $\mathcal{D}_n$ is identical to the alternative joint distribution of $\mathbf{Y}',\mathbf{W}'$ under $\mathcal{D}_n'$. All estimators in this paper are functions of $\mathbf{Y},\mathbf{W}$. So the researcher can always calculate based on the factual data what the value of any of the estimators in this paper would have been under the alternative DGP $(\mathcal{D}_n',\mathbf{Y}')$---and the conclusions of Theorems \ref{thm:optimalrate}-\ref{thm:varest} hold under alternative DGP $(\mathcal{D}_n',\mathbf{Y}')$.

\section{Experimental Designs with Cross-Cluster Treatment Dependence}\label{sec:stratification}

This appendix provides results for experimental designs that do not assign treatment to clusters independently, as in Equation \ref{eq:design}. First we consider a broad class of designs where the cluster treatments are allowed to be correlated with one another so long as (1) treatments within small spatial balls are not close to being linearly dependent and (2) far-apart treatments are independent. In Section \ref{sec:correlated_designs} we formalize these two requirements with Assumption \ref{assum:blockrand} and show that under it: (i) a modified OLS estimator converges to the GATE at the same rate as in Theorem \ref{thm:ols_bstar}, and that (ii) a CLT for this estimator holds.

Then we show that the IPW estimator requires stronger conditions on the experimental design in order to be consistent. We consider two special cases of Assumption \ref{assum:blockrand}. Subsection \ref{sec:stratified_bigclust} considers the special case of a stratified experiment, for which IPW can be consistent. In contrast, Subsection \ref{subsec:random_saturation} shows that IPW is inconsistent in the special case of random saturation designs, i.e. those in which there are no clusters and instead units are assigned to spatial groupings and each unit within group $g$ is then treated with group-specific probability $p_g \in (0,1)$ as in, e.g. \cite{Baird}.


\subsection{Regressions for Correlated Designs}\label{sec:correlated_designs}

Since linear regressions are more tolerant of incompletely treated or untreated neighborhoods, they will work under looser assumptions on the experimental design. Here we adjust the OLS estimator to allow for designs with meaningfully correlated treatment assignments, including most spatially stratified designs. In order for the GATE to be consistently estimable, we will require those correlations to not be too strong over small distances. Specifically, define $\Omega(i)$ as the $\phi(i,\kappa_n)\times \phi(i,\kappa_n)$ covariance matrix of the cluster treatment assignments $W_c$ for clusters $c$ within distance $\kappa_n$ of unit $i$ (i.e. the same cluster set as $B_i$). The key condition on the design and estimator jointly is that, for the sequence of radii $\kappa_n$ we want to choose, all of the $\Omega(i)$ are invertible and the column sums of $\Omega(i)$ and their inverses do not explode:

\begin{assumption}\label{assum:blockrand}
    The researcher chooses a sequence of radii $\kappa_n=o(n^{1/d})$. Given those radii, the following hold:
    \begin{enumerate}[label=(\alph*)]
         \item Each $\Omega(i)$ is invertible and: 
         $$ \limsup_{n\to \infty}\max_{i\in \mathcal{N}_n}\max\{||\Omega(i)^{-1}||_{1},||\mathbb{E}[\mathbf{V}\mathbf{V}']||_{1}\} < \infty $$
        \item  There is a sequence of numbers $r_n$ such that $r_n \sim \kappa_n+g_n$ and for any two disjoint sets of units $\mathcal{U}_1,\mathcal{U}_2 \subset \mathcal{N}_n$ such that $\min_{i\in \mathcal{U}_1,j\in\mathcal{U}_2}\rho(i,j)>r_n$, the $\sigma$-algebra generated by the treatments of units in $\mathcal{U}_1$ is independent of the $\sigma$-algebra generated by the treatments of units in $\mathcal{U}_2$.
    \end{enumerate}
\end{assumption}

Assumption \ref{assum:blockrand} requires that (1) no cluster's treatment status is close to a linear function of the treatment statuses of other clusters within $\kappa_n+g_n$ and (2) that clusters are independent of very faraway neighbors.  A leading example of this is a stratified experiment where the strata all contain sufficiently many clusters that the maximum fraction of clusters that any $\kappa_n$ neighborhood can intersect is upper-bounded, and strata do not grow too fast. Example \ref{ex:stratified_example} discusses this. In practice, a researcher might consider this assumption implausible if several units have $\kappa_n$ neighborhoods that contain an entire stratum or nearly an entire stratum. In this case, the practitioner would be advised to reduce $\kappa_n$ until this is no longer the case. This requirement can be dropped when treatment probabilities are randomized by stratum, as in EHMNW and as discussed in Example \ref{ex:ehmnw_design} below.


\begin{example}\label{ex:stratified_example}{\bf Clustered-randomized designs with fixed cluster treatment probabilities}\normalfont

 Assumption \ref{assum:blockrand} nests standard stratified designs, so long as strata are geographic groupings and the number of clusters per stratum is nontrivial and bounded above. Specifically, suppose that clusters are partitioned into mutually exclusive groups called strata. Let stratum $g$ contain $m_g>d$ spatially adjacent clusters. Suppose that each stratum is contained within a ball of radius $K\kappa_n$ where $K>0$ is a universal constant. Within each stratum $g$ exactly fraction $p_g$ of clusters are treated at random with all $p_g$ uniformly bounded away from zero and one. Suppose that $\kappa_n$ grows slowly enough that no neighborhood $\mathcal{N}(i,\kappa_n)$ intersects more than fraction $\overline{p}$ of clusters in any one stratum and $\overline{p} < \min_g\min\{p_g,1-p_g\}$.\footnote{If $m_g < 3$ this can be difficult to satisfy because any meaningful radius could contain an entire stratum.} Then, each $\Omega(i)$ is strictly diagonally dominant with diagonal elements bounded away from the absolute row sums. By Corollary 1 of \cite{VARAH19753}, this diagonal dominance guarantees that Assumption \ref{assum:blockrand}.a is satisfied. Since cluster treatments are independent across strata and strata members are contained within balls that grow no faster than $\kappa_n$, Assumption \ref{assum:blockrand}.b is satisfied.
\end{example}

\begin{example}\label{ex:ehmnw_design}{\bf Clustered-randomized designs with random cluster treatment probabilities}\normalfont

    When strata are assigned treatment probabilities at random,  Assumption \ref{assum:blockrand} can be satisfied even when $\kappa_n$ contains entire strata. Consider a two-stage design: first it assigns saturation levels of either $p_1\in(0,1)$ or $p_2\in(0,1)$ with $p_1\neq p_2$ at random to strata. In the second stage, the design treats the fraction of clusters in strata corresponding to its assigned saturation level. The unconditional covariance between the treatment statuses of any two distinct clusters $c,d$ in the same stratum that contains $m_g$ clusters is:
    \begin{equation}\label{eq:Omega_example_randomsat}
        \text{Cov}(W_c,W_d)=\frac{(p_1-p_2)^2}{4}-\frac{p_1(1-p_1) +p_2(1-p_2)}{2(m_g-1)}
    \end{equation}
    \noindent The proof is in Appendix \ref{proof:example:ehmnw_design}. $\Omega(i)$ is always invertible because no treatment is almost surely a linear function of other treatments within $\kappa_n$. When all $m_g,p_1,p_2$ are uniformly bounded above and below for all $n$, the $1$-norms of $\Omega(i)$ and its inverse must also be uniformly bounded by continuity. So Assumption \ref{assum:blockrand}.a is satisfied. If the strata are spatial groupings that grow no faster than $\kappa_n+g_n$,  Assumption \ref{assum:blockrand}.b will be satisfied. A ``random saturation design" is a special case of this where there are no clusters and therefore within-grouping randomization is at the unit level. 
\end{example}

When Assumption \ref{assum:blockrand} is satisfied, the matrices $\Omega(i)$ are invertible and we can define a $m_n\times m_n$ matrix $\Gamma(i)$ as follows. When clusters $c,\ell$ are both within $\kappa_n$ of unit $i$, let $\Gamma(i)_{c,\ell}$ equal the corresponding element of $\Omega(i)^{-1}$. Otherwise let  $\Gamma(i)_{c,\ell}=0$. Thus, $\Omega(i)^{-1}$ is a submatrix of $\Gamma(i)$. Now define the regressor $\widetilde{X}_{i} $.
\begin{align*}
     \widetilde{X}_i &\equiv\frac{1}{\overline{\phi}} \sum_{c=1}^{m_n}\sum_{m=1}^{m_n}B_{ic}(W_m-p)\Gamma(i)_{cm} = \frac{1}{\overline{\phi}} B_i\Gamma(i)\mathbf{V}
\end{align*}

We can show in Appendix \ref{proof:waldratio} that:
\begin{equation}\label{eq:wald_denominator}
   \frac{1}{n}\sum_{i=1}^n \mathbb{E}[X_i\widetilde{X}_i ] =\frac{1}{\overline{\phi}}
\end{equation}

We also show in Appendix \ref{proof:waldratio} that $Y_i\widetilde{X}_i$ recovers the original GATE up to $\kappa_n$:
\begin{equation}\label{eq:wald_numerator}
  \frac{1}{n}\sum_{i=1}^n  \mathbb{E}\left[Y_i\widetilde{X}_i \right] =  \frac{1}{\overline{\phi}}\frac{1}{n}\sum_{i=1}^n\sum_{j=1}^n  {A}_{ij}+\mathcal{O}\left(\frac{\kappa_n^{-\gamma}}{\overline{\phi}}\right)
\end{equation}

Therefore the ratio of expectations converges to the GATE with bias term $[\mathcal{B}]_{\text{STRAT}}$ arising from the long-range spillovers.
\begin{equation}\label{eq:bias_wald}
   \bar{\theta}_{n,\kappa_n}^{STRAT}\equiv  \frac{\frac{1}{n}\sum_{i=1}^n \mathbb{E}\left[Y_i\widetilde{X}_i \right] }{ \frac{1}{n}\sum_{i=1}^n\mathbb{E}\left[X_i\widetilde{X}_i \right]} =\theta_n+\mathcal{O}\left(\kappa_n^{-\gamma}\right)
\end{equation}

This motivates the following Wald regression where the adjusted $\widetilde{X}_i $ instruments for the unadjusted $X_i$. When clusters are treated i.i.d. this is equivalent to an OLS regression on $X_i$.
\begin{equation}
    \widehat{\theta}_{n,\kappa_n}^{STRAT} = \frac{\frac{1}{n}\sum_{i=1}^n Y_i\widetilde{X}_i  - \left(\frac{1}{n}\sum_{i=1}^n Y_i\right)\left(\frac{1}{n}\sum_{i=1}^n \widetilde{X}_i \right) }{\frac{1}{n}\sum_{i=1}^n X_i\widetilde{X}_i  - \left(\frac{1}{n}\sum_{i=1}^n X_i\right)\left(\frac{1}{n}\sum_{i=1}^n \widetilde{X}_i \right)}
\end{equation}

To see why $ \widehat{\theta}_{n,\kappa_n}^{STRAT}$ is consistent and asymptotically normal like $ \widehat{\theta}_{n,\kappa_n}^{OLS}$,  we can characterize the estimation error in a similar way to Proposition \ref{prop:ols_asymptotic_decomp}. This yields the decomposition in Proposition \ref{prop:ols_decomp_correlated_designs}:

\begin{proposition}\label{prop:ols_decomp_correlated_designs}
Let all the conditions of Proposition \ref{prop:ols_asymptotic_decomp} hold, except that clusters may have nonzero covariance. Assume that, given the sequence of radii $\kappa_n$, Assumption  \ref{assum:blockrand} holds and $\frac{\kappa_n^{2d}+g_n^{2d}}{ng_n^d} \to 0$. Then: 
    $$ \widehat{\theta}_{n,\kappa_n}^{STRAT} -  \:\theta_n =    \underbrace{\frac{\overline{\phi}}{n}\sum_{i=1}^n \widetilde{X}_i\left(\mathbb{E}[Y_i|\mathcal{F}_{i,r_n}]- \frac{1}{n}\sum_{j=1}^n \mathbb{E}[Y_j] -X_i\bar{\theta}_{n,\kappa_n}^{STRAT}\right)}_{[\mathcal{A}]_{\text{STRAT}}}+\underbrace{\bar{\theta}_{n,\kappa_n}^{STRAT}-\theta_n}_{[\mathcal{B}]_{\text{STRAT}}} + o_p\left(\frac{\kappa_n^{d}+g_n^d}{g_n^{d/2}n^{1/2}}\right)$$

    Proof: Section \ref{proof:eq:stratified_decomp}
\end{proposition}

The only difference between $[\mathcal{A}]_{\text{STRAT}} $ and $[\mathcal{A}]$ is the multiplication by $\widetilde{X}_i$ instead of $X_i$.  Assumption \ref{assum:blockrand}.a prevents the variance of $\widetilde{X}_i$ from being of a larger order than $X_i$ and Assumption \ref{assum:blockrand}.b prevents the degree of the dependency graph of $\widetilde{X}_i$ from being of a larger order than the graph for $X_i$. Using these facts we can verify that the asymptotic properties are not changed when we account for correlations in treatment assignment.

Theorem \ref{thm:ols_strat} demonstrates that the stratified estimator converges at the same rate as OLS would if the clusters were treated i.i.d.

\begin{theorem}{\label{thm:ols_strat}}
    Assume that the conditions of Theorem \ref{thm:ols_bstar} hold, except that clusters may have nonzero covariance. Assume also that, given the sequence of radii $\kappa_n$, Assumption  \ref{assum:blockrand} holds. Then, 
      \begin{align*}
\left| \widehat{\theta}_{n,\kappa_n}^{STRAT} -\theta_n\right|&\lesssim  \frac{\kappa_n^{d}+g_n^{d}}{g_n^{d/2}n^{1/2}} +\kappa_n^{-\gamma}
\end{align*}

    Proof: Section \ref{proof:thm:ols_strat}
\end{theorem}

 Theorem \ref{thm:clt_stratified} provides a CLT for OLS under designs satisfying Assumption  \ref{assum:blockrand}.

\begin{theorem}\label{thm:clt_stratified}
    Let the conditions of Theorem \ref{thm:clt} hold,  except that clusters may have nonzero covariance. Assume also that, given the sequence of radii $\kappa_n$, Assumption  \ref{assum:blockrand} holds. Then: $$\frac{[\mathcal{A}]_{\text{STRAT}}}{ \sqrt{\mathbb{V}\left[[\mathcal{A}]_{\text{STRAT}}\right]}} \to_d N(0,1)$$

    Proof: Section \ref{proof:thm:clt_stratified}
\end{theorem}

Variance estimation is the same as Section \ref{sec:inference}, with one change. The adjacency matrix $\Lambda$ should connect two units if any two members of their $\kappa_n$-neighborhoods have dependent treatment statuses (in the case of EHMNW, $\Lambda$ connects two units if their neighborhoods share a common saturation group, rather than just a common cluster). Assumptions \ref{assum:euclidean_space}.b and \ref{assum:blockrand}.b together guarantee that the degree of $\Lambda$ remains $\mathcal{O}\left(\kappa_n^d+g_n^d\right)$. Similarly, we replace the block-diagonal adjacency matrix $Q$ with the block-diagonal matrix $Q^*$ that connects two units if their own treatment statuses are dependent (in the case of our empirical application, $Q^*$ connects members of the same saturation group rather than just members of the same cluster).\footnote{The cluster-robust standard error estimator is unchanged.}

\subsection{IPW for Stratified Designs}\label{sec:stratified_bigclust}

The IPW estimator in general requires stronger conditions on the experimental design than the OLS estimator. The class of designs satisfying Assumption \ref{assum:blockrand} contains some designs for which IPW is consistent. For instance, Theorem \ref{thm:ipw_blocks} shows that the stratified designs described in Example \ref{ex:stratified_example} above do allow for IPW so long as $\kappa_n$ is small enough. Specifically,  the IPW estimator is consistent for the GATE provided that the probability that all clusters within distance $\kappa_n$ of any unit are (un)treated is uniformly bounded below. In order to guarantee enough overlap, $\kappa_n$ must not be so large that any one unit's $\kappa_n$-neighborhood contains a large fraction of any one stratum, which is guaranteed by the setup of Example \ref{ex:stratified_example}.

\begin{theorem}\label{thm:ipw_blocks}
    Suppose that all the conditions of Theorem \ref{thm:rateipw} hold, except that the randomization is via the stratified design described in Example \ref{ex:stratified_example} above. Then the conclusions of Theorem \ref{thm:rateipw} hold.
\end{theorem}
\begin{proof}
    The bias properties are unchanged from Theorem \ref{thm:rateipw}. Since the degree of the dependency graph is of the same order as before, to control the variance we need only verify that $\mathbb{P}[T_{i1}=1],\mathbb{P}[T_{i0}=1]$ are uniformly bounded away from zero. Lemma \ref{lem:phi} guarantees that since $\kappa_n\sim g_n$, the number of clusters that $\mathcal{N}(i,\kappa_n)$ intersects is uniformly bounded above by a constant. Thus the number of strata that intersect $\mathcal{N}(i,\kappa_n)$ is also uniformly bounded above by a constant $c_1$. Let $m_{b,i}$ denote the number of clusters in stratum $b$ that  intersect $\mathcal{N}(i,\kappa_n)$, and let $M_b$ be the total  number of clusters in that stratum. By Lemma \ref{lem:phi} and since $\kappa_n\sim g_n$, we have $\limsup_{n\to \infty}\max_{b,i}m_{b,i} \equiv M^* < \infty$.  By the setup of Example \ref{ex:stratified_example}, $m_{b,i} < \min\{p_b,\,1-p_b\} M_b$. So, under complete randomization within each stratum, the probability that all $m_{b,i}$  members of stratum $b$ that intersect $\mathcal{N}(i,\kappa_n)$ are treated (or untreated) is bounded below by a constant  $c_2 \in (0,1)$ depending only on $(M^*,\overline{p})$.  Thus, $\mathbb{P}[T_{i1}=1],\mathbb{P}[T_{i0}=1] \geq c_2^{c_1}>0$. So the IPW weights are uniformly bounded above.
\end{proof}

\subsection{Failure of IPW for Random Saturation Designs}
\label{subsec:random_saturation}  

Not every design that satisfies Assumption \ref{assum:blockrand} allows for polynomial-rate IPW estimation; in particular, the clusters must still grow. This appendix provides a negative result for IPW under a random saturation design as discussed in Remark \ref{rem:random_saturation} and studied in \cite{Baird}.\footnote{What \cite{Baird} call ``clusters" we call ``spatial groupings" or ``strata." In this paper, two units are said to be in the same ``cluster" if they are always treated together.} Random saturation designs are similar to Example \ref{ex:ehmnw_design} above, except they additionally require that clusters are units---i.e. that the design is not cluster-randomized.  

Randomized saturation designs will usually make the IPW estimator inconsistent for the GATE, because they yield few neighborhoods that are either entirely treated or entirely not. Mathematically, this design guarantees that if $\kappa_n \to \infty$ and $p_1,p_2 \in (0,1)$, then $\mathbb{P}\left[T_{i0}=1\right],\mathbb{P}\left[T_{i1}=1\right] \to 0$ rapidly, causing the variance to diverge. Theorem \ref{thm:fail_ipw_rand_sat} states this result, generalized to allow clusters of any fixed size (including single units).

\begin{theorem}\label{thm:fail_ipw_rand_sat}
    Suppose that all the conditions of Theorem \ref{thm:rateipw} hold, except that the randomization follows Example \ref{ex:ehmnw_design} above and $\limsup_{n\to \infty}g_n<\infty$. Then, for every $a>0$ there is some sequence of sets of potential outcomes satisfying Assumptions \ref{assum:boundedoutcomes} and \ref{assum:ANI} such that:

    $$ n^a\left|\widehat{\theta}_{n,\kappa_n}^{IPW}-\theta_n\right| \to_p \infty $$
\end{theorem}
\begin{proof}
  Without loss of generality let $\overline{Y}\geq c$. For any $a>0$, let the potential outcomes be $$Y_i = c n^{-a/2}\mathbf{1}\left\{\text{$\phi(i,\kappa_n) \geq \overline{\phi}$ and all units within distance $n^{a/(2\gamma)}$ of $i$ are treated}\right\}$$
  
  By Lemma \ref{lem:phi}, $\max_i\phi(i,\kappa_n)/\overline{\phi} < \infty$. Therefore the fraction of units with $\phi(i,\kappa_n) \geq \overline{\phi}$ does not go to zero. Therefore $\liminf_{n\to \infty}\theta_n n^{a/2}>0$. Since $\limsup_{n\to \infty}g_n<\infty$, there is a constant $c_g>0$ such that for each unit $i$ with $\phi(i,\kappa_n)\geq \overline{\phi}$: $$\mathbb{P}[\text{all units within distance $n^{a/(2\gamma)}$ of $i$ are treated}] \lesssim \max\{p_1,p_2\}^{\phi(i,\kappa_n)} \lesssim  \max\{p_1,p_2\}^{n^{ad/(2\gamma)}c_g}$$ 
  Therefore: $\mathbb{P}[Y_i \neq 0] \lesssim\max\{p_1,p_2\}^{n^{ad/(2\gamma)}c_g}$. By the union bound:  $\mathbb{P}[\max_i |Y_i| \neq 0] \lesssim n\max\{p_1,p_2\}^{n^{ad/(2\gamma)}c_g}$. So $\mathbb{P}[\max_i |Y_i| = 0] \to 1$ and $\mathbb{P}[\widehat{\theta}_{n,\kappa_n}^{IPW} = 0] \to 1$ and therefore $n^a\widehat{\theta}_{n,\kappa_n}^{IPW}\to_{a.s.} 0$. Since $\liminf_{n\to \infty}\theta_n n^{a/2}>0$, then $ n^a\left|\widehat{\theta}_{n,\kappa_n}^{IPW}-\theta_n\right| \to_p \infty $.
  
\end{proof}

\section{Inference for IPW}\label{sec:ipw_clt}

\cite{Leung22} and \cite{leung2025crosscluster} provide a CLT and variance estimator for IPW when $\gamma > d$. Here we generalize the CLT to $\gamma < d$. The first step toward the CLT is to break up the IPW estimator into three pieces: a variance term called $[\mathcal{A}]_{IPW}$ with expectation zero that does not include long-range spillovers, a nonstochastic term called $[\mathcal{B}]_{IPW} \:\text{(bias)}$, and a dominated term. Proposition \ref{prop:ols_asymptotic_decomp} above has already done this for OLS. Proposition \ref{prop:ipw_asymptotic_decomp} now provides a similar decomposition for IPW. 
\begin{proposition}\label{prop:ipw_asymptotic_decomp}
   Define $\overline{\mu}_1 = \frac{1}{n}\sum_{j=1}^n\mathbb{E}\left[Y_j|T_{j1}=1\right]$ and $\overline{\mu}_0 = \frac{1}{n}\sum_{j=1}^n\mathbb{E}\left[Y_j|T_{j0}=1\right]$. Under the conditions of Theorem \ref{thm:rateipw} and assuming that $\frac{\kappa_n^d}{n}\to 0$:
    \begin{multline*} 
    \widehat{\theta}_{n,\kappa_n}^{IPW} -\theta_n = \underbrace{\frac{1}{n}\sum_{i=1}^n\left[ \left(\mathbb{E}\left[Y_i|T_{i1}=1\right]-\overline{\mu}_1\right)\frac{T_{i1}}{\mathbb{P}\left[T_{i1}=1\right]} -\left(\mathbb{E}\left[Y_i|T_{i0}=1\right]-\overline{\mu}_0\right)\frac{T_{i0}}{\mathbb{P}\left[T_{i0}=1\right]}\right] }_{[\mathcal{A}]_{IPW}\: \text{(variance)}}\\
    +\underbrace{\theta_n - (\overline{\mu}_1-\overline{\mu}_0)}_{[\mathcal{B}]_{IPW} \:\text{(bias)}} +o_p\left(\sqrt{\frac{\kappa_n^d}{n}}+\kappa_n^{-\gamma}\right)
    \end{multline*}

    Proof: Section \ref{proof:prop:ipw_asymptotic_decomp}
\end{proposition}

Next we show that the variance term  $[\mathcal{A}]_{IPW}$ is asymptotically normal:

\begin{theorem}{\bf Central Limit Theorem for IPW}\label{thm:clt_ipw}

    Suppose that the researcher uses a Scaling Clusters design such that  $\frac{g_n^d}{n}\to 0$. If, in addition, the conditions of Theorem \ref{thm:rateipw} are satisfied, and $\liminf_{n\to \infty}{\mathbb{V}\left([\mathcal{A}]_{IPW}\right)}\frac{n}{\kappa_n^d}>0$ , then:
       \begin{align*} 
        [\mathcal{A}]_{IPW}/\sqrt{\mathbb{V}\left([\mathcal{A}]_{IPW}\right)} \to_d N(0,1)
    \end{align*}
    
    Proof: Section \ref{proof:thm:clt_ipw}.
\end{theorem}

To estimate the variance $\mathbb{V}\left([\mathcal{A}]_{IPW}\right)$, we recommend the standard errors from \cite{leung2025crosscluster}.

Even with a CLT and a consistent variance estimator, inference for IPW can be very challenging in practice. Theorem \ref{thm:ipw_bias_dominates} shows that when the treatment clusters grow more slowly than the optimal rate, IPW's bias will usually dominate its variance.

\begin{theorem}{\bf Bias of IPW  for Small Clusters}\label{thm:ipw_bias_dominates}

        Let the conditions of Theorem \ref{thm:smallclusters} hold. Then, for any $q_2>0$ such that $\widehat{\theta}_{n,\kappa_n}^{IPW}-\theta_n \to_p 0$, there exist potential outcomes (which, additionally, will satisfy Assumptions \ref{assum:boundedoutcomes}-\ref{assum:linearity}) such that:

    $$ |[\mathcal{B}]_{IPW}|/\sqrt{\mathbb{V}\left([\mathcal{A}]_{IPW}\right)} \to \infty $$

    Proof: Section \ref{proof:thm:ipw_bias_dominates}
\end{theorem}

When bias dominates variance, inference becomes very difficult because confidence intervals cannot asymptotically cover the GATE, and adjusting the critical value does not help.

\section{Additional results for the OLS estimator}\label{sec:additional_ols}

This section provides additional results for OLS estimators under cluster-randomized experimental designs with independent cluster treatments, i.e. those satisfying Equation (\ref{eq:design}).

\subsection{Incorrect Estimand when Regressing on Fraction of Treated Units}\label{sec:ols_unit_regression} 

Suppose that instead of regressing the outcome $Y$ on the share $X$ of nearby treated clusters, as defined in Section \ref{sec:linearity}, the researcher instead regresses $Y$ on the fraction $\bar{X}$ of nearby treated units:\footnote{Centering  $\bar{X}_i$   by subtracting $p$ has algebraically zero effect on $\widehat{\theta}^{\text{OLS}}_{\text{Units}} $.} 
\begin{equation}
    \bar{X}_i \equiv \frac{1}{|\mathcal{N}(i,\kappa_n)|} \sum_{j \in \mathcal{N}(i,\kappa_n) } (D_j-p)
\end{equation}
The OLS regression of $Y$ on $\bar{X}$ can be written as: \begin{equation}
    \widehat{\theta}^{\text{OLS}}_{\text{Units}} \equiv \frac{\frac{1}{n}\sum_{i=1}^n Y_i\bar{X}_i - \left(\frac{1}{n}\sum_{i=1}^n Y_i\right)\left(\frac{1}{n}\sum_{i=1}^n \bar{X}_i\right) }{\frac{1}{n}\sum_{i=1}^n \bar{X}_i^2 - \left(\frac{1}{n}\sum_{i=1}^n \bar{X}_i\right)^2}
\end{equation}

We can write $\bar{X}_i$ in terms of a $n\times m_n$ matrix $\widetilde{B}$ where $\widetilde{B}_{ic} = \frac{\sum_{j=1}^n\mathbf{1}\left\{j\in \mathcal{N}\left(i,\kappa_n\right)\cap \mathcal{P}_c\right\}}{|\mathcal{N}(i,\kappa_n)|} $ and the $m_n\times 1$ vector of cluster treatment assignments $\mathbf{W}$ and $\widetilde{B}_i$ is the $i$th row: $\bar{X}_i = \widetilde{B}_i(\mathbf{W}-p)=\widetilde{B}_i\mathbf{V}$. By Lemma \ref{lem:pot_outcomes_linearity}: $Y_i = (\beta_0+p\theta_n)+ (AC)_i\mathbf{V} +p(A_i\mathbf{1}-\theta_n)+\epsilon_i$. By Lemma \ref{lem:covar2}, the target of $\widehat{\theta}^{\text{OLS}}_{\text{Units}} $ can be expressed as:
\begin{equation}
   \widehat{\theta}^{\text{OLS}}_{\text{Units}} =  \frac{\mathbb{E}\left[\frac{1}{n}\sum_{i=1}^n Y_i\bar{X}_i - \left(\frac{1}{n}\sum_{i=1}^n Y_i\right)\left(\frac{1}{n}\sum_{i=1}^n \bar{X}_i\right)\right] }{\mathbb{E}\left[\frac{1}{n}\sum_{i=1}^n \bar{X}_i^2 - \left(\frac{1}{n}\sum_{i=1}^n \bar{X}_i\right)^2\right]}+o_p(1) =  \frac{\text{tr}\left(\widetilde{B}'AC\right)}{\text{tr}\left(\widetilde{B}'\widetilde{B}\right)}+o_p(1) 
\end{equation}

Computing this ratio of traces:
\begin{align*}
  \frac{tr(\widetilde{B}'AC)}{n}
    &= \frac{1}{n}\sum_{i=1}^n\sum_{c=1}^{{m_n}} \frac{\sum_{j=1}^{n}  \mathbbm{1}\left\{j \in \mathcal{N}(i,\kappa_n) \cap \mathcal{P}_c\right\}}{ \left|\mathcal{N}(i,\kappa_n)\right|} \sum_{k=1}^n A_{ik}\mathbbm{1}\left\{k\in \mathcal{P}_c\right\} \\
    &= \frac{1}{n}\sum_{i=1}^n\sum_{k=1}^n A_{ik}\sum_{c=1}^{{m_n}} \mathbbm{1}\left\{k\in \mathcal{P}_c\right\} \frac{\sum_{j=1}^{n}  \mathbbm{1}\left\{j \in \mathcal{N}(i,\kappa_n) \cap \mathcal{P}_c\right\}}{ \left|\mathcal{N}(i,\kappa_n)\right|} \\
      \frac{tr(\widetilde{B}'\widetilde{B})}{n}
    &= \frac{\sum_{i=1}^n\sum^{m_n}_{c=1} (\widetilde{B}_{ic})^2}{n}\\
    &=\frac{1}{n}  \sum_{i=1}^n\sum^{m_n}_{c=1} \left(\frac{\sum_{j=1}^{n}  \mathbbm{1}\left\{j \in \mathcal{N}(i,\kappa_n) \cap \mathcal{P}_c\right\}}{ \left|\mathcal{N}(i,\kappa_n)\right|} \right)^2 
\end{align*}

So the estimand when OLS is run with $\widetilde{B}$ is:
\begin{align*}
      \frac{tr(\widetilde{B}'AC)}{tr(\widetilde{B}'\widetilde{B})} &=\sum_{i=1}^n\sum_{k=1}^n A_{ik} \left( \frac{\sum_{c=1}^{{m_n}} \mathbbm{1}\left\{k\in \mathcal{P}_c\right\} \frac{\sum_{j=1}^{n}  \mathbbm{1}\left\{j \in \mathcal{N}(i,\kappa_n) \cap \mathcal{P}_c\right\}}{ \left|\mathcal{N}(i,\kappa_n)\right|}}{\sum_{\ell=1}^n\sum^{m_n}_{c=1} \left(\frac{\sum_{j=1}^{n}  \mathbbm{1}\left\{j \in \mathcal{N}(\ell,\kappa_n) \cap \mathcal{P}_c\right\}}{ \left|\mathcal{N}(\ell,\kappa_n)\right|} \right)^2} \right)
\end{align*}

In contrast, recall that under Assumption \ref{assum:linearity} the GATE is simply:
\begin{align*}
      \theta_n &= \frac{1}{n}\sum_{i=1}^n\sum_{k=1}^n A_{ik}
\end{align*}

So $\widehat{\theta}^{\text{OLS}}_{\text{Units}}$ applies unnecessary weights to the treatment effects. While non-negative, these weights apply to alter-ego pairs and depend on the specifics of the experimental design.

\subsection{Consistency of OLS for Nonlinear Outcomes}\label{sec:nonlinear_outcomes} 

When the ``outer crusts" of the clusters make up a shrinking fraction of the population and clusters grow, OLS is consistent even without Assumption \ref{assum:linearity}. The intuition is that when treatment clusters become large, nonlinearities contribute diminishing bias since most units are surrounded by neighbors of the same treatment status as themselves.

\begin{proposition}\label{prop:ols_nonlinear}
    Assume that all the conditions of Theorem \ref{thm:ols_bstar} hold except for Assumption \ref{assum:linearity}. Assume also that the researcher uses a Scaling Clusters design, $g_n\to \infty$, and  $\frac{1}{n}\sum_{c=1}^{m_n}|\mathcal{P}_c \setminus \mathcal{N}(q_c,g_n)|\to 0$. Then,
    $$ \widehat{\theta}^{OLS}_{n,\kappa_n}-\theta_n \to_p 0 $$
\end{proposition}
\begin{proof}

Consider the OLS estimate where we replace $Y_i$ with $\mathbb{E}[Y_i|D_i]$:
$$ \widetilde{\theta}_{n,\kappa_n}^{*} \equiv \frac{\frac{1}{n}\sum_{i=1}^n \mathbb{E}[Y_i|D_i]X_i - \left(\frac{1}{n}\sum_{i=1}^n \mathbb{E}[Y_i|D_i]\right)\left(\frac{1}{n}\sum_{i=1}^n X_i\right) }{\frac{1}{n}\sum_{i=1}^n X_i^2 - \left(\frac{1}{n}\sum_{i=1}^n X_i\right)^2}  $$
Notice that $\mathbb{E}[Y_i|D_i]$ is (trivially) a linear function of $D_i$. So by Theorem \ref{thm:ols_bstar}: $\widetilde{\theta}_{n,\kappa_n}^{*}   - \frac{1}{n}\sum_{i=1}^n (\mathbb{E}[Y_i|D_i=1] - \mathbb{E}[Y_i|D_i=0]) \to_p 0$. By Assumption \ref{assum:ANI}, $\left|Y_i(\mathbf{1}) -\mathbb{E}[Y_i|D_i=1]\right|\leq c\rho_i^{-\gamma}$ and  $\left|Y_i(\mathbf{0}) -\mathbb{E}[Y_i|D_i=0]\right|\leq c\rho_i^{-\gamma}$ where $\rho_i$ is the distance from unit $i$ to the nearest neighbor from another cluster. So $|\widetilde{\theta}_{n,\kappa_n}^{*}   - \theta _n|  =\mathcal{O}_p\left( \frac{1}{n}\sum_{i=1}^n \rho_i^{-\gamma}\right)+o_p(1)$. Also by Assumption \ref{assum:ANI}, $\left|Y_i-\mathbb{E}[Y_i|D_i]\right|=\left|Y_i-\mathbb{E}[Y_i|\mathcal{F}_{i,\rho_i}]\right|\leq c\rho_i^{-\gamma}$. So $|\widetilde{\theta}_{n,\kappa_n}^{*}  -\widehat{\theta}_{n,\kappa_n}^{OLS}| =\mathcal{O}_p\left( \frac{1}{n}\sum_{i=1}^n \rho_i^{-\gamma}\right)$. By the triangle inequality:
$$  \widehat{\theta}_{n,\kappa_n}^{OLS} -\theta_n =\mathcal{O}_p\left(\frac{1}{n}\sum_{i=1}^n \rho_i^{-\gamma}\right) +o_p(1) $$

The proof of Theorem \ref{thm:varest} showed already that $ \frac{1}{n}\sum_{i=1}^n \rho_i^{-\gamma} \to 0$ whenever $\frac{1}{n}\sum_{c=1}^{m_n}|\mathcal{P}_c \setminus \mathcal{N}(q_c,g_n)|\to 0$. So OLS is consistent. 
 
\end{proof}

\clearpage
\section{Additional exhibits}\label{app:additional_exhibits}

\begin{table}[hp]
\begin{center}
\caption{ Alternative OLS Estimators, by estimator radius\label{table:rings}}
\vspace{1em}


{
\renewcommand{\arraystretch}{1.3}
\newcolumntype{Y}{>{\centering\arraybackslash}X}
\begin{tabularx}{\textwidth}{p{6cm}YYYYYYY}
  \toprule
  & \multicolumn{7}{c}{Radius ($\kappa_n$)} \\
  \cmidrule(lr){2-8}
 & 0m & 250m & 500m & 750m & 1000m & 1500m & 2000m \\  
  \midrule
  \addlinespace[6pt]

  $\widehat{\theta}^{\text{OLS,STRAT}}$ & 307 & 365 & 445 & 391 & 446 & 389 & 366 \\ 
  & (58) & (74) & (90) & (122) & (131) & (170) & (260) \\ 
 
  \addlinespace[6pt]
\midrule
\addlinespace[6pt]
    $\widehat{\theta}$ ignore strata& 306 & 365 & 446 & 395 & 459 & 414 & 392 \\ 
 $\widehat{\theta}$ ignore strata, own-cluster dummy  & 366 & 443 & 374 & 436 & 390 & 368 \\ 
 $\widehat{\theta}$  ignore strata, treated unit share  & 306 & 340 & 398 & 444 & 488 & 365 & 317 \\ 
   \addlinespace[6pt]
\midrule
\addlinespace[6pt]
  $\overline{\phi}$ & 1.00 & 1.70 & 2.90 & 4.50 & 6.30 & 10.60 & 15.70 \\
  \bottomrule 
\end{tabularx}
}

\end{center}
\small
{ Estimates of the GATE on annual household consumption expenditure in PPP US dollars using data from \cite{EggeretalGE}. Units are households and clusters are villages. Columns indicate the estimator radius $\kappa_n$ in meters. $\widehat{\theta}^{\text{OLS,STRAT}}$ indicates estimates from our recommended regression on the fraction of nearby treated clusters (Equation \ref{eq:ols_recommended}), i.e. replicating the results at top of Table \ref{table:application_results}. The remaining rows report results from alternative estimators, all of which omit the Wald weights defined in Appendix \ref{sec:stratification} and thus ignore the stratification in the design. The second includes an indicator for own-cluster treatment as an additional regressor. The third regresses on the share of nearby treated units, rather than the share of nearby treated clusters. All estimates use the original sampling weights. The bottom row reports, as a diagnostic, the mean number $\overline{\phi}$ of villages within the given radius.}
\end{table}

\begin{figure}[tp]
    \caption{Simulation Results: $\gamma = 1.5$}
    \label{fig:simulations_gamma_1_5}
    \vspace{-1em}
    \begin{subfigure}[t]{0.48\textwidth}
        \centering
        \includegraphics[width=\textwidth]{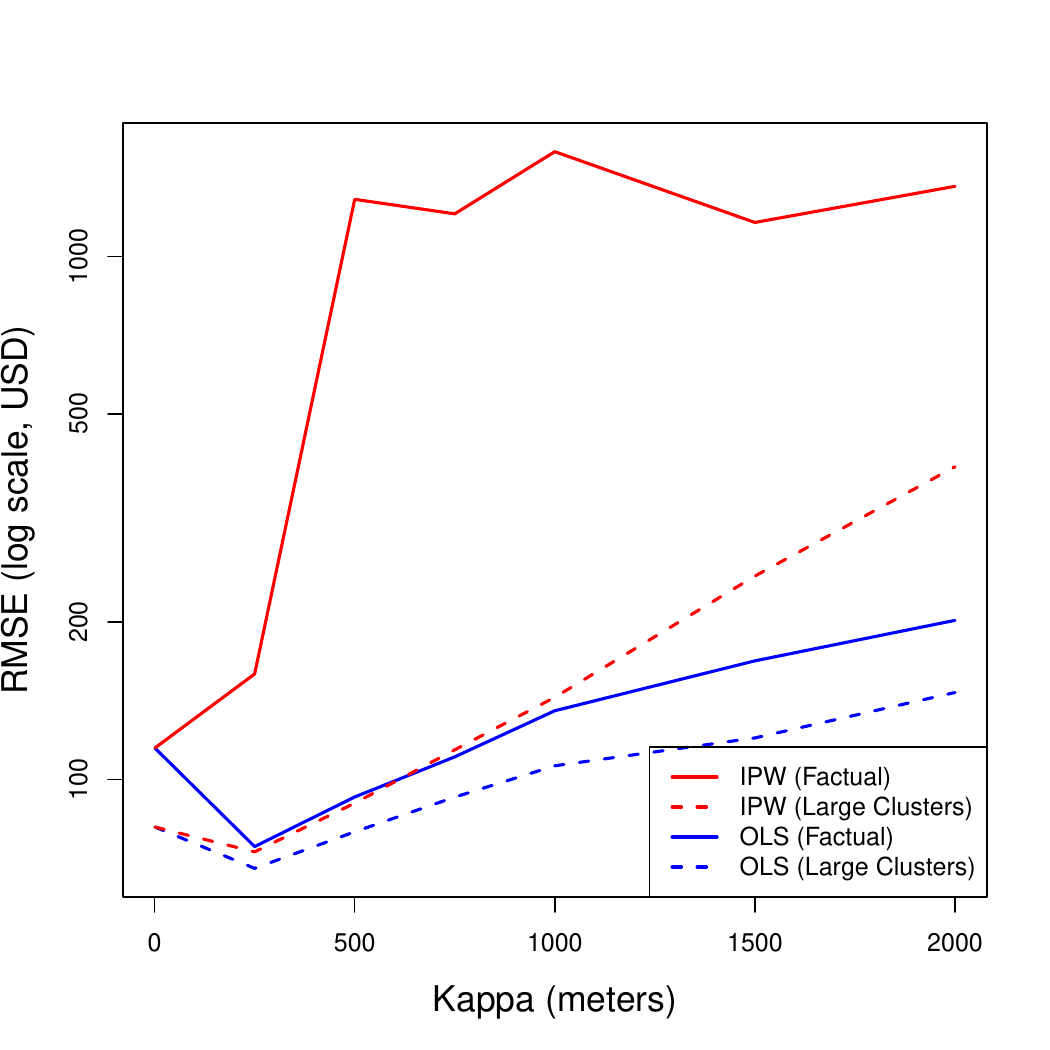}
    \end{subfigure}
    \hfill
    \begin{subfigure}[t]{0.48\textwidth}
        \centering
        \includegraphics[width=\textwidth]{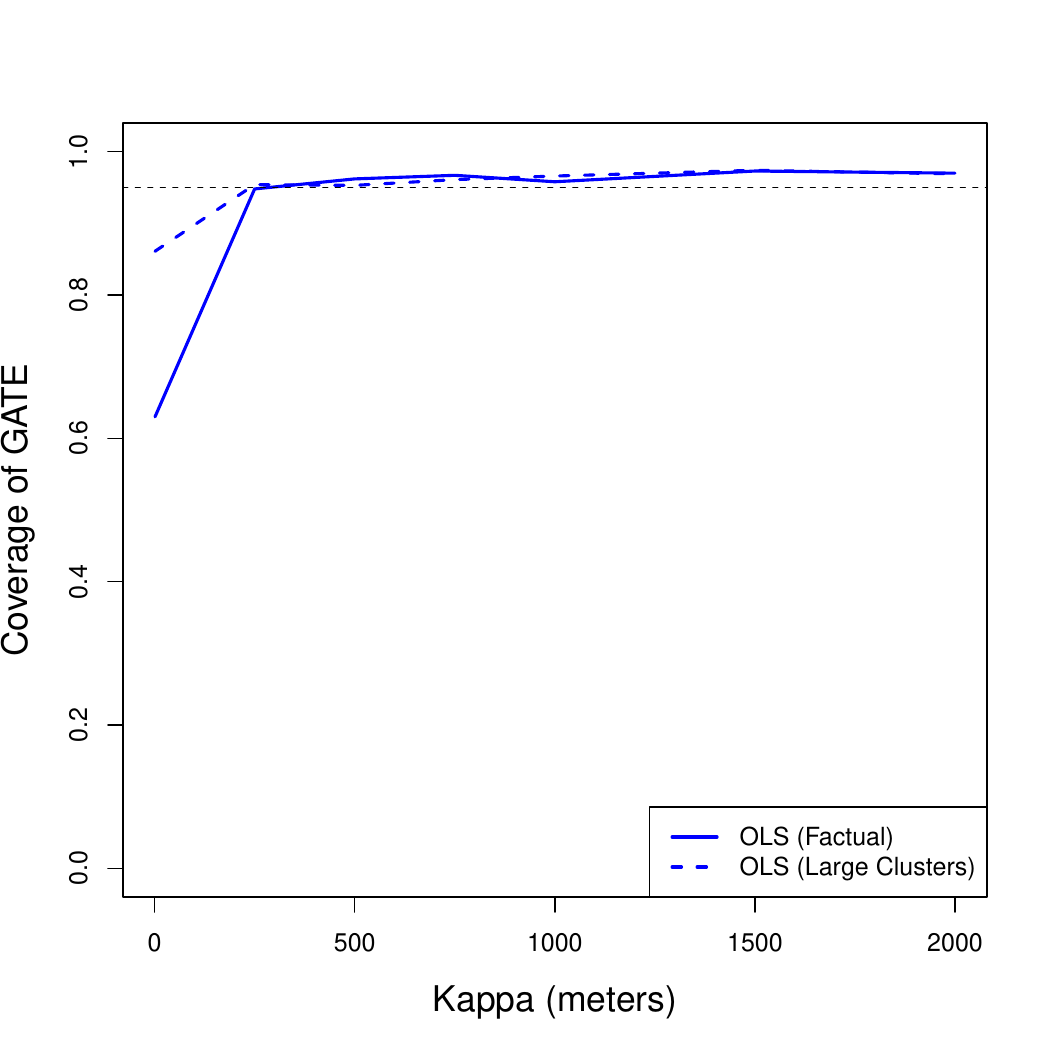}
    \end{subfigure}
    \vspace{-1em}

    {\footnotesize This figure plots the root mean squared error (left-hand panel) and coverage of OLS confidence intervals (right-hand panel) in simulation. Solid lines use the factual, small-cluster design and dashed lines use a counterfactual, large-cluster design. Blue lines are OLS and red lines are IPW. The horizontal line in the right-hand panel indicates the nominal coverage level, 95\%. We set $\gamma = 1.5$, given which the true GATE is $\$401$; for comparison, the outcome standard deviation is $\$1,802$.}

        
\end{figure}

\begin{figure}[tp]
    \caption{Simulation Results for $\gamma = 0.50$}
    \label{fig:simulations_gamma_0.50}
    \vspace{-1em}
    \begin{subfigure}[t]{0.48\textwidth}
        \centering
        \includegraphics[width=\textwidth]{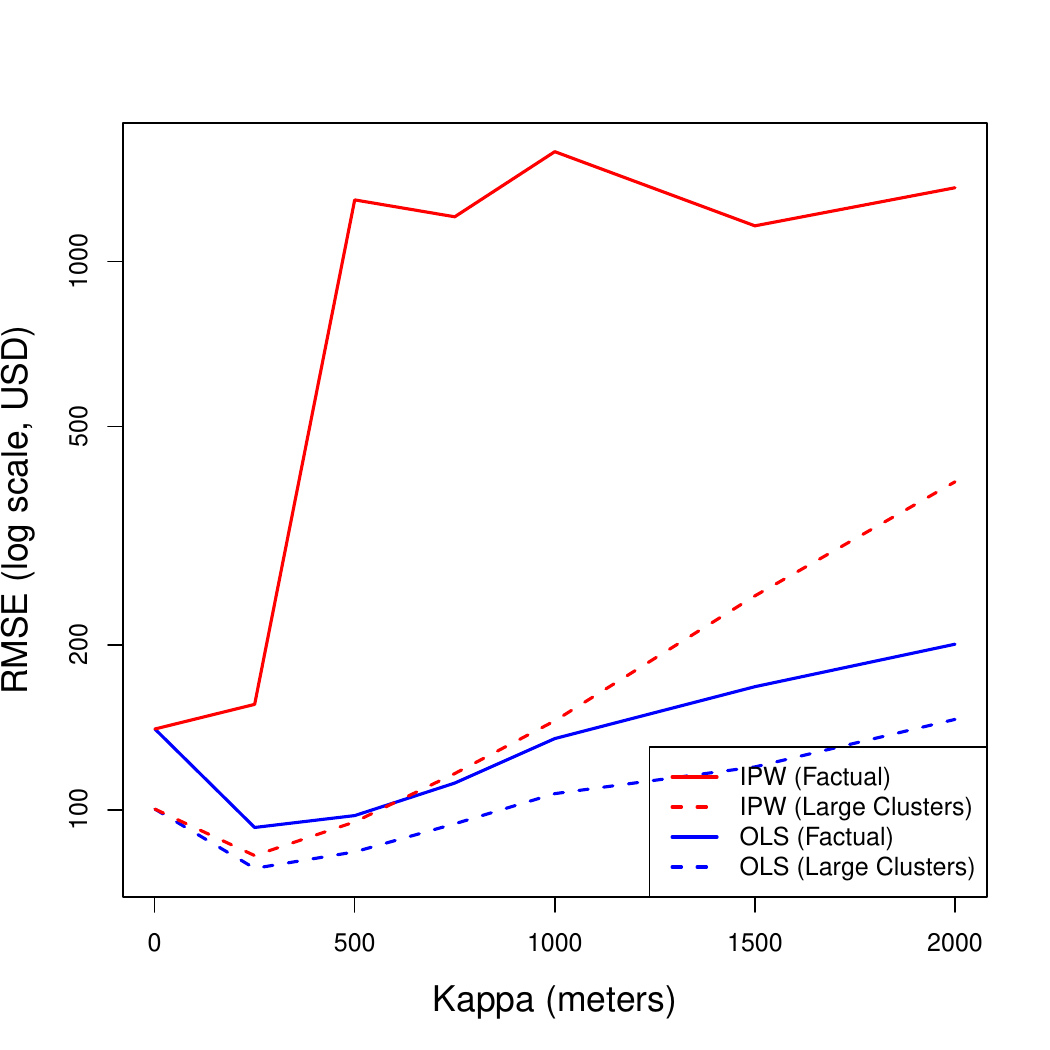}
        \label{fig:simulations_gamma_0.25:left_panel}
    \end{subfigure}
    \hfill
    \begin{subfigure}[t]{0.48\textwidth}
        \centering
        \includegraphics[width=\textwidth]{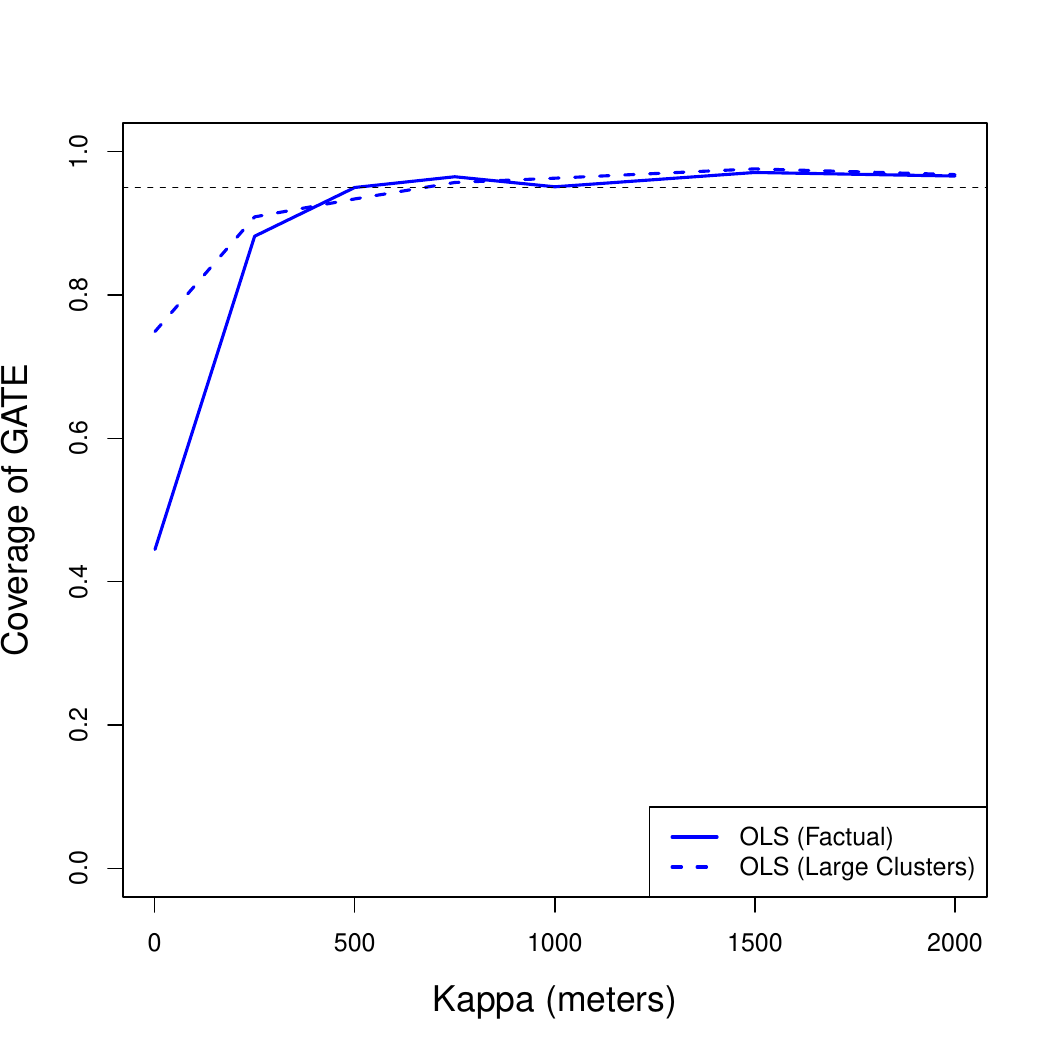}
        \label{fig:simulations_gamma_0.50_right_panel}
    \end{subfigure}
    \vspace{-1em}

    
    {\footnotesize This figure plots the root mean squared error (left-hand panel) and coverage of OLS confidence intervals (right-hand panel) in simulation. Solid lines use the factual, small-cluster design and dashed lines use a counterfactual, large-cluster design. Blue lines are OLS and red lines are IPW. The horizontal line in the right-hand panel indicates the nominal coverage level, 95\%. We set $\gamma = 0.50$, given which the true GATE is $\$430$; for comparison, the outcome standard deviation is $\$1,802$.} 
\end{figure}

\section{Extended Proofs of Appendix Results}

\subsection{Proof of Equations (\ref{eq:wald_denominator}) and (\ref{eq:wald_numerator})}\label{proof:waldratio}  

It is helpful to rewrite everything in vector notation and use the ``trace trick." Let $B$ be the $n\times m_n$ matrix of ones and zeros where $B_{ic}$ indicates whether cluster $c$ is within $\kappa_n$ of unit $i$. Let $V_c \equiv W_c-\mathbb{E}[W_c]$. This means that $X_i = \overline{\phi}^{-1}B_i\mathbf{V}$ and $\widetilde{X}_i = \frac{1}{\overline{\phi}}B_i\Gamma(i)\mathbf{V}$. The key step is to see that $\Gamma(i)\mathbb{E}\left[\mathbf{V}\mathbf{V}'\right]B_i' = B_i'$  because $\Gamma(i)$ is zero except on a submatrix which is the inverse of the corresponding submatrix of $\mathbb{E}\left[\mathbf{V}\mathbf{V}'\right]$ and the support of $B_i$ is exactly the indices of that submatrix. Now we  can verify Equation (\ref{eq:wald_denominator}):
\begin{align*}
      \mathbb{E}[X_i\widetilde{X}_i]  &= \frac{1}{\overline{\phi}^2} \mathbb{E}\left[\mathbf{V}'B_i'B_i\Gamma(i)\mathbf{V}\right]
      = \frac{1}{\overline{\phi}^2} \mathbb{E}\left[\text{tr}\left(\mathbf{V}'B_i'B_i\Gamma(i)\mathbf{V}\right)\right] \\
       &= \frac{1}{\overline{\phi}^2} \mathbb{E}\left[\text{tr}\left(\Gamma(i)\mathbf{V}\mathbf{V}'B_i'B_i\right)\right] =\frac{1}{\overline{\phi}^2} \text{tr}\left(\Gamma(i)\mathbb{E}\left[\mathbf{V}\mathbf{V}'\right]B_i'B_i\right)
         =\frac{1}{\overline{\phi}^2}\text{tr}\left(B_i'B_i\right)=  \phi_i\overline{\phi}^{-2}
\end{align*}
Taking the mean: $\frac{1}{n}\sum_{i=1}^n\phi_i\overline{\phi}^{-2} = \overline{\phi}^{-1}  $.

 Next we verify (\ref{eq:wald_numerator}) which says that the numerator of the Wald estimator has the desired target using the trace trick. Recall that $C$ is the $n\times m_n$ matrix where $C_{ic}$ indicates whether unit $i$ belongs to cluster $c$. Below we exploit the fact that $(\widetilde{A}C)_i$ has support only on $B_i$, so $(\widetilde{A}C)_i\mathbb{E}\left[\mathbf{V}\mathbf{V}'\right]\Gamma(i)'B_i' = (\widetilde{A}C)_iB_i'$.
\begin{align*}
     \mathbb{E}\left[\widetilde{X}_iY_i\right] &=\overline{\phi}^{-1} \mathbb{E}\left[\mathbf{V}'\Gamma(i)'B_i'(AC)_i\mathbf{V}\right]\\
     &=    \overline{\phi}^{-1}\mathbb{E}\left[\text{tr}\left(\mathbf{V}'\Gamma(i)'B_i'(AC)_i\mathbf{V}\right)\right]
     = \overline{\phi}^{-1}\text{tr}\left((AC)_i\mathbb{E}\left[\mathbf{V}\mathbf{V}'\right]\Gamma(i)'B_i'\right)\\
     &= \overline{\phi}^{-1}\text{tr}\left((\widetilde{A}C)_i\mathbb{E}\left[\mathbf{V}\mathbf{V}'\right]\Gamma(i)'B_i'\right)+\frac{1}{\overline{\phi}}\text{tr}\left((AC-\widetilde{A}C)_i\mathbb{E}\left[\mathbf{V}\mathbf{V}'\right]\Gamma(i)'B_i'\right)
     \end{align*}

     Now by Assumption \ref{assum:ANI}: $\max_i||((A-\widetilde{A})C)_i||_1\lesssim \kappa_n^{-\gamma}$. Notice that $||\Omega(i)^{-1}||_\infty,||\mathbb{E}[\mathbf{V}\mathbf{V}']||_\infty \lesssim 1$ by Assumption \ref{assum:blockrand} (and the fact that these are symmetric matrices). Moreover $||\Gamma(i)||_\infty \leq ||\Omega(i)^{-1}||_\infty$ and $||B_i||_\infty \leq 1$. So we use Holder's Inequality to conclude that $ (AC-\widetilde{A}C)_i\mathbb{E}\left[\mathbf{V}\mathbf{V}'\right]\Gamma(i)'B_i'  \leq ||((A-\widetilde{A})C)_i||_1 ||\mathbb{E}\left[\mathbf{V}\mathbf{V}'\right]\Gamma(i)'B_i'||_\infty \lesssim \kappa_n^{-\gamma}$. So we can control the residual:
     \begin{align*}
     \mathbb{E}\left[\widetilde{X}_iY_i\right] &= \overline{\phi}^{-1}(\widetilde{A}C)_iB_i'+\mathcal{O}\left(\kappa_n^{-\gamma}\overline{\phi}^{-1}\right)\\
     &=\overline{\phi}^{-1} \sum_{j=1}^n  \widetilde{A}_{ij}\mathbf{1}\left\{\mathcal{P}_{c(j)}\cap \mathcal{N}(i,\kappa_n) \neq \emptyset\right\}+\mathcal{O}\left(\kappa_n^{-\gamma}\overline{\phi}^{-1}\right)=\frac{1}{\overline{\phi}} \sum_{j=1}^n  \widetilde{A}_{ij}+\mathcal{O}\left(\kappa_n^{-\gamma}\overline{\phi}^{-1}\right)\\
 \frac{1}{n}\sum_{i=1}^n     \mathbb{E}\left[\widetilde{X}_iY_i\right] &=   \overline{\phi}^{-1} \frac{1}{n}\sum_{i=1}^n \sum_{j=1}^n  \widetilde{A}_{ij} +\mathcal{O}\left(\kappa_n^{-\gamma}\overline{\phi}^{-1}\right)=\overline{\phi}^{-1} \frac{1}{n}\sum_{i=1}^n \sum_{j=1}^n  {A}_{ij} +\mathcal{O}\left(\kappa_n^{-\gamma}\overline{\phi}^{-1}\right)
\end{align*}

\subsubsection{Lemma \ref{lem:control_x_correlated_designs} and proof} 

\begin{lemma}\label{lem:control_x_correlated_designs}
Let the conditions of  Proposition \ref{prop:ols_decomp_correlated_designs} hold. If $\frac{\kappa_n^d+g_n^d}{n}\to 0$, then the following  statements also hold:
$$\max_{i\in\mathcal{N}_n}\mathbb{E}\left[\widetilde{X}_i^2\right]\lesssim \overline{\phi}^{-1},\quad \max_{i\in\mathcal{N}_n}\mathbb{E}\left[{X}_i^2\right]\lesssim \overline{\phi}^{-1},\quad \mathbb{E}\left[\overline{X}^2\right] \lesssim {\frac{\kappa_n^d+g_n^d}{n\overline{\phi}}}, \quad \mathbb{E}\left[\overline{\widetilde{X}}^2\right] \lesssim {\frac{\kappa_n^d+g_n^d}{n\overline{\phi}}} $$
and: $\mathbb{V}\left[\frac{\overline{\phi}}{n} \sum_{i=1}^n X_i\widetilde{X}_i\right]\lesssim \frac{\kappa_n^{2d}+g_n^{2d}}{ng_n^d}$.
\end{lemma}

{\bf Proof:} To prove claim 1, notice that $\widetilde{X}_i =  \overline{\phi}^{-1}  B_i\Gamma(i)\mathbf{V}$. By Assumption \ref{assum:blockrand}.a, $||\Gamma(i)||_1 \lesssim 1$. By Lemma \ref{lem:phi} $\max_i||B_i||_1\leq \max_i\phi(i,\kappa_n) \lesssim \overline{\phi}$ and $\max_i||B_i||_\infty =1$. By construction, the sub-matrix of $\mathbb{E}[\mathbf{V}\mathbf{V}']\Gamma(i)$  corresponding to the support of $B_i$ is the identity matrix. Putting these facts together: $\mathbb{E}\left[\widetilde{X}_i^2\right] = \overline{\phi}^{-2}\mathbb{E}\left[B_i\Gamma(i)\mathbf{V}\mathbf{V}'\Gamma(i)B_i'\right] =\overline{\phi}^{-2}B_i\Gamma(i)B_i' \lesssim \overline{\phi}^{-2}||\Gamma(i)||_1 ||B_i||_1||B_i||_\infty \lesssim \overline{\phi}^{-1}$.

To prove claim 2, recall that  ${X}_i =  \overline{\phi}^{-1}   B_i\mathbf{V}$. Taking the expectation accounting for covariances and using Assumption \ref{assum:blockrand}: $\mathbb{E}\left[{X}_i^2\right] = \overline{\phi}^{-2}\mathbb{E}\left[B_i \mathbf{V}\mathbf{V}'B_i'\right]= \overline{\phi}^{-2} \mathbf{1}'\Omega(i)\mathbf{1}\leq \overline{\phi}^{-2}||\Omega(i)||_1\overline{\phi}\lesssim \overline{\phi}^{-1}$.

To prove claim 3, notice that by Assumption \ref{assum:blockrand}.b, each $\widetilde{X}_i$ is uncorrelated with $\widetilde{X}_j$ for units $j$ farther than $r_n\sim \kappa_n+g_n$ away from unit $i$. So each  $\widetilde{X}_i$ is correlated with at most $\mathcal{O}\left(\kappa_n^d+g_n^d\right)$ other $\widetilde{X}_j$  (and the same for each $X_i$). Moreover $\mathbb{E}[X_i]=0$ because $\mathbb{E}[V_c]=0$ for all $c$. So $\mathbb{E}\left[\overline{\widetilde{X}}^2\right]=\mathbb{V}\left[\overline{\widetilde{X}}\right] \lesssim \frac{\kappa_n^d+g_n^d}{n}\max_i\mathbb{V}[\widetilde{X}_i]\lesssim {\frac{\kappa_n^d+g_n^d}{\overline{\phi}n}}$. An identical argument yields the next claim as well.

Now for the last claim. Since Assumption \ref{assum:blockrand}.a guarantees that $\max_i||\Gamma(i)||_1\lesssim 1$, $\max_i|\widetilde{X}_i|\lesssim \frac{\phi(i,\kappa_n)}{\overline{\phi}}$. The same is true for $X_i$ by construction. Now using the first two claims:\begin{align*}
    \max_{i\in \mathcal{N}_n}\mathbb{V}\left[X_i\widetilde{X}_i\right] &\leq  \max_{i\in \mathcal{N}_n}\mathbb{E}\left[X_i^4\right]+\mathbb{E}\left[\widetilde{X}_i^4\right]\leq  \max_{i\in \mathcal{N}_n} \frac{\phi(i,\kappa_n)^2}{\overline{\phi}^2}\left(\mathbb{E}\left[X_i^2\right]+\mathbb{E}\left[\widetilde{X}_i^2\right]\right)\lesssim \overline{\phi}^{-1}
\end{align*}

Since each $X_i\widetilde{X}_i$ is correlated with at most $\mathcal{O}\left(\kappa_n^d+g_n^d\right)$ other  $X_j\widetilde{X}_j$ by Assumption \ref{assum:blockrand}.b, we can bound the variance of the mean: $\mathbb{V}\left[\frac{\overline{\phi}}{n} \sum_{i=1}^n X_i\widetilde{X}_i\right]\lesssim  \overline{\phi}^2\frac{\kappa_n^d+g_n^d}{n} \max_{i\in \mathcal{N}_n}\mathbb{V}\left[X_i\widetilde{X}_i\right]\lesssim  \overline{\phi}\frac{\kappa_n^{d}+g_n^{d}}{n}\lesssim \frac{\kappa_n^{2d}+g_n^{2d}}{ng_n^d}$.

\subsection{Proof of Proposition \ref{prop:ols_decomp_correlated_designs}}\label{proof:eq:stratified_decomp}  

These arguments mostly follow the proof of Proposition \ref{prop:ols_asymptotic_decomp}. Starting with the estimator:
$$ \widehat{\theta}_{n,\kappa_n}^{STRAT} \equiv \frac{\frac{1}{n}\sum_{i=1}^n Y_i\widetilde{X}_i - \left(\frac{1}{n}\sum_{i=1}^n Y_i\right)\left(\frac{1}{n}\sum_{i=1}^n \widetilde{X}_i\right) }{\frac{1}{n}\sum_{i=1}^n X_i\widetilde{X}_i - \left(\frac{1}{n}\sum_{i=1}^n X_i\right)\left(\frac{1}{n}\sum_{i=1}^n \widetilde{X}_i\right)} = \frac{\overline{\phi}\left(\frac{1}{n}\sum_{i=1}^n Y_i\widetilde{X}_i - \left(\frac{1}{n}\sum_{i=1}^n Y_i\right)\left(\frac{1}{n}\sum_{i=1}^n \widetilde{X}_i\right)\right) }{\overline{\phi}\left(\frac{1}{n}\sum_{i=1}^n X_i\widetilde{X}_i - \left(\frac{1}{n}\sum_{i=1}^n X_i\right)\left(\frac{1}{n}\sum_{i=1}^n \widetilde{X}_i\right)\right)}$$

First we address the  denominator and then the numerator. By Lemma \ref{lem:control_x_correlated_designs}, $\left(\frac{1}{n}\sum_{i=1}^n X_i\right)^2,\left(\frac{1}{n}\sum_{i=1}^n \widetilde{X}_i\right)^2 \lesssim \frac{\kappa_n^d+g_n^d}{\overline{\phi}n} $. So $\overline{\phi}\left(\frac{1}{n}\sum_{i=1}^n X_i\widetilde{X}_i - \left(\frac{1}{n}\sum_{i=1}^n X_i\right)\left(\frac{1}{n}\sum_{i=1}^n \widetilde{X}_i\right)\right)= \frac{\overline{\phi}}{n}\sum_{i=1}^n X_i\widetilde{X}_i + \mathcal{O}_p\left(\frac{\kappa_n^d+g_n^d}{n}\right)$. Next we simplify the numerator below: 
\begin{align}
  \frac{ \overline{\phi}}{n}\sum_{i=1}^n Y_i\widetilde{X}_i -  \left(\frac{1}{n}\sum_{i=1}^n Y_i\right)\left(\frac{\overline{\phi}}{n}\sum_{i=1}^n \widetilde{X}_i\right) &= \frac{\overline{\phi}}{n}\sum_{i=1}^n \left(Y_i- \frac{1}{n}\sum_{j=1}^n Y_j\right)\left(\widetilde{X}_i-\frac{1}{n}\sum_{j=1}^n \widetilde{X}_j\right) \\
    &=\frac{\overline{\phi}}{n}\sum_{i=1}^n \left(Y_i- \frac{1}{n}\sum_{j=1}^n \mathbb{E}\left[Y_j\right]\right)\left(\widetilde{X}_i-\frac{1}{n}\sum_{j=1}^n \widetilde{X}_j\right) \\
    &= \frac{\overline{\phi}}{n}\sum_{i=1}^n \widetilde{X}_i\left(Y_i- \frac{1}{n}\sum_{j=1}^n \mathbb{E}\left[Y_j\right]\right) +o_p\left(\sqrt{\overline{\phi}\frac{\kappa_n^d+g_n^d}{n}}\right)\label{eq:strat_needs_lem}
\end{align}
\noindent Equation (\ref{eq:strat_needs_lem}) above holds because by Lemma \ref{lem:ybar} (whose proof is unchanged when the cluster treatments are possibly dependent but satisfy Assumption \ref{assum:blockrand}) and then Lemma \ref{lem:control_x_correlated_designs}: $$\frac{\overline{\phi}}{n}\sum_{i=1}^n \widetilde{X}_i \left(\frac{1}{n}\sum_{j=1}^nY_j- \frac{1}{n}\sum_{j=1}^n \mathbb{E}\left[Y_j\right]\right) =o_p\left(\frac{\overline{\phi}}{n}\sum_{i=1}^n \widetilde{X}_i \right)=o_p\left(\sqrt{\overline{\phi}\frac{\kappa_n^d+g_n^d}{n}}\right)$$ 

By Lemma \ref{lem:phi}: $\sqrt{\overline{\phi}\frac{\kappa_n^d+g_n^d}{n}} \sim \frac{\kappa_n^d+g_n^d}{g_n^{d/2}n^{1/2}}$.  Putting the numerator and denominator together:

$$ \widehat{\theta}_{n,\kappa_n}^{STRAT}  = \frac{\frac{\overline{\phi}}{n}\sum_{i=1}^n \widetilde{X}_i\left(Y_i- \frac{1}{n}\sum_{j=1}^n \mathbb{E}\left[Y_j\right]\right)  +o_p\left(\frac{\kappa_n^d+g_n^d}{g_n^{d/2}n^{1/2}}\right)}{\frac{\overline{\phi}}{n}\sum_{i=1}^n X_i\widetilde{X}_i+\mathcal{O}_p\left(\frac{\kappa_n^d+g_n^d}{n}\right)}$$

 Lemma  \ref{lem:control_x_sharp} and Equation (\ref{eq:wald_denominator}) guarantee that 
$\frac{\overline{\phi}}{n}\sum_{i=1}^n X_i\widetilde{X}_i\to_p 1$. Since $g_n^d\to 0$, we have that $\frac{\kappa_n^d+g_n^d}{n} = o\left(\frac{\kappa_n^d+g_n^d}{g_n^{d/2}n^{1/2}}\right)$. Therefore:
$$ \widehat{\theta}_{n,\kappa_n}^{STRAT}  = \frac{\frac{\overline{\phi}}{n}\sum_{i=1}^n \widetilde{X}_i\left(Y_i- \frac{1}{n}\sum_{j=1}^n \mathbb{E}\left[Y_j\right]\right)  +o_p\left(\frac{\kappa_n^d+g_n^d}{g_n^{d/2}n^{1/2}}\right)}{\frac{\overline{\phi}}{n}\sum_{i=1}^n X_i\widetilde{X}_i}$$

By the same arguments as Proposition \ref{prop:ols_asymptotic_decomp}, we can use Lemma \ref{lem:ols_decomp_generic} and then Equation (\ref{eq:bias_wald}) to conclude:
\begin{equation}\label{eq:strat_readyforlongrage}
     \widehat{\theta}_{n,\kappa_n}^{STRAT}-\theta_n =     \frac{\overline{\phi}}{n}\sum_{i=1}^n \widetilde{X}_i\left(Y_i- \frac{1}{n}\sum_{j=1}^n \mathbb{E}[Y_j] -X_i\bar{\theta}_{n,\kappa_n}^{STRAT}\right)\quad+\quad\bar{\theta}_{n,\kappa_n}^{STRAT}-\theta_n  +o_p\left(\frac{\kappa_n^{d}+g_n^d}{g_n^{d/2}n^{1/2}}\right)
\end{equation}

{\bf Long-Range Spillovers:} Showing that we can replace $Y_i$ with $\mathbb{E}[Y_i|\mathcal{F}_{i,\kappa_n}]$ in (\ref{eq:strat_readyforlongrage}), i.e.  that the long-range spillovers contribute only negligibly to the variance, is involved. In the arguments below, we will show that: 
\begin{equation}
    \frac{\overline{\phi}}{n}\sum_{i=1}^n \widetilde{X}_i(Y_i-\mathbb{E}[Y_i|\mathcal{F}_{i,r_n}]) =o_p\left(\frac{\kappa_n^{d}+g_n^d}{g_n^{d/2}n^{1/2}}\right)
\end{equation}

First we find the expectation. By Assumption \ref{assum:blockrand}, $\widetilde{X}_i$ depends only on treatments within distance $r_n \lesssim \kappa_n+g_n$ of unit $i$. By the law of iterated expectations:
\begin{align*}
    \mathbb{E}\left[\frac{\overline{\phi}}{n}\sum_{i=1}^n \widetilde{X}_i(Y_i-\mathbb{E}[Y_i|\mathcal{F}_{i,r_n}])\right]=0
\end{align*}

Now we control the variance of the sum:
\begin{align*}
    \mathbb{V}\left[\frac{\overline{\phi}}{n}\sum_{i=1}^n \widetilde{X}_i(Y_i-\mathbb{E}[Y_i|\mathcal{F}_{i,r_n}])\right] &= \frac{\overline{\phi}^2}{n^2}\sum_{i=1}^n\sum_{j:\rho(i,j)>r_n}^n\mathbb{E}\left[ \widetilde{X}_i\widetilde{X}_j(Y_i-\mathbb{E}[Y_i|\mathcal{F}_{i,r_n}])(Y_j-\mathbb{E}[Y_j|\mathcal{F}_{j,r_n}])\right]\\
    &+\frac{\overline{\phi}^2}{n^2}\sum_{i=1}^n\sum_{j:\rho(i,j)\leq r_n}^n\mathbb{E}\left[ \widetilde{X}_i\widetilde{X}_j(Y_i-\mathbb{E}[Y_i|\mathcal{F}_{i,r_n}])(Y_j-\mathbb{E}[Y_j|\mathcal{F}_{j,r_n}])\right]
\end{align*}
 We now  bound the sum of covariances over the near-pairs $\rho(i,j)\leq r_n$. First, we use Assumption \ref{assum:ANI} to bound $|Y_i-\mathbb{E}[Y_i|\mathcal{F}_{i,r_n}]|$, then use the above fact that $\mathbb{V}\left[\overline{\phi}\widetilde{X}_i\right]\lesssim \overline{\phi}$, then  use Lemma \ref{lem:phi}, and finally use the fact that $r_n \lesssim \kappa_n+g_n$ but $r_n\to \infty$:
\begin{align*}
    \frac{\overline{\phi}^2}{n^2}\sum_{i=1}^n\sum_{\rho(i,j)\leq r_n}^n\mathbb{E}\left[ \widetilde{X}_i\widetilde{X}_j(Y_i-\mathbb{E}[Y_i|\mathcal{F}_{i,r_n}])(Y_j-\mathbb{E}[Y_j|\mathcal{F}_{j,r_n}])\right] &\leq \frac{r_n^d}{n}r_n^{-2\gamma}\overline{\phi}^2\mathbb{V}[X_i]\\
    &\lesssim \frac{r_n^d}{n}r_n^{-2\gamma} \overline{\phi} =o\left(\frac{\kappa_n^{2d}+g_n^{2d}}{g_n^{d}n}\right)
\end{align*}

Now we bound the covariances over the ``far" pairs $i,j$ where $\rho(i,j)>r_n$. Adding and subtracting:
\begin{align*}
   \label{eq:expecfactorizes} \mathbb{E}\left[ \widetilde{X}_i\widetilde{X}_j(Y_i-\mathbb{E}[Y_i|\mathcal{F}_{i,r_n}])(Y_j-\mathbb{E}[Y_j|\mathcal{F}_{j,r_n}])\right] &= \mathbb{E}\left[ \widetilde{X}_i\widetilde{X}_j(Y_i-\mathbb{E}[Y_i|\mathcal{F}_{i,r_n},\mathcal{F}_{j,r_n}])(Y_j-\mathbb{E}[Y_j|\mathcal{F}_{j,r_n}])\right]\\
    &+\mathbb{E}\left[ \widetilde{X}_i\widetilde{X}_j(\mathbb{E}[Y_i|\mathcal{F}_{i,r_n},\mathcal{F}_{j,r_n}]-\mathbb{E}[Y_i|\mathcal{F}_{i,r_n}])(Y_j-\mathbb{E}[Y_j|\mathcal{F}_{j,r_n}])\right]
\end{align*}

By Assumption \ref{assum:blockrand} and the Scaling Clusters design, if $\rho(i,j) > r_n$, then the set of clusters influencing $\widetilde{X}_i$ and $\widetilde{X}_j$ are disjoint and independent, so $\widetilde{X}_i \independent \widetilde{X}_j$. By Assumption \ref{assum:linearity}: $$Y_i-\mathbb{E}[Y_i|\mathcal{F}_{i,r_n},\mathcal{F}_{j,r_n}]= \sum_{c=1}^{m_n}(AC)_{ic}V_c\mathbf{1}\left\{\mathcal{P}_c\cap\left(\mathcal{N}(i,r_n)\cup \mathcal{N}(j,r_n)\right) = \emptyset\right\}$$

So $Y_i-\mathbb{E}[Y_i|\mathcal{F}_{i,r_n},\mathcal{F}_{j,r_n}]$ depends only on treatments farther than $r_n$ away from $i$ and $j$.  So  $\widetilde{X}_j$ is independent of $\widetilde{X}_i$, $(Y_i-\mathbb{E}[Y_i|\mathcal{F}_{i,r_n},\mathcal{F}_{j,r_n}])$, and $(Y_j-\mathbb{E}[Y_j|\mathcal{F}_{j,r_n}])$ and has expectation zero.  So $\mathbb{E}\left[ \widetilde{X}_i\widetilde{X}_j(Y_i-\mathbb{E}[Y_i|\mathcal{F}_{i,r_n}])(Y_j-\mathbb{E}[Y_j|\mathcal{F}_{j,r_n}])\right] =0$. This leaves us with:
\begin{align*}
      \mathbb{E}\left[ \widetilde{X}_i\widetilde{X}_j(Y_i-\mathbb{E}[Y_i|\mathcal{F}_{i,r_n}])(Y_j-\mathbb{E}[Y_j|\mathcal{F}_{j,r_n}])\right] &= \mathbb{E}\left[ \widetilde{X}_i\widetilde{X}_j(\mathbb{E}[Y_i|\mathcal{F}_{i,r_n},\mathcal{F}_{j,r_n}]-\mathbb{E}[Y_i|\mathcal{F}_{i,r_n}])(Y_j-\mathbb{E}[Y_j|\mathcal{F}_{j,r_n}])\right]
\end{align*}

By an identical argument for the $Y_j$:
\begin{align*}
      &\mathbb{E}\left[ \widetilde{X}_i\widetilde{X}_j(Y_i-\mathbb{E}[Y_i|\mathcal{F}_{i,r_n}])(Y_j-\mathbb{E}[Y_j|\mathcal{F}_{j,r_n}])\right]\\
      &= \mathbb{E}\left[ \widetilde{X}_i\widetilde{X}_j(\mathbb{E}[Y_i|\mathcal{F}_{i,r_n},\mathcal{F}_{j,r_n}]-\mathbb{E}[Y_i|\mathcal{F}_{i,r_n}])(\mathbb{E}[Y_j|\mathcal{F}_{i,r_n},\mathcal{F}_{j,r_n}]-\mathbb{E}[Y_j|\mathcal{F}_{j,r_n}])\right]\\
      &= \mathbb{E}\left[ \widetilde{X}_i(\mathbb{E}[Y_j|\mathcal{F}_{i,r_n},\mathcal{F}_{j,r_n}]-\mathbb{E}[Y_j|\mathcal{F}_{j,r_n}])\right]\mathbb{E}\left[ \widetilde{X}_j(\mathbb{E}[Y_i|\mathcal{F}_{i,r_n},\mathcal{F}_{j,r_n}]-\mathbb{E}[Y_i|\mathcal{F}_{i,r_n}])\right]\\
      &=  \overline{\phi}^{-2} \left(\sum_{k\in \mathcal{N}(i,r_n)}A_{jk}\right)\left(\sum_{k\in \mathcal{N}(j,r_n)}A_{ik}\right)
\end{align*}

So we have:
\begin{align*}
     &\frac{\overline{\phi}^2}{n^2}\sum_{i=1}^n\sum_{\rho(i,j)>r_n}^n\mathbb{E}\left[ \widetilde{X}_i\widetilde{X}_j(Y_i-\mathbb{E}[Y_i|\mathcal{F}_{i,r_n}])(Y_j-\mathbb{E}[Y_j|\mathcal{F}_{j,r_n}])\right]\\
     &=\frac{1}{n^2 }\sum_{i=1}^n\sum_{j:\rho(i,j)>r_n}^n\left(\sum_{k\in \mathcal{N}(i,r_n)}A_{jk}\right)\left(\sum_{k\in \mathcal{N}(j,r_n)}A_{ik}\right)\\
      &\lesssim \frac{r_n^{-\gamma}}{n^2 }\sum_{i=1}^n\sum_{j:\rho(i,j)>r_n}^n\left(\sum_{k\in \mathcal{N}(j,r_n)}A_{ik}\right)
    \end{align*}

The last sum can be bounded using the fact that $\sum_{j:\rho(i,j)>r_n}^n\left(\sum_{k\in \mathcal{N}(j,r_n)}A_{ik}\right)$ adds up each element $|A_{ik}|$ at most $\max_k| \mathcal{N}(k,r_n)|\lesssim r_n^d$ times and $\sum_k |A_{ik}|\leq \overline{Y}$ by Assumption \ref{assum:boundedoutcomes}:  
\begin{align*}
     \frac{r_n^{-\gamma}}{n^2 }\sum_{i=1}^n\sum_{j:\rho(i,j)>r_n}^n\left(\sum_{k\in \mathcal{N}(j,r_n)}A_{ik}\right)  &\lesssim  \frac{r_n^{-\gamma}}{n }r_n^d=o\left(\frac{\kappa_n^d+g_n^d}{n}\right)
\end{align*}

Putting the covariance bounds on the sums over near and far pairs together: 
$$\mathbb{V}\left[ \frac{\overline{\phi}}{n}\sum_{i=1}^n \widetilde{X}_i(Y_i-\mathbb{E}[Y_i|\mathcal{F}_{i,r_n}])\right] =o\left(\frac{\kappa_n^{2d}+g_n^{2d}}{g_n^{d}n}\right) $$

And therefore by Chebyshev's Inequality:
$$ \frac{\overline{\phi}}{n}\sum_{i=1}^n \widetilde{X}_i(Y_i-\mathbb{E}[Y_i|\mathcal{F}_{i,r_n}]) = o_p\left(\frac{\kappa_n^d+g_n^d}{g_n^{d/2}n^{1/2}}\right)$$

So when we exchange the $Y_i$ for $\mathbb{E}[Y_i|\mathcal{F}_{i,r_n}]$ the resulting error is small: 
$$  \widehat{\theta}_{n,\kappa_n}^{STRAT}-\theta_n =     \frac{\overline{\phi}}{n}\sum_{i=1}^n \widetilde{X}_i\left(\mathbb{E}[Y_i|\mathcal{F}_{i,r_n}]- \frac{1}{n}\sum_{j=1}^n \mathbb{E}[Y_j] -X_i\bar{\theta}_{n,\kappa_n}^{STRAT}\right)\quad+\quad\bar{\theta}_{n,\kappa_n}^{STRAT}-\theta_n  +o_p\left(\frac{\kappa_n^{d}+g_n^d}{g_n^{d/2}n^{1/2}}\right)   $$

\subsection{Proof of Theorem \ref{thm:ols_strat}}\label{proof:thm:ols_strat} 

Taking the ratio of Equations (\ref{eq:wald_denominator}) and (\ref{eq:wald_numerator}) proven above and using the fact that $\theta_n = \frac{1}{n}\sum_{i=1}^n\sum_{j=1}^nA_{ij}$, Equation (\ref{eq:bias_wald}) is immediate:

\begin{align*}
       \frac{\frac{1}{n}\sum_{i=1}^n \mathbb{E}\left[Y_i\widetilde{X}_i \right] }{ \frac{1}{n}\sum_{i=1}^n\mathbb{E}\left[X_i\widetilde{X}_i \right]}&=\frac{1}{n}\sum_{i=1}^n\sum_{j=1}^n \widetilde{A}_{ij}\\
          \frac{\frac{1}{n}\sum_{i=1}^n \mathbb{E}\left[Y_i\widetilde{X}_i \right] }{ \frac{1}{n}\sum_{i=1}^n\mathbb{E}\left[X_i\widetilde{X}_i \right]} -\theta_n = [\mathcal{B}]_{\text{STRAT}} &= \frac{1}{n}\sum_{i=1}^n\sum_{j=1}^n (\widetilde{A}_{ij}-A_{ij})\lesssim \kappa_n^{-\gamma}
\end{align*}

Substituting into Proposition \ref{prop:ols_decomp_correlated_designs}:
\begin{align*}
        \widehat{\theta}_{n,\kappa_n}^{STRAT}-\theta_n  
    =  [\mathcal{A}]_{\text{STRAT}}+[\mathcal{B}]_{\text{STRAT}} +o_p\left(  \frac{\kappa_n^{d}+g_n^d}{g_n^{d/2}n^{1/2}}\right) 
\end{align*}

We already showed that $[\mathcal{B}]_{\text{STRAT}}\lesssim \kappa_n^{-\gamma}$, so we need only verify that the variance of $[\mathcal{A}]_{\text{STRAT}}$ can be bounded in the same way as $[\mathcal{A}]$. By Assumption  \ref{assum:blockrand}.b, the maximum degree of the dependency graph of $\widetilde{X}_i$ is no larger than $r_n^d \sim \kappa_n^d+g_n^d$ (just like $X_i$ with independent clusters).

By Lemma \ref{lem:control_x_correlated_designs} and then  Lemma \ref{lem:phi}: $\max_i\mathbb{V}\left[\overline{\phi}\widetilde{X}_i\right]\lesssim \overline{\phi}\lesssim \frac{\kappa_n^d+g_n^d}{g_n^d}$. Each summand of $[\mathcal{A}]_{\text{STRAT}}$ has the form $\widetilde{X}_iZ_i$ with $Z_i$ bounded, so the maximum variance over all summands is  $\mathcal{O}\left(\frac{\kappa_n^d+g_n^d}{g_n^d}\right)$. By Assumption \ref{assum:blockrand}.b, each summand of $[\mathcal{A}]_{\text{STRAT}}$  is correlated with at most $\max_i|\mathcal{N}(i,r_n)|\lesssim \kappa_n^d+g_n^d$ other summands. So  $[\mathcal{A}]_{\text{STRAT}} \lesssim \frac{\kappa_n^{d}+g_n^{d}}{g_n^{d/2}n^{1/2}}$ and the theorem is proven. 

\subsection{Proof of Theorem \ref{thm:clt_stratified}}\label{proof:thm:clt_stratified}

To show that  $[\mathcal{A}]_{\text{STRAT}} $ is asymptotically normal, we use a very similar argument to Theorem \ref{thm:clt}. Notice that the degree of the dependency graph of $\widetilde{X}_i$ is not of a larger order than $X_i$ by Assumption \ref{assum:blockrand}.b. We now need only bound the moments of the summands of  $[\mathcal{A}]_{\text{STRAT}} $ . Let $\widetilde{U}_i\equiv \frac{\overline{\phi}}{n}\widetilde{X}_i\left(\mathbb{E}[Y_i|\mathcal{F}_{i,r_n}]- \frac{1}{n}\sum_{j=1}^n \mathbb{E}[Y_j] -X_i\bar{\theta}_{n,\kappa_n}\right)$ be the summands of $[\mathcal{A}]_{\text{STRAT}}$. Define $\sigma_{n}^2 \equiv \mathbb{V}\left[[\mathcal{A}]_{\text{STRAT}}\right]$. Define $\Psi_n$ to be the maximum degree of the dependency graph among the summands $\widetilde{U}_i$. The dependency graph has degree $\mathcal{O}\left(\kappa_n^d+g_n^d\right)$ because the spillovers in the matrix $\widetilde{A}$ are all local and Assumption \ref{assum:blockrand}.b allows clusters to be dependent only within a ball. Now we will use a loose but sufficient bound on the third and fourth moments. By Lemma \ref{lem:phi}: $|\widetilde{U}_i| \lesssim |\overline{\phi}\widetilde{X}_i/n| $. We can uniformly bound $\overline{\phi}\widetilde{X}_i$ by using  Assumption \ref{assum:blockrand}.a: $\overline{\phi}\widetilde{X}_i =\mu_2\mathbf{1}'B_i\Gamma(i)\mathbf{V} \leq  ||B_i||_1||\Gamma(i)||_1||\mathbf{V}||_\infty$. So $\max_i|\overline{\phi}\widetilde{X}_i|\lesssim \overline{\phi}$. Moreover, we already showed in Section \ref{proof:thm:ols_strat} that $\mathbb{E}\left[(\overline{\phi}\widetilde{X}_i)^2\right]\lesssim \overline{\phi}$. Therefore:
\begin{align*}
    \max_{i\in \mathcal{N}_n}\mathbb{E}[|\widetilde{U}_i|^3]&\leq \max_{i\in \mathcal{N}_n}\mathbb{E}[|\overline{\phi}\widetilde{X}_i|^3]n^{-3}\overline{Y}^3 \lesssim n^{-3}\max_{i\in \mathcal{N}_n}\mathbb{E}[|\overline{\phi}\widetilde{X}_i|^2]\max_{j\in \mathcal{N}_n}|\overline{\phi}\widetilde{X}_j|\lesssim n^{-3}\overline{\phi}^2\lesssim \frac{\kappa_n^{2d}+g_n^{2d}}{g_n^{2d}n^{3}}
\end{align*}

Similarly: $\max_i\mathbb{E}[|\widetilde{U}_i|^4]\lesssim \frac{\kappa_n^{3d}+g_n^{3d}}{g_n^{3d} n^4}$. 

So we can use   Theorem 3.6 from \cite{RossSteins} to bound the Wasserstein distance $d\left([\mathcal{A}]_{{STRAT}}/\sigma_{n},\mathcal{Z}\right) $ between a standardized version of the distribution of $[\mathcal{A}]_{\text{STRAT}}$ and the standard normal distribution. Using the assumption that $\liminf_{n\to \infty}\sigma_n \frac{g_n^{d/2}n^{1/2}}{\kappa_n^{d}+g_n^{d}} >0 $:
\begin{align*}
    d\left([\mathcal{A}]_{\text{STRAT}}/\sigma_{n},\mathcal{Z}\right) &\leq \frac{\Psi_n^2}{\sigma_{n}^3}\sum_{i=1}^n \mathbb{E}[|\widetilde{U}_i|^3] +\frac{\sqrt{28}\Psi_n^{3/2}}{\sqrt{\pi}\sigma_{n}^2}\sqrt{\sum_{i=1}^n \mathbb{E}[\widetilde{U}_i^4]}
   \\&
   \lesssim \frac{\kappa_n^{2d}+g_n^{2d}}{\sigma_n^3}\frac{\kappa_n^{2d}+g_n^{2d}}{n^2g_n^{2d}}+\frac{\kappa_n^{(3/2)d}+g_n^{(3/2)d}}{\sigma_n^{2}}\frac{\kappa_n^{(3/2)d}+g_n^{(3/2)d}}{g_n^{(3/2)d}n^{3/2}} \lesssim \frac{\kappa_n^{d}+g_n^d}{n^{1/2}g_n^{d/2}}+\frac{\kappa_n^{d}+g_n^{d}}{n^{1/2}g_n^{d/2}} \to 0
\end{align*}

\subsection{Proof of Equation (\ref{eq:Omega_example_randomsat})}\label{proof:example:ehmnw_design}

For each saturation group consisting of $m$ clusters, let $p_1,p_2 \in (0,1)$ be numbers such that $p_1m$ and $p_2m$ are integers. With probability $\frac{1}{2}$ treat exactly fraction $q=p_1$ of the clusters in the saturation group and with probability $\frac{1}{2}$ treat instead exactly fraction $q=p_2$ of the clusters in the saturation group.  Thus, conditional on the random saturation level $q$, each grouping treats exactly $K=qm$ clusters at random. For distinct clusters $c,d$ in the same grouping, the covariance of the treatment indicators is:
\begin{align*}
   \mathbb{E}[W_c|K] &= q=\frac{K}{m}\\
   \text{Cov}(W_c,W_d|q) &= \frac{\binom{K}{2}}{\binom{m}{2}} -\frac{K^2}{m^2} = \frac{K(K-1)}{m(m-1)} -\frac{K^2}{m^2}=\frac{K}{m}\left(\frac{K-1}{m-1}-\frac{K}{m}\right)\\
   &= \frac{K}{m}\left( K-1-K\frac{m-1}{m} \right)\frac{1}{m-1}=-\frac{K}{m}\left( 1-\frac{K}{m}\right)\frac{1}{m-1}\\
   &=  -\frac{q(1-q)}{m-1}
\end{align*}

By the law of total covariance:
\begin{align*}
    \text{Cov}(W_c,W_d) &= \mathbb{E}\left[ Cov(W_c,W_d|K)\right]+  \text{Cov}\left(\mathbb{E}[W_c|K],\mathbb{E}[W_d|K]\right)\\
  \mathbb{E}\left[ Cov(W_c,W_d|K)\right]  &=  -\frac{p_1(1-p_1) +p_2(1-p_2)}{2(m-1)}\\
   \text{Cov}\left(\mathbb{E}[W_c|K],\mathbb{E}[W_d|K]\right) &= \mathbb{E}[q^2]-\mathbb{E}[q]^2\\
   &= \frac{2p_1^2+2p_2^2}{4} - \frac{(p_1+p_2)^2}{4}=\frac{p_1^2+p_2^2-2p_1p_2}{4} = \frac{(p_1-p_2)^2}{4}\\
     \text{Cov}(W_c,W_d) &= -\frac{p_1(1-p_1) +p_2(1-p_2)}{2(m-1)} +\frac{(p_1-p_2)^2}{4}
\end{align*}
In cases like EHMNW, $p_1 = \lceil \frac{m}{3}\rceil / m$ and $p_2 =  \lceil \frac{2m}{3}\rceil / m$ and the saturation groups contain 3-24 clusters. If $m$ is divisible by three, this reduces to:  $\frac{(m-9)}{36(m-1)}$. 

So we can write the matrix $\Omega(i) = \left(\frac{1}{4}-f(m)\right)I+f(m)J$ where $I$ is the identity matrix and $J$ is zero except on an $m\times m$ submatrix consisting of all ones and $f(m) = -\frac{\lceil \frac{m}{3}\rceil(1-\lceil \frac{m}{3}\rceil) + \lceil \frac{2m}{3}\rceil(1- \lceil \frac{2m}{3}\rceil)}{2(m-1)} +\frac{(\lceil \frac{m}{3}\rceil-\lceil \frac{2m}{3}\rceil)^2}{4}$. $\Omega(i)$ must be invertible because no subset of treatments are linearly dependent. This is because the fraction of treated units in any locally proximate set has nontrivial variance.

\subsection{Proof of Proposition \ref{prop:ipw_asymptotic_decomp} }\label{proof:prop:ipw_asymptotic_decomp} 

First we show  a well-known result: that the Hajek IPW estimator is asymptotically equivalent to the residualized Horvitz-Thompson estimator. 
\begin{align}
  \frac{1}{n}\sum_{i=1}^n\frac{Y_iT_{i1}}{\mathbb{P}\left[T_{i1}=1\right] \frac{1}{n}\sum_{j=1}^n\frac{T_{j1}}{\mathbb{P}\left[T_{j1}=1\right]}}-\overline{\mu}_1&=\frac{1}{n}\sum_{i=1}^n(Y_i-\overline{\mu}_{1})\frac{T_{i1}}{\mathbb{P}\left[T_{i1}=1\right] \frac{1}{n}\sum_{j=1}^n\frac{T_{j1}}{\mathbb{P}\left[T_{j1}=1\right]}} \\
  &=  \frac{1}{n}\sum_{i=1}^n \left(Y_i-\overline{\mu}_{1} \right)\frac{T_{i1}}{\mathbb{P}\left[T_{i1}=1\right]} \left(\frac{1}{ \frac{1}{n}\sum_{j=1}^n\frac{T_{j1}}{\mathbb{P}\left[T_{j1}=1\right]}}\right)\\
  &=  \frac{1}{n}\sum_{i=1}^n \left(Y_i-\overline{\mu}_{1} \right)\frac{T_{i1}}{\mathbb{P}\left[T_{i1}=1\right]} +o_p\left(\sqrt{\frac{\kappa_n^d}{n}}\right) \label{eq:ipw_lastiline}
\end{align}
To see the last equality (\ref{eq:ipw_lastiline}) above, note the following. Theorem \ref{thm:rateipw} already showed that $\mathbb{P}[T_{i1}=1]$ are uniformly bounded below. Since each term of the sum is correlated with at most $\mathcal{O}\left(\kappa_n^d\right)$ other terms: $ \frac{1}{n}\sum_{j=1}^n\frac{T_{j1}}{\mathbb{P}\left[T_{j1}=1\right]} -1=\mathcal{O}_p\left(\sqrt{\frac{\kappa_n^d}{n}}\right)$. Taking a first-order Taylor Expansion: $\left(\frac{1}{ \frac{1}{n}\sum_{j=1}^n\frac{T_{j1}}{\mathbb{P}\left[T_{j1}=1\right]}}-1\right) = \mathcal{O}_p\left(\sqrt{\frac{\kappa_n^d}{n}}\right)$. The proof of Theorem \ref{thm:rateipw} already showed that the Horvitz-Thompson sums are consistent: $ \frac{1}{n}\sum_{i=1}^n \left(Y_i-\overline{\mu}_{1} \right)\frac{T_{i1}}{\mathbb{P}\left[T_{i1}=1\right]}=o_p(1)$. So: $$\frac{1}{n}\sum_{i=1}^n \left(Y_i-\overline{\mu}_{1} \right)\frac{T_{i1}}{\mathbb{P}\left[T_{i1}=1\right]} \left(\frac{1}{ \frac{1}{n}\sum_{j=1}^n\frac{T_{j1}}{\mathbb{P}\left[T_{j1}=1\right]}}-1\right) = o_p\left(\sqrt{\frac{\kappa_n^d}{n}}\right)$$

This proves (\ref{eq:ipw_lastiline}). Next we will use Assumption \ref{assum:ANI} to show that replacing $Y_i$ with its conditional expectation incurs only a dominated error. 
\begin{align*}
     \frac{1}{n}\sum_{i=1}^n\frac{ \left(Y_i-\overline{\mu}_{1} \right)T_{i1}}{\mathbb{P}\left[T_{i1}=1\right]} &= \frac{1}{n}\sum_{i=1}^n \left(\mathbb{E}\left[Y_i|T_{i1}=1\right]-\overline{\mu}_{1} \right)\frac{T_{i1}}{\mathbb{P}\left[T_{i1}=1\right]}+\frac{1}{n}\sum_{i=1}^n \left(Y_i-\mathbb{E}\left[Y_i|T_{i1}=1\right] \right)\frac{T_{i1}}{\mathbb{P}\left[T_{i1}=1\right]}
\end{align*}

Our next task is to bound the second sum: $\frac{1}{n}\sum_{i=1}^n \left(Y_i-\mathbb{E}\left[Y_i|T_{i1}=1\right] \right)\frac{T_{i1}}{\mathbb{P}\left[T_{i1}=1\right]}$.  Let $\mathcal{F}_i^{+}$ denote the $\sigma$-algebra generated by all treatment statuses of all clusters at most distance $\kappa_n^{1+\delta}$ away from unit $i$, where $\delta=\gamma/d$. Adding and subtracting $\frac{1}{n}\sum_{i=1}^n \mathbb{E}\left[Y_i|\mathcal{F}_i^+\right] \frac{T_{i1}}{\mathbb{P}\left[T_{i1}=1\right]}$ yields:
\begin{align*}
    &\frac{1}{n}\sum_{i=1}^n \left(Y_i-\mathbb{E}\left[Y_i|T_{i1}=1\right] \right)\frac{T_{i1}}{\mathbb{P}\left[T_{i1}=1\right]}\\
    &= \frac{1}{n}\sum_{i=1}^n \left(Y_i-\mathbb{E}\left[Y_i|\mathcal{F}_i^+\right] \right)\frac{T_{i1}}{\mathbb{P}\left[T_{i1}=1\right]}+\frac{1}{n}\sum_{i=1}^n \left( \mathbb{E}\left[Y_i|\mathcal{F}_i^+\right] -\mathbb{E}\left[Y_i|T_{i1}=1\right]\right)\frac{T_{i1}}{\mathbb{P}\left[T_{i1}=1\right]}
\end{align*}

By Assumption \ref{assum:ANI}, $\max_i|Y_i - \mathbb{E}[Y_i|\mathcal{F}_{i}^+]|\lesssim \kappa_n^{-\gamma(1+\delta)}$ and therefore the first sum is $\mathcal{O}_p\left(\kappa_n^{-\gamma(1+\delta)}\right)=o_p\left(\kappa_n^{-\gamma}\right)$. For the second sum, all terms have expectation zero and by Assumption \ref{assum:ANI} are almost-surely bounded in absolute value by $\kappa_n^{-\gamma}$. Moreover, all terms are uniformly bounded and correlated with at most $\mathcal{O}\left(\kappa_n^{d(1+\delta)}\right)$ other terms. So, by Chebyshev's Inequality the second term is $\mathcal{O}_p\left(\sqrt{\frac{\kappa_n^{d(1+\delta)-2\gamma}}{n}}\right)=o_p\left(\sqrt{\frac{\kappa_n^{d}}{n}}\right)$. So we have shown that:
\begin{align}
    \frac{1}{n}\sum_{i=1}^n\frac{ \left(Y_i-\overline{\mu}_{1} \right)T_{i1}}{\mathbb{P}\left[T_{i1}=1\right]} &= \frac{1}{n}\sum_{i=1}^n \left(\mathbb{E}\left[Y_i|T_{i1}=1\right]-\overline{\mu}_{1} \right)\frac{T_{i1}}{\mathbb{P}\left[T_{i1}=1\right]}+o_p\left(\sqrt{\frac{\kappa_n^d}{n}}+\kappa_n^{-\gamma}\right)  \label{eq:ipw_longrange}
\end{align}

Combining (\ref{eq:ipw_longrange}) with (\ref{eq:ipw_lastiline}), we have shown that:
$$  \frac{1}{n}\sum_{i=1}^nY_i\frac{T_{i1}}{\mathbb{P}\left[T_{i1}=1\right] \frac{1}{n}\sum_{j=1}^n\frac{T_{j1}}{\mathbb{P}\left[T_{j1}=1\right]}}-\overline{\mu}_1  =\frac{1}{n}\sum_{i=1}^n \left(\mathbb{E}\left[Y_i|T_{i1}=1\right]-\overline{\mu}_{1} \right)\frac{T_{i1}}{\mathbb{P}\left[T_{i1}=1\right]} +o_p\left(\sqrt{\frac{\kappa_n^d}{n}}+\kappa_n^{-\gamma}\right)  $$

Thus, applying an identical argument for $T_{i0}$ yields: $  \widehat{\theta}_{n,\kappa_n}^{IPW} -\theta_n = [\mathcal{A}]_{IPW}+[\mathcal{B}]_{IPW}+o_p\left(\sqrt{\frac{\kappa_n^d}{n}}+\kappa_n^{-\gamma}\right)$.

\subsection{Proof of Theorem \ref{thm:clt_ipw} }\label{proof:thm:clt_ipw} 

We show here the CLT for IPW. The argument for IPW is largely the same as  the proof in \cite{Leung22} but allows $\gamma < d$. Let $Z_i$ be the centered summands of $[\mathcal{A}]_{IPW}$. By the scaling clusters design, the bounded spatial density in Assumption \ref{assum:euclidean_space}.b, and $g_n\sim \kappa_n$, the $Z_i$ have dependency graph with maximum degree $\Psi_n \lesssim \kappa_n^d$. The probability weights are bounded because $\mathbb{P}[T_{i1}=1] = p^{\phi(i,\kappa_n)}$ and $\phi(i,\kappa_n) \lesssim \frac{\kappa_n^d}{g_n^d}\sim 1$ by Lemma \ref{lem:phi}. Since the outcomes are uniformly bounded by Assumption \ref{assum:boundedoutcomes}, the $Z_i$ are uniformly bounded and their moments can be upper bounded: $\mathbb{E}[|Z_i|^3] \lesssim \frac{1}{n^3}$ and $\mathbb{E}[Z_i^4]\lesssim \frac{1}{n^4}$. Define $ \sigma_{n}^2 \equiv \mathbb{V}\left([\mathcal{A}]_{IPW}\right)$. We can use Lemma B.2 from \cite{Leung22} (or Theorem 3.6 from \cite{RossSteins}) to bound the Wasserstein distance $d\left([\mathcal{A}]_{IPW}/\sigma_{n},\mathcal{Z}\right) $ between a standardized version of the distribution of $[\mathcal{A}]_{IPW}$ and the standard normal distribution.  
\begin{align*}
    d\left([\mathcal{A}]_{IPW}/\sigma_{n},\mathcal{Z}\right) &\leq \frac{\Psi_n^2}{\sigma_{n}^3}\sum_{i=1}^n \mathbb{E}[|Z_i|^3] +\frac{\sqrt{28}\Psi_n^{3/2}}{\sqrt{\pi}\sigma_{n}^2}\sqrt{\sum_{i=1}^n \mathbb{E}[Z_i^4]}
    \lesssim \frac{\kappa_n^{2d}}{\sigma_n^3}\frac{1}{n^2}+\frac{\kappa_n^{(3/2)d}}{\sigma_n^2}\frac{1}{n^{3/2}} 
\end{align*}

The right hand side converges to zero whenever $\liminf_{n\to \infty}\frac{\sigma_nn^{2/3}}{\kappa_n^{(2/3)d}}\to \infty$ and $\liminf_{n\to \infty}\frac{\sigma_nn^{3/4}}{\kappa_n^{(3/4)d}}\to \infty $. Since $\kappa_n^d/n\to 0$ by assumption, the first limit implies the second. So the main condition on $\sigma_n$ is:
$ \liminf_{n\to \infty}\frac{\sigma_nn^{2/3}}{\kappa_n^{(2/3)d}}\to \infty$ which is satisfied because we assumed that $\liminf_{n\to \infty}\sigma_n \sqrt{\frac{n}{\kappa_n^d}} >0$ and $\frac{\kappa_n^d}{n}\to 0$.

\subsection{Proof of Theorem \ref{thm:ipw_bias_dominates}}\label{proof:thm:ipw_bias_dominates} 

The assumptions of this theorem match Theorem \ref{thm:smallclusters}, so we can conclude from step 1 of its proof in Section \ref{proof:thm:smallclusters} that in order for IPW to be consistent, $q_2\leq q_1$. Moreover we can also conclude from step 2 that there is a set of potential outcomes satisfying Assumptions \ref{assum:boundedoutcomes}-\ref{assum:linearity} such that $[\mathcal{B}]_{IPW}\sim n^{-q_1\gamma}$. The variance  $\mathbb{V}([\mathcal{A}]_{IPW})$ converges with the inverse of the number of clusters and can be upper-bounded by  $\mathbb{V}([\mathcal{A}]_{IPW}) \lesssim {g_n^d/n}\sim n^{-(1-dq_1)} $. Since by assumption $q_1 < \frac{1}{2\gamma +d}$, then $\gamma q_1 <  (1-dq_1)/2$. So the square root of the variance of IPW must converge faster than the bias.

\end{document}